\documentclass[prc,showpacs,twocolumn,superscriptaddress,nofootinbib]{revtex4-2}

\usepackage{dsfont}
\usepackage{graphicx,amsmath}
\usepackage{hyperref}
\usepackage{xcolor}
\newcommand{\lsim}{\stackrel{\scriptstyle <}{\phantom{}_{\sim}}}
\newcommand{\gsim}{\stackrel{\scriptstyle >}{\phantom{}_{\sim}}}

\def\rmF{{\rm F}}
\def\rmd{{\rm d}}
\def\om{\omega}
\begin{document}
\title{Abnormal Dense and Dilute Nuclear Systems}
\homepage{In 2026 there will be the 100th anniversary of Tsung-Dao Lee and the 115th anniversary of Arkady Migdal. Their ideas sparked extensive theoretical and experimental research programs investigating anomalous dense and dilute nuclear states in the Universe and in heavy-ion collisions. Our review is devoted to developments of this field.}
\author{E. E. Kolomeitsev}
\affiliation{Laboratory of Theoretical Physics, Joint Institute for Nuclear Research, RU-141980 Dubna, Russia}
\affiliation{National Center of Physics and Mathematics, RU-607328 Sarov, Russia}
\affiliation{Matej Bel University, SK-97401 Banska Bystrica, Slovakia}
\author{D. N. Voskresensky}
\affiliation{Laboratory of Theoretical Physics, Joint Institute for Nuclear Research, RU-141980 Dubna, Russia}
\affiliation{National Research Nuclear University (MEPhI), 115409 Moscow, Russia}

%=================================================================%
\date{}
\begin{abstract}
Researchers have long been interested in exotic states of matter. In the early 1970s, Migdal proposed  existence of metastable or stable nuclei containing a pion condensate, and Bodmer posited  existence of collapsed nuclei with a quark core. Lee and Wick proposed that compressed nuclear matter could undergo a phase transition into a scalar condensate state with effectively very light baryons. In the mid-1980s, Witten pointed out that strange quark matter may be an absolutely stable form of matter that can exist in the form of quark  strangelets. These hypotheses stimulated large experimental efforts to detect these states in heavy-ion collisions and in terrestrial samples. Additionally, the cosmos is the arena for a broad search for traces of exotic matter in compact objects and other energetic phenomena. We review various approaches and arguments for a possibility of abnormal states of matter, including metastable and stable dense objects with condensates, as well as dilute nuclear objects. Among these, we discuss scalar condensate, p-wave and s-wave pion condensations in dense nuclear systems and   formation of a $\Delta$ resonance matter. Then we consider a scalar mode condensation triggered by the Pomeranchuk instability  in dilute nuclear matter, and  a clustering. We also mention recent ideas that a relativistic rotation and a dark-matter component may stabilize abnormal nuclear systems. Finally, we list some observational anomalies which  cannot be yet appropriately explained by conventional physics.
\end{abstract}

% \pacs{
%{21.65.-f}{},  % Nuclear matter
%{71.10.Ay}{},  % Fermi liquid theory
%{71.45.-d}{}   % Collective effects
%}
%\keywords{Nuclear matter, Fermi liquid, sound-like excitations, Bose condensation}
\maketitle
%=======================================================================

\tableofcontents

\section{Introduction}

A search for \emph{abnormal} forms of matter accompanied the development of modern physics during the 20th century.
In year 1931 Lev Landau wrote a paper~\cite{Landau:1932uwv}, in which he derived an upper limit of star mass consisting of an ultra-relativistic gas. He  hypothesized that  stars heavier than   $\sim 1.5\, M_\odot$ may possess regions, in which the laws of quantum mechanics (and therefore of quantum statistics) are violated. In such ``unheimliche Sterne'' (\emph{Germ}: eerie stars) as they were dubbed by Landau, see \cite{Yakovlev:2012rd}, electroneutral mixture of protons and electrons, the only particles known by then, would be stable against particle annihilation. The discovery of neutron by James Chadwick in 1932 lead Dmitri Ivanenko to the idea~\cite{Ivanenko1932} that nuclei may consist of protons and neutrons instead of protons and electrons as thought before. In~\cite{HeisenbergNuclTh-1,HeisenbergNuclTh-2,HeisenbergNuclTh-3} Werner Heisenberg introduced the theory of an atomic nucleus consisting of nucleons, --- strongly interacting particles with the putative isospin 1/2, which manifest themselves as protons and neutrons depending of the nucleon isospin projections. In 1934 Walter Baade and Fritz Zwicky \cite{Baade1934} proposed that sufficiently massive stars end their life undergoing a transition (supernova event) into compact object consisting mainly of neutrons and called, therefore, neutron stars.  More than three decades the neutron stars were considered as a theoretical exotica. Only in  1967 Jocelyn Bell and Anthony Hewish reported \cite{Hewish1968} the first observation of a radio pulsar. By now neutron stars are not considered as exotica, and more than 3000 pulsars are observed, presently.

What is the ground state of the world around us has been one of the `big' questions since the beginning of physics and then at the discussion of the role of the  either. If there is some underlying symmetry behind the construction of our world, would it be possible to change the ground state?
In condensed matter physics this concept was proposed by Landau in 1937~\cite{Landau:1937obd-1,Landau:1937obd-2} for the description of phase transitions. He pointed out that any phase transition may be characterized by a change in the value and/or the symmetry structure of some \emph{order parameter}, which can be used as an effective degree of freedom in the thermodynamic potential. This order parameter can have a rather complicated relation to the fundamental degrees of freedom of the considered system. Generalization to the case of an order parameter interacting with the magnetic field was done by Vitaly Ginzburg and Lev Landau in 1950 \cite{Ginzburg1950}.

Addressing the question of a possible transformation of the ground state of our physical world it is natural to suggest a simplest choice  that the order parameter should have quantum number of vacuum $0^+$ (spin and parity), i.e., correspond to some scalar field. There were no scalar particles known in the mid 60s. Nevertheless, Julian Schwinger tried to formulate a theory of strong interactions using the spin 0, spin 1, and spin 1/2 fields~\cite{Schwinger1957}.
The first attempts to build a model of interacting hadrons using a symmetry between vector and axial currents, which enter equally in weak interactions,
were undertaken in works by John Polkinghorn, Murrey Gell-Mann and Maurice L\'evy \cite{Polkinghorne1958,GellMannLevy}. The so-called $\sigma$-model, including scalar and pseudoscalar mesons coupled to nucleons, was formulated. The Lagrangian of this model contains the part symmetric with respect to local transformations keeping parity of the states
(the chiral invariant part) and the part, which is not invariant under the parity mixing transformations.  Our physical ground state ($0^+$) is chosen as a result of \emph{the spontaneous symmetry breaking} in the invariant part of the Lagrangian whereas the symmetry breaking part is treated as a rather weak perturbation.
A dynamical mechanism of the nucleon mass generation through the mechanism of spontaneous breaking of the chiral symmetry was elaborated in independent works by Valentin Vaks and Anatoly Larkin~\cite{VL1960} and Yoichiro Nambu~\cite{Nambu1960}, applying superconductivity methods to a Heisenberg type four-fermion two-component Lagrangian.

We should certainly mention a role played by the spontaneous symmetry breaking in the unified electroweak theory of Abdul Salam, Sidney Glashow and Steven Weinberg, and the role played by the Higgs scalar boson in it~\cite{Glashow:1959wxa,Glashow:1961tr,Salam:1959zz,Salam:1961en,Weinberg:1967tq}.
The study of the properties of superdense matter described by unified gauge theories began in 1972 with the work of David Kirzhnits~\cite{Kirzhnits:1972iw}, who showed that the classical scalar field responsible for symmetry breaking disappears at a high enough temperature. In 1974  first grand unified theory was proposed by Howard Georgi and Sidney Glashow~\cite{Georgi:1974sy}.

Now let us come to the subject of this  review, i.e., discussion of yet considered as exotica, possibilities of existence of various abnormal nuclear systems.
Soon after prediction of quarks by Murrey Gell-Mann and George Zweig in 1964, Dmitry Ivanenko and Dmitry Kurdgelaidze suggested that a neutron star may transform in a quark star~\cite{Ivanenko:1965dg}, see also~\cite{Itoh:1970uw}. In 1971  Arnold Bodmer proposed that also atomic nuclei may undergo a transition in a superdense quark state~\cite{Bodmer:1971we}.

In 1971 Arkady Migdal put forward the idea that ground state of atomic nuclei may contain a pseudoscalar condensate ($0^-$) of pions. He demonstrated occurrence of the pion condensation in a sufficiently deep potential well in scalar and electric fields forming deep potential wells for pions and suggested possibility of superdense stable or metastable nuclei glued by the pion condensate~\cite{Migdal1971}. Three types of abnormal $\pi$ condensate nuclei: superdense, neutron and supercharged ones were discussed in~\cite{Migdal1974,Migdal:1976wm}, more details see in reviews~\cite{Migdal1978,Migdal:1990vm}.

In 1974 Tsung-Dao Lee and Gian-Carlo Wick suggested possible existence of a superdense state of nuclear matter characterized by a value of the scalar  (e.g., $\sigma$) field different from its mean value in atomic nuclei~\cite{Lee:1974jk,LeeWick1974}. This idea was reviewed in~\cite{Lee1974b,Lee1976}. The simplest variant of the model  contains only two fields, the neutral scalar meson field and the isospinor nucleon field. As the $\pi$ condensate nuclei, the $\sigma$   abnormal
nuclear state  can be either stable or metastable. In the former case the normal nuclei after a while would undergo the phase transition to the new state.  In first works the mean field approximation was used in spite of the assumed strong interaction of the scalar meson with the nucleons. To somehow justify this assumption the authors  noted  analogy between the abnormal state and the Bose--Einstein condensation one. Then Lee and Margulies argued~\cite{LeeMargulies1975,Lee1979} that quantum fluctuations may not change the  statement about the existence of a super-dense state of nuclear matter.

At the workshop at Bear Mountain, to north from New York City, on ``BeV/nucleon collisions of heavy ions'' Lee presented his results on the possible vacuum reconfiguration~\cite{Lee:1974jk} and noted~\cite{BearMountain74} that
``\emph{in order to produce the abnormal nuclear state, we must consider reactions in
which (i) a large number of nucleons is involved and (ii) the nucleon density can be increased by a significant factor. One is, therefore, led to considerations of high energy collisions between heavy ions, say ${\rm U} + {\rm U}\to Ab+\dots$, where $Ab$ denotes the abnormal nuclear state.}''
The event at Bear Mountain was pivotal in the conception of heavy ion physics.
1974 there were the Synchrophasotron at the Joint Institute for Nuclear Research in Dubna and a high-energy heavy-ion accelerator, the Bevalac, being completed at the Lawrence Berkeley Laboratory in Berkeley. Ideas about possibility of the  production of abnormal nuclei in heavy-ion collisions gave a push to active experimental investigations. One suggested also the look for collapsed quark nuclei~\cite{Bodmer:1971we} and superdense nuclei with pion condensate in heavy-ion collisions, see Refs.~\cite{Aleshin:1976ww,GalitskyUFN}.
The possibility of achieving a high degree of nuclear matter compression in shock waves, which could contribute to the formation of anomalously dense droplets in heavy ion collisions, was investigated in Refs.~\cite{Scheid:1974zz,Sobel:1975bq}. Limits on the yields of abnormal nuclei in heavy-ion collisions and on their possible terrestrial concentrations were discussed in Refs.~\cite{Holt:1976af,Karnaukhov1977,Avdeev:1982xd,Anikina:1983ypp,Anikina:1983ypp,Avdeyev1988}.
The interpretation of the data collected from heavy-ion collisions so far does not require the presence of abnormal nuclei. Feasibly, one of reasons  is that  compressed nuclear matter is produced in  heavy ion collisions  at a higher temperature  than the  binding energy of abnormal nuclei. Nevertheless, further dedicated experimental and theoretical analysis is needed.

Cosmos was another domain to look for an anomalous state of matter. Yakov Zeldovich liked to claim that the Universe is a poor man's
accelerator: experiments don't need to be funded, and all we have to do is to collect the experimental data and interpret them properly. Years later ideas of spontaneous symmetry breaking penetrated in cosmological models, cf.~\cite{Linde:1990flp}.

Probably in 1975, Migdal returned from a conference and told one of us (D.N.V.) that during his talk, in which he mentioned a possibility of the existence of  anomalous supercharged nuclei with a pion condensate, Gell-Mann asked him about a role of electrons in such highly-charged systems. Migdal suggested to think about this problem. In works~\cite{Migdal:1976wx, Migdal:1977rn} the authors developed a relativistic semiclassical treatment of the Dirac equation in deep electric potentials and the many-particle approximation for electrons occupying levels of the lower continuum. The occupation of the vacuum levels by electrons and negative muons in the deep electric potential well was named ``a fermion condensation''. The presence of the upper bound on the electric potential was  discussed  by Berndt M\"uller and Johann Rafelski in~\cite{Muller:1974fh}. Preparing this manuscript, we realized that Lee also discussed a role of electrons in abnormal nuclei. He wrote in Ref.~\cite{Lee1976} that
``...the abnormal state can create $e^+e^-$ pairs. The $e^+$ will be sent to infinity, but a fair fraction of the $e^-$ will be kept within the abnormal nucleus. As $Z$ increases, the number of $e^-$ also increases. The interplay between the added Fermi energy of $e^-$ and the Coulomb energy may eventually bring the abnormal state to the point of instability when $A$ reaches $\approx 10^4$.'' References~\cite{Voskresensky:1977mz,Voskresensky:1978uf,Voskresensky1977} described the charged condensate pions and vacuum electrons  in  strong electric fields and  suggested a possibility of existence of ``nuclei-stars'' of arbitrary size (up to the compact star size) glued by the charged  $\pi$ condensate produced in the dense nuclear matter and the electrons produced from the vacuum.

In 1979 Mikhail Troitsky and Nikolai Chekunaev discussed in~\cite{Troitsky:1979ch} that at densities above the nuclear saturation density, $n_0=0.16$\,fm$^{-3}$, the gain in the energy due to the pion condensation can exceed the energy loss associated with the transformation of nucleons into heavier baryon species like hyperons and $\Delta$ isobars. The similar idea was applied to quarks by Edward Witten in 1984. In Ref.~\cite{Witten1984}, he argued  that
in deconfined quark matter, as the baryon density increases, it becomes energetically favorable to occupy the $s$-quark Fermi sea alongside the $u$- and $d$-quark Fermi seas. Moreover, he claimed that the proposed `strange matter', which contains comparable numbers of $u$, $d$, and $s$ quarks may be absolutely stable, i.e. more stable than an iron atomic nucleus. The  Migdal hypothesis of abnormal nuclei with a pion condensate, the Lee-Wick hypothesis of abnormal nuclei with a $\sigma$ condensate and the Witten strange matter hypothesis may play an important role in cosmology, compact stars, cosmic ray physics, and relativistic heavy-ion collisions. Developing Witten's idea, Edward~Farhi and Robert~Jaffe examined the strange matter clusters with an intermediate baryon number $10^2 <A<10^7$, which are in equilibrium with respect to weak interactions~\cite{Farhi:1984qu}. Later Alvaro~De Rujula and Sidney~Glashow suggested the possibility of the existence of quark  strangelets, large-size nuclearites and strange stars of a larger mass (when gravity comes into play), see Refs.~\cite{DeRujula:1984axn,Rujula1985}. The charge distribution in the (strangelets and nuclearites) and the strange stars was considered in~\cite{Alcock1986} in a similar manner as was done for supercharged nuclei in Refs.~\cite{Migdal:1977rn,Voskresensky:1977mz,Voskresensky:1978uf}. The surface layer of strange objects (strangelets, quark nuclearites and strange stars) has typical thickness $\sim 10$ fm.
Also it was suggested that there may exist stars with a hadronic shell of a macroscopic size and quark core --hybrid stars~\cite{Shapiro1983,Bethe:1987sv}.

First investigations were done ignoring possibility of color superconductivity. As a detailed review one may use Ref.~\cite{Madsen1998}. The phenomenon of the Cooper pairing was known to exist in quark matter starting from works~\cite{Barrois1977,Bailin1984}, where the pairing gaps were estimated to be $\lsim$\,MeV, as for nucleons.
Then, an interest to a possibility of quark stars revived after arguments were given in~\cite{Shuryak,Rajagopal} in favor of large values of the quark-quark pairing gaps, $\sim 100$\,MeV, see Ref.~\cite{Alford2008} and references therein.

The axion quark nuggets (AQN) model was proposed in Refs.~\cite{Zhitnitsky:2002qa,Zhitnitsky:2021iwg,Zhitnitsky:2024ydy}. The latter model incorporates axion domain walls formed during the QCD phase transition. The domain walls act as stabilizing structures, alleviating the need for a first-order phase transition as in the original Witten model. A generic feature of the quark nuggets is the electrosphere. In case of antimatter axion quark nuggets the electrosphere of the $\sim 10^{-8}$\,cm depth contains positrons. The axion quark nuggets behave as conventional dark matter components in the low density environment and become strongly interacting macroscopically large objects in the relatively high-density environment. The available energy due to matter-antimatter annihilation can be as high as $\sim 2$\,GeV per baryon charge.

If dark matter particles were superheavy (with mass $\gsim $\,TeV) and negatively charged, they could compensate the Coulomb field being embedded inside a normal nuclear matter drop that leads to the stability of nuclear drops of a large size~\cite{Gani:2018mey}.

Although the initial studies~\cite{Lee:1974jk, LeeWick1974, Migdal1974} discussed the possibility of the sigma or pion condensation occurring at densities both greater than and less than $n_0$, it soon became evident that the latter possibility was unrealistic. On the other hand various models of equation of state of the dilute nearly isospin-symmetric nuclear matter demonstrate the presence of region of the first-order phase transition and spinodal instability at nucleon densities  $0.3 n_0\lsim n\lsim 0.7 n_0$~\cite{Schulz1982,Curtin1983,BS1983,SVB1983,Borderie}. In this density region the compressibility of nuclear matter is negative and the scalar Landau-Migdal parameter of the Fermi liquid $f_0(n)$ is smaller than $-1$, that causes the Pomeranchuck instability. In \cite{KV2016} it was shown that the Pomeranchuck instability may lead to a condensation of an effective scalar field in a Fermi liquid. The presence of such condensate would remove instabilities in first and zeroth sounds. Owing to the condensate, a novel metastable state might exist at nuclear subsaturation density for $N\sim Z$, where $N$ is the number of neutrons and $Z$ is the number of protons. In \cite{MV2019} it was realized that the stability region remains also in case of significantly isospin-asymmetric matter.

It was found in Refs,~\cite{Voskresensky:2023znr,vosk25-111-036022} that a rapidly rotating drop of nuclear matter, being not bound in the rest frame, may become bound in the rotation frame because of the formation of a charged pion condensate  giant vortex. This observation invites looking for the nuclear drops with high angular momenta.

A few words about the notion ``abnormal'' that we use in the title are in order. Often, a new concept seems `abnormal' but is considered as normal after a while. Therefore, we should clarify beforehand what we mean by the term abnormal nuclear systems. Various definitions were used in the literature: the scalar-condensate abnormal nuclei; the pion-condensate abnormal nuclei and nuclei-stars; nuggets; strangelets; nuclearites;  strange stars and hybrid stars; quark stars and some others. Different authors have given different meanings to these terms. We will basically follow the definitions introduced by pioneers in this field: The term \emph{abnormal nuclei} was used by Migdal to describe hypothetical superdense (with a density larger than several $n_0$) stable or metastable nuclei glued together by a pion condensate as well as by Lee who applied it to hypothetical superdense nuclei glued together by a scalar condensate. Farhi and Jaffe in 1984 introduced term \emph{strangelets} to describe hypothetical stable  quark systems  of a microscopic size bound more strongly than $^{57}$Fe, which should contain a similar number of up, down, and strange quarks.
Hypothetical quark objects of a larger (macroscopic) size De Ruhula and Glashow in 1984 named nuclearites. Below all abnormal stable hypothetical nuclear systems of such a size will be called \emph{nuclearites}. For such objects the gravity contributes  still much weaker than nuclear forces. When the baryon number becomes $A \gsim 10^{56}$, the gravity starts to contribute significantly to the binding, and we are dealing with \emph{nuclei-stars}. For nuclei-stars consisting of the strange quark matter (bound even at smaller $A$) we will use the term \emph{strange stars}. Compact stars with quark interiors (which are not necessarily self-bound without gravitational attraction) and hadron exteriors of a macroscopic size are usually called \emph{hybrid stars}.

The existence of stable or metastable nuclear systems of various sizes would have numerous important consequences in physics. Therefore, it would be worthwhile to continue theoretical studies and, especially, experimental searches of such objects.

The paper has the following structure.
In Sect.~\ref{sec:Lee-Wick} we focus on the Lee-Wick model of abnormal nuclei with a scalar (e.g. a $\sigma$ field) condensate. We discuss the relationship between the Lee-Wick model and the standard RMF models, as well as the RMF models with $\sigma$-scaled hadron masses and coupling constants. Section~\ref{sec:Anomalous} considers various aspects of the model of supercharged nuclei held together by light charged bosons, if the latter existed. In reality the role of the light boson could be played by an ordinary pion, which can become effectively lighter in a dense nuclear medium and/or in external fields, for example in rapidly rotating nuclear matter. In Sect.~\ref{sec:Superheavy} we discuss density dependence of the p-wave pion-nucleon-nucleon and pion-nucleon-$\Delta$ isobar interactions which may result in charged and neutral pion condensates and a possibility of existence of superdense abnormal nuclei as proposed by Migdal. These phenomena can be modeled using the $\sigma$ model. Then we turn our attention to the subtleties associated with the description of the s-wave pion-nucleon interaction and the s-wave pion condensation. In Sect.~\ref{sec:Delta-sec} we consider $\Delta$-resonance matter. The uncertainty in the $\Delta$ optical potential allows for a wide variation in its value. For certain values,  metastable or even stable resonance matter and abnormal nuclei may exist. The concept of $\Delta$-resonance matter is somewhat similar to Witten's idea of strange quark matter. In the latter case the occupation of the strange quark Fermi sea results in a decrease of the energy. The energy of the state also decreases with a decrease of the strange quark mass and the bag constant, the values of which are not well known.  In Sect.~\ref{sec:Strangelets-sec} we focus on the hypothesis of strangelets and strange and hybrid stars. Ordinary neutron stars and stars with exotic matter have different mass-radius relations, and their cooling histories are different. These topics will be discussed. Section~\ref{sec:dilute-sec} presents the idea of the scalar field condensation and possible existence of dilute metastable nuclear states. A scalar condensate field, determined by the value of the Landau-Migdal parameter in the scalar channel, may appear owing to the Pomeranchuk instability occurring in a dilute nuclear matter. Thereby we apply the idea of Lee and Wick of the possible existence of dense abnormal nuclei held together by a scalar field. Then, we consider the possibility of the formation of a self-bound dilute Bose-Einstein condensate state, e.g.,of $\alpha$ particles in the $\beta$-equilibrated system described with the help of a scalar condensate field. Next, we discuss anomalous states in nonequilibrium nuclear systems. Section~\ref{sec:observation-sec} discusses anomalies that defy conventional physics interpretations, part of which could be related to abnormal nuclear systems. Conclusions are drawn in Sect.~\ref{conclusion-sec}.

Throughout the text we use the system of units where $\hbar=c=1$.

\section{The Lee-Wick model}\label{sec:Lee-Wick}

In this section we present the idea of Lee and Wick in a nutshell. We consider the Lagrangian density of nucleons interacting with a scalar field $\phi$:
\begin{eqnarray}
\mathcal{L}=\frac{1}{2}\partial_\mu \phi\partial^\mu \phi - U(\phi) + \bar{\psi}(p\!\!\!/-m_N+g\phi)\psi\,,
\label{Lag-Lee}
\end{eqnarray}
where $m_N=939$\,MeV is the nucleon mass, $\psi$ is the nucleon bispinor field, $g$ is the strong interaction coupling constant, and $U(\phi)$ is the scalar field potential. The scalar field $\phi$ will be treated on the mean-field level. Then the nucleon spectrum is $\sqrt{m_N^{*2} + p^2}$ with the effective nucleon mass $m_N^*=m_N-g\phi$. While in the vacuum $\phi=0$, Lee and Wick suggested that in the nucleon medium at some critical density the effective nucleon mass $m_N^*(n)$ can reach zero. It would occur for $\phi_c =m_N/g$. The energy density gain at this transition can be estimated as
\begin{align}
\Delta E(n)\approx U(\phi_c)+\nu \frac{p_{\rm F}^4}{4\pi^2}-m_N \,n - \frac{3p_\rmF^2}{10 m_N} n ,
\end{align}
where $p_\rmF=(3\pi^2n /\nu)^{1/3}$ is the nucleon Fermi momentum and $\nu=1$ for the neutron matter and $\nu=2$ for the  isospin-symmetric nuclear matter.
Then the critical density can be estimated as
\begin{align}
n_c &= n_{c,0}\Big(1+\frac{3p_{\rmF,c}}{4 m_N}+\frac{9 p_{\rmF,c}^2}{20 m_N^2}+O(p_{\rmF,c}^3/m_N^3)\Big),\quad
\nonumber\\
n_{c,0} &= \frac{U(\phi_c)}{m_N} = \nu \frac{p^3_{\rmF,c}}{3\pi^2}\,.
\label{Leecrit}
\end{align}
For a simplistic estimate we can take $U(\phi)=\frac12 m^2_\phi \phi^2$, where $m_\phi$ is the mass of the scalar meson. Then, neglecting the correction $O(p_{\rmF,c}/m_N)$ we find
\begin{align}
n_c\simeq n_{c,0}= \frac{m^2_\phi m_N}{2g^2}.
\label{nc-simple}
\end{align}
Using typical values for the strong interaction constant $g\sim 10$ and the mass of the scalar  meson, $m_\phi\sim m_\sigma\sim 700$ MeV, we obtain $n_c\sim 2n_0$. The energy gain per nucleon for $n>n_c$ and  for $(n-n_c)/n_c \lsim 1$ can be estimated as $\mathcal{E}=\Delta E/n \simeq -(1-n_c/n)m_N$\,. These rough estimates show to a possibility that the ordinary nuclei may be exist in a metastable state and after a while they may undergo the first-order phase transition to the superdense ground state. The energy of each nucleon inside the abnormal nucleus is $p$, and  outside $\sqrt{m_N^2+p^2}$. Thus, there is a potential barrier holding the nucleons
in the given abnormal state.

\begin{figure}
\centering
\includegraphics[width=8cm]{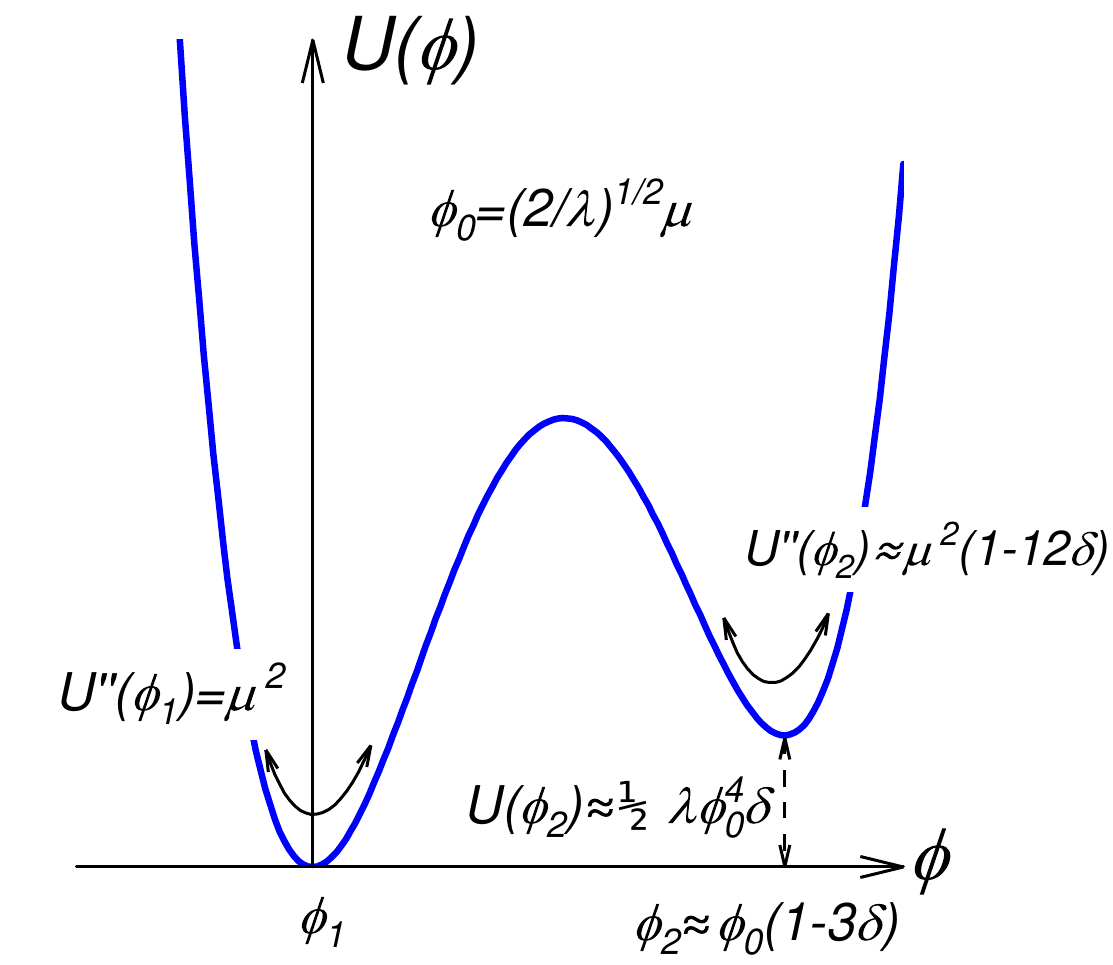}
\caption{Potential (\ref{Lee-U}) for $\mu>0$, $\lambda>0$ and $0<\delta <1 - 2 \sqrt{2}/3$.
\label{fig:Lee-potential}
}
\end{figure}

The main shortcoming of the above simplified argumentation is that we did not consider the equation of motion for the $\phi$ field following from the Lagrangian (\ref{Lag-Lee}) or perform the minimization of the energy density with respect to $\phi$. To remove this shortcoming one has to consider potential $U(\phi)$, which has the spontaneous symmetry breaking pattern. Let us take the potential in the form
\begin{align}
U(\phi)=\frac12\mu^2 \phi^2 -\frac{\mu \sqrt{\lambda}}{\sqrt 2}(1-\delta)\phi^3 + \frac14\lambda\phi^4 \,,\,\, \mu,\lambda>0\,,
\label{Lee-U}
\end{align}
which for $0<\delta <1 - 2 \sqrt{2}/3\approx 0.057$ has two local minima: one at $\phi_1=0$ and another one at
\begin{align}
 \phi_2 &= \frac{\phi_0}{4}(3(1- \delta)+\sqrt{9 (1-\delta)^2-8})
\nonumber\\
 & =\phi_0 (1-3\delta +O(\delta)), \,\,
 \mbox{where}\quad \phi_0=\sqrt{2}\frac{\mu}{\sqrt{\lambda}}.
\label{phi-min-2}
\end{align}
The global minimum of the potential is realized at $\phi=0$ that corresponds to the ground state. The second minimum is higher,
\begin{align}
U(\phi_2) \approx \frac12\lambda \phi_0^4 (\delta+O(\delta^2))\,,
\label{Umin2}
\end{align}
and the parameter $\delta$ is responsible for lifting the symmetry of two minima.
In both minima the potential has the similar curvature $m_\phi^2=U''(\phi_1)=\mu^2$ and $U''(\phi_2)=\mu^2(1-12 \delta+O(\delta^2))$.
The potential is illustrated in Fig.~\ref{fig:Lee-potential}. 

\begin{figure}
\centering
\includegraphics[width=8cm]{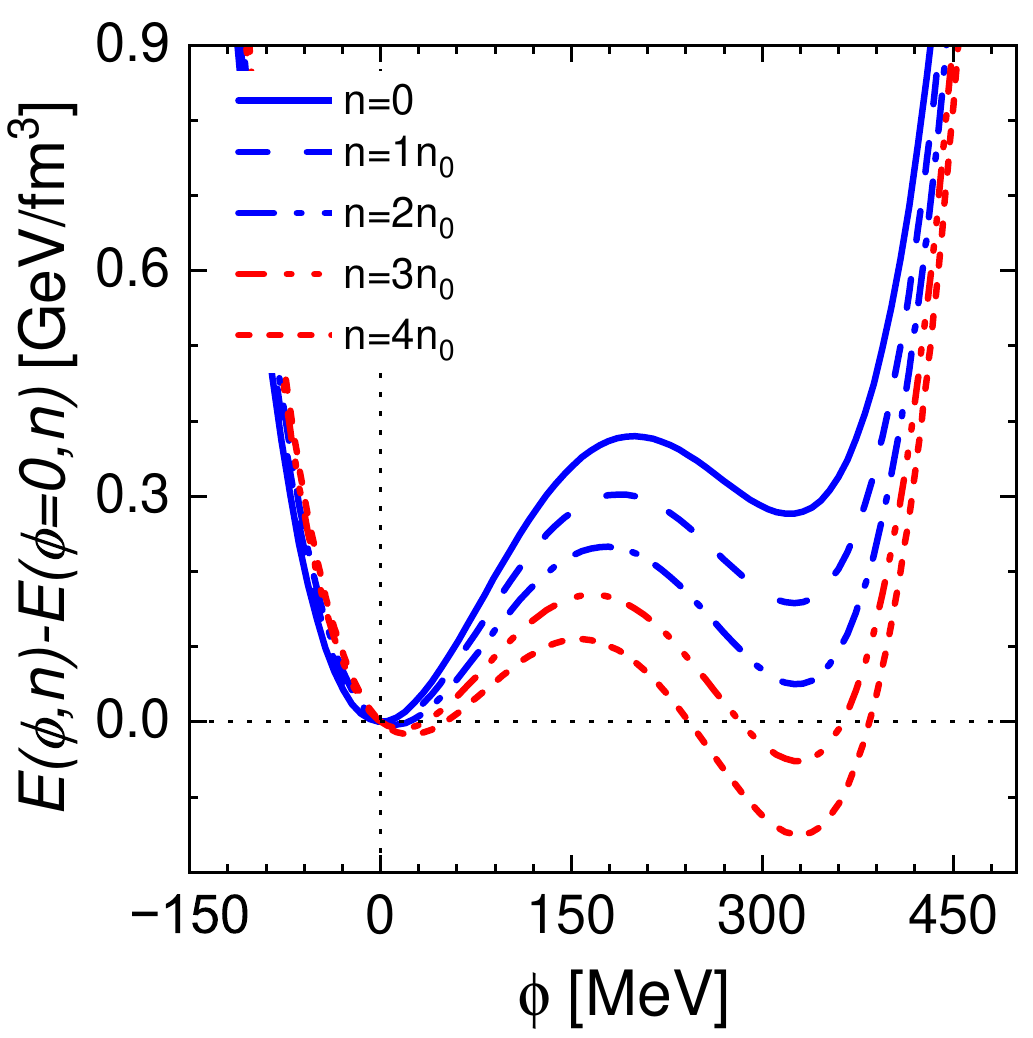}
\caption{Difference in the energy density (\ref{Lee-Edens}) due to the development of the scalar field for ISM as function of $\phi$ for various nucleon densities  calculated with the model parameters (\ref{Lee-param}).
\label{fig:DeltaE-n}
}
\end{figure}

\begin{figure}
\centering
\includegraphics[width=8cm]{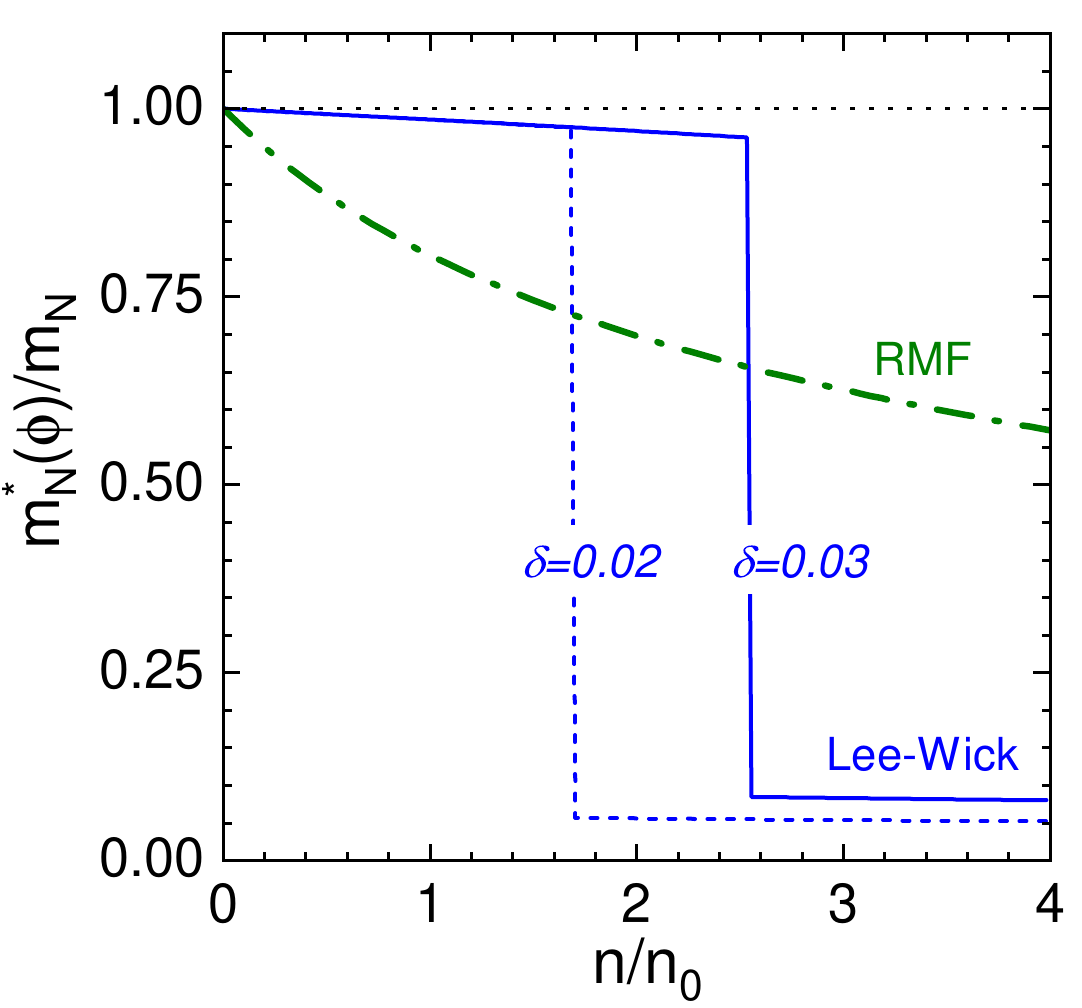}
\caption{Effective nucleon mass for ISM as a function of the nucleon density. Solid line is calculated for the Lee-Wick model with parameters (\ref{Lee-param}). Dashed line is drawn for $\delta=0.02$. Dash-dotted line corresponds to the RMF model (\ref{Lee-W}) with parameters (\ref{Lee-W-param}).
\label{fig:Lee-mass}
}
\end{figure}

The coupling of the scalar field to nucleons is defined as previously so that the nucleon mass becomes very small closely to the new minimum $\phi\sim\phi_2$,
\begin{align}
g=\frac{m_N}{\phi_0} = \frac{m_N}{\mu}\sqrt{\frac{\lambda}{2}} = \frac{m_N}{m_\phi}\sqrt{\frac{\lambda}{2}}\,.
\label{Lee-g-def}
\end{align}
The critical density $n_{c,0}$ in (\ref{Leecrit}) can be estimated now with the help of Eq.~(\ref{Umin2}) as
\begin{align}
n_{c,0}=\frac{\lambda\phi_0^4}{2m_N}\delta=\frac{2m_\phi^3}{\lambda m_N}\delta = \frac{m_\phi^2m_N}{g^2}\delta \,,
\label{nc0}
\end{align}
where to get the second and third equalities we used Eqs.~(\ref{phi-min-2}) and (\ref{Lee-g-def}). In contrast to the more  rough  estimate (\ref{nc-simple}) the critical density (\ref{nc0}) is determined by the symmetry-breaking parameter $\delta$. Let us now consider the energy density of the system taking into account the nucleon kinetic energy
\begin{align}
E(\phi,n)=U(\phi) + \nu\intop_0^{p_\rmF}\frac{\rmd p p^2}{\pi^2} \sqrt{(m_N-g\phi)^2+p^2}
\label{Lee-Edens}
\end{align}
and plot $\Delta E(\phi,n)=E(\phi,n)-E(0,n)$ as a function of $\phi$ for various values of the nucleon density.
In Fig.~\ref{fig:DeltaE-n} we present the results obtained for isospin-symmetric matter (ISM), $\nu=2$, and for the set of parameters
\begin{align}
\mu=m_\phi=800\,{\rm MeV}\,,\quad \lambda=10\,,\quad \delta=0.03\,.
\label{Lee-param}
\end{align}
We see that with a density increase the relative height of the second  minimum (at $\phi\neq 0$) decreases and at some density it becomes a global minimum.
At this density the nucleon mass changes abruptly as shown in Fig.~\ref{fig:Lee-mass} by the solid line. The dashed line presents the result obtained with a  value  $\delta=0.02$. The transition density decreases, respectively.

\begin{figure}
\centering
\includegraphics[width=8cm]{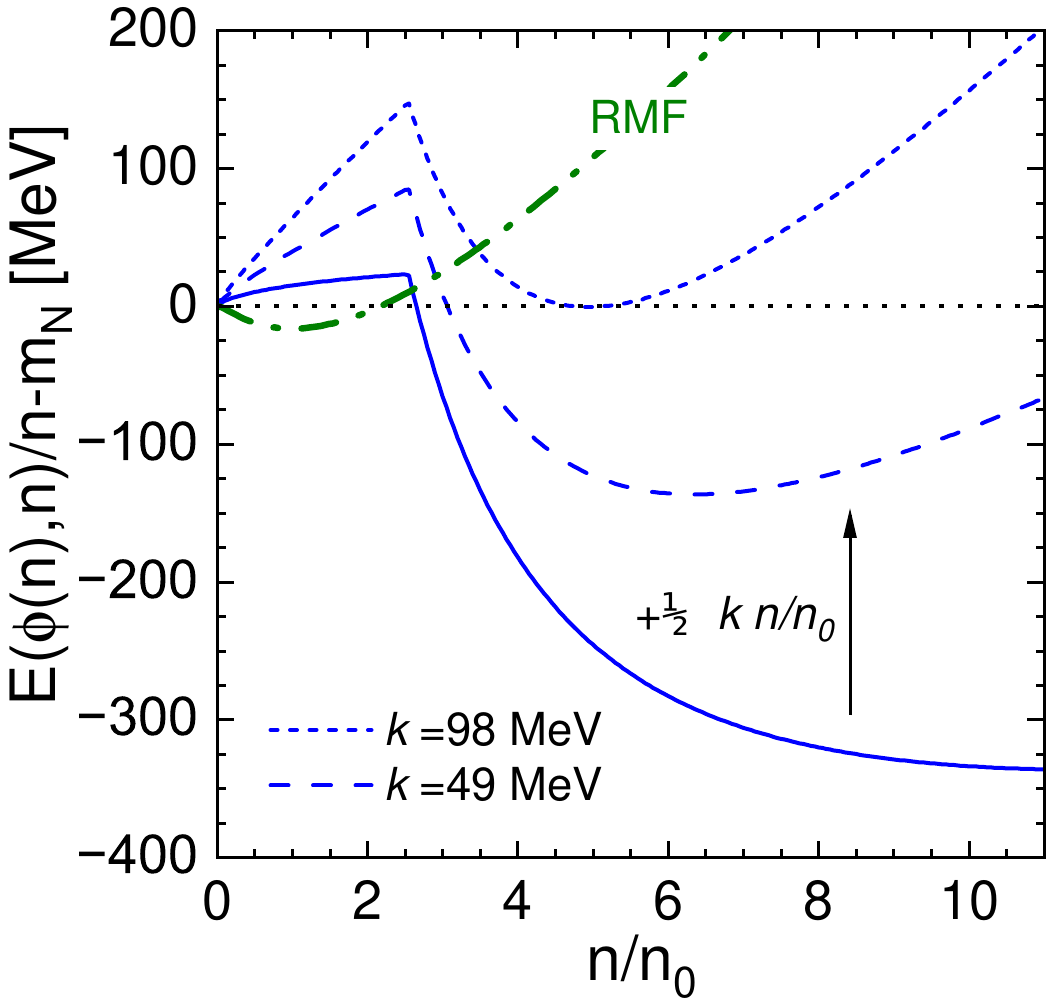}
\caption{Binding energy per nucleon (\ref{Ebind-def}) for ISM
as a function of nucleon density. Difference in the energy density (\ref{Lee-Edens}) due to the development of the scalar field is calculated for parameters (\ref{Lee-param}). Solid line shows the result of  the Lee-Wick model. Dashed lines include nucleon-nucleon interaction term  (\ref{k-correl}). The dash-dotted curve shows the RMF calculation with parameters (\ref{Lee-W-param}).
\label{fig:Lee-Ebind}
}
\end{figure}

In Fig.~\ref{fig:Lee-Ebind} we plot the binding energy per nucleon,
\begin{align}
\mathcal{E}(n) = \frac{1}{n} E(\min\phi(n),n) -m_N\,.
\label{Ebind-def}
\end{align}
We see, that for the parameter set (\ref{Lee-param}), the vanishing nucleon mass leads to tremendous binding of the nuclear matter with a minimum realized at $\sim 12\,n_0$ with the binding energy of $-340$\,MeV, see the solid line in Fig.~\ref{fig:Lee-Ebind}.  To reduce the binding one can take into account he nucleon-nucleon repulsion. To simulate it one can include a term quadratic in the nucleon density in expression (\ref{Lee-Edens}),
\begin{align}
E(\phi,n) \to E(\phi,n) +\frac12 k \frac{n^2}{n_0},\label{k-correl}
\end{align}
where $k$ is treated as a parameter. An increase of $k$ reduces binding, which disappears completely at the critical value, $k=98$\,MeV, see the short-dashed line in Fig.~\ref{fig:Lee-Ebind}. Reducing the value of $k$, we find that the minimum in the energy appears at $n\sim 5\,n_0$, and at the half critical value, $k=49$\,MeV  the minimum is located at the density $\sim 6.4\,n_0$ and has the depth of $-136$\,MeV, see the dashed line in Fig.~\ref{fig:Lee-Ebind}.

The same year when Lee and Wick proposed their model, Walecka formulated a relativistic mean-field model (RMF)~\cite{Walecka1974,Walecka1975}. Here one should also  mention   previous works by Johnson and Teller, and by D\"urr \cite{Johnson:1955zz,Duerr:1956zz}, where the nuclear saturation arises due to the difference between scalar and vector nucleon densities.
The Walecka model includes interaction of nucleons with mean fields of the scalar $\sigma$  and vector $\omega$ mesons allowing the $\sigma$ field dependence of the nucleon mass.  Coupling constants of $\sigma$  and vector $\omega$ mesons with nucleons were fitted to describe binding energy and saturation density  of nuclear matter. Incompressibility at nuclear saturation proved to be unrealistically high and effective nucleon mass low.  To improve the fit Boguta included possible scalar field self-interactions depending on two extra fitting parameters \cite{Boguta:1977xi}. Also $\rho$ meson mean field was included. These changes made the RMF models more flexible, see \cite{Serot:1997xg}.
Doing in Eq.~(\ref{Lee-Edens}) replacements
\begin{align}
\phi &\to\sigma\,,\quad \mu \to m_\sigma\,,\quad g\to g_\sigma\,,\quad
\nonumber\\
\lambda &\to c g_\sigma^4\, \quad \delta\to 1+\frac{\sqrt{2} }{3 \sqrt{c} m_\sigma }b g_\sigma m_N \,,
\end{align}
and adding the repulsive contribution due to the $\omega$ meson exchange with the mass $m_\omega$ and the coupling constant $g_\om$, we obtain the RMF energy density
\begin{align}
E(\sigma,n) &= \frac{m_\sigma ^2 \sigma^2}{2}
+ \frac{b}{3} m_N (g_\sigma\sigma)^3
+ \frac{c}{4} (g_\sigma\sigma)^4
\nonumber\\
& + \nu\intop_0^{p_\rmF}\frac{\rmd p p^2}{\pi^2} \sqrt{(m_N-g_\sigma\sigma)^2+p^2} + \frac{g_\om^2 n^2}{2m_\om^2} \,.
\label{Lee-W}
\end{align}
The magnitude of the scalar field follows from the equation $\partial E(\sigma,n)/\partial \sigma =0$.

As example, taking the parameter of binding energy per nucleon $\mathcal{E}_0=16$\,MeV, compressibility modulus $K=275$\,MeV and the effective nucleon mass $m_N^*(n_0)=0.805$ at the saturation density $n_0$, as we did in Ref.~\cite{KV2005}, we obtain parameters
\begin{align}
&\frac{g_\sigma^2 m_N^2}{m_\sigma^2} = 184.356\,,\quad \frac{g_\om^2 m_N^2}{m_\om^2}=87.5996 \,,\quad
\nonumber\\
& b =5.53871\times 10^{-3}\,,\quad
c=2.29759\times 10^{-2}\,.
\label{Lee-W-param}
\end{align}

The density dependence of the effective nucleon mass of the ISM calculated within the RMF model is shown by the dash-dotted line in Fig.~\ref{fig:Lee-mass}. The mass decreases smoothly with the density increase without any jumps in difference to the Lee-Wick model. The binding energy per nucleon (\ref{Ebind-def}) for the nuclear matter ($\nu=2$) as a function of nucleon density is depicted in Fig.~\ref{fig:Lee-Ebind}, solid line for Lee-Wick model. Long and short dashed lines include the nucleon-nucleon interaction term  (\ref{k-correl}). The dash-dotted line  shows the RMF calculation with parameters (\ref{Lee-W-param}). The Lee-Wick abnormal state does not appear in the latter model.

Various modifications of the original Walecka   model including $\rho$ mesons, a non-linear $\sigma$ field dependent non-linear  potential, possible $\omega-\sigma$ interaction,  etc, are  widely and rather successfully  used to describe equation of state of the hadron matter in compact stars, heavy-ion collisions and finite nuclei, cf. \cite{Serot:1997xg,Geng:2005yu} and references therein. Usually, one assumes that there is no pseudoscalar mean-field of pions in the spin balanced system. However, including pionic excitations beyond the mean-field level improves the description of the equation of state of nuclear matter at finite temperatures, e.g., in application to heavy-ion collisions~\cite{V1993}. At low temperatures, which are of main our interest below, the contribution of pions to the equation of state is rather small.

\begin{figure}
\includegraphics[width=8cm]{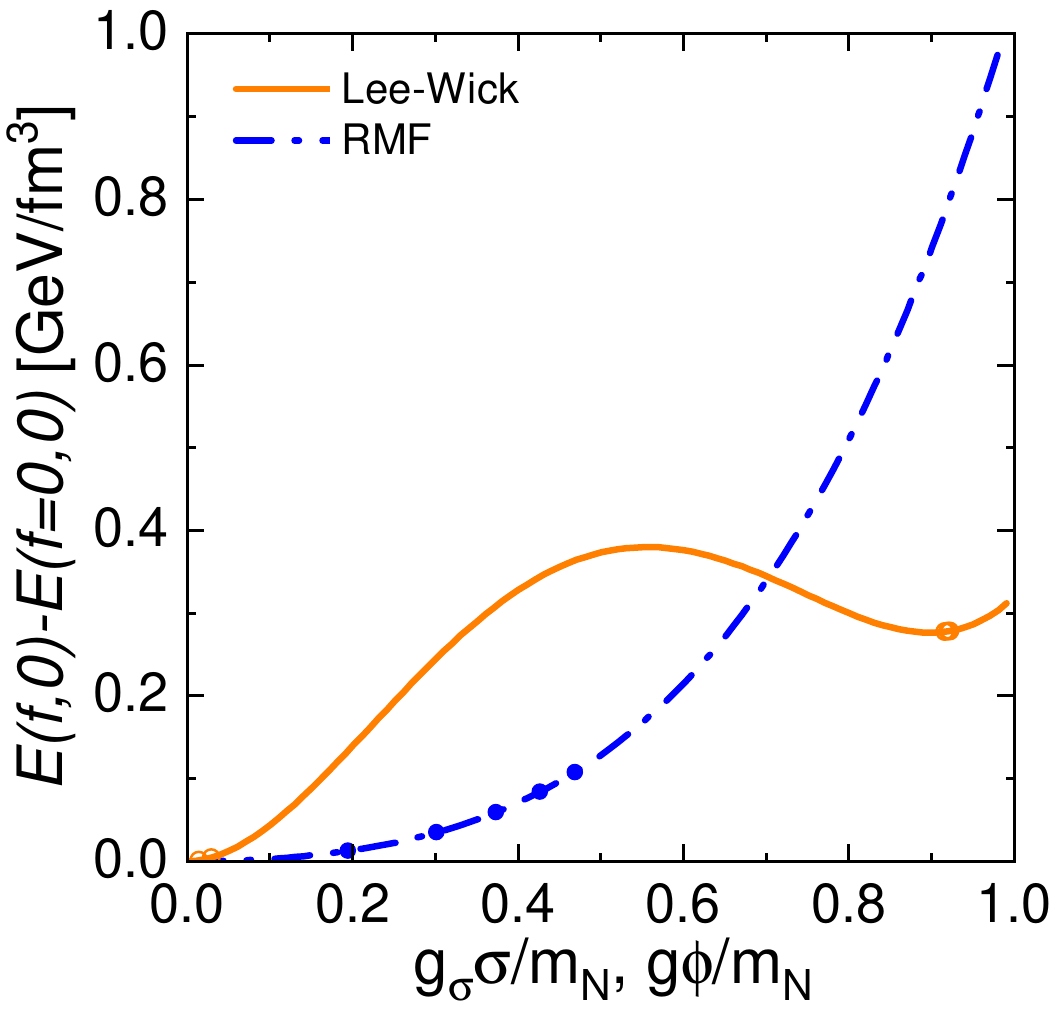}
\caption{Potentials of the scalar field for the Lee-Wick model (\ref{Lee-U}) with parameters (\ref{Lee-param}) and for the RMF model with parameters (\ref{Lee-W-param}) as functions of the dimensionless scalar field. Dots show the values of the dimensionless magnitude of the scalar field for five values of the nucleon density from $1\,n_0$ till $5\,n_0$ with the step $1\,n_0$.
\label{fig:Pot-comp} }
\end{figure}

In spite of the success in describing various phenomena at densities $\lsim 2\,n_0$, RMF models should be considered as merely phenomenological extrapolations at densities $\gg n_0$. In Fig.~\ref{fig:Pot-comp} we compare the potentials of the scalar field for the  Lee-Wick model and the RMF model. Dots show the values of the dimensionless magnitude of the scalar field for five values of the nucleon density from $1\,n_0$ till $5\,n_0$ with the step $1\,n_0$.
For parameters fitted to the properties of the nuclear system at $n=n_0$, the Lee-Wick abnormal state does not appear at larger densities in the RMF model. As seen in Fig.~\ref{fig:Pot-comp}, no information about the possible restoration of the chiral symmetry is included in the standard RMF models, whereas a mechanism for a phase transition to a new state with a small nucleon mass is build in the Lee-Wick model by construction.
Although the latter model cannot describe the properties of the ordinary atomic nuclei and nuclear matter at $n\lsim n_0$, it demonstrates the principle possibility of the existence of a new stable (or metastable) abnormal nuclear state at $n$ significantly above $n_0$.

The possibility of the Lee-Wick state can be studied within an RMF model if the latter is augmented with the concept of partial restoration of the chiral symmetry. Such idea was realized in a series of works~\cite{KV2005,KTV2007,KTV2008,MKVPhysRev2016,MKV2016,KMV2017}, where the hadron masses and coupling constants were taken to be dependent on the scalar $\sigma$ field, so that not only baryon masses but also masses of other hadrons ($\omega$, $\rho$ and the $\sigma$ mesons) decrease with an increase of the $\sigma$ mean field.  This approach differs from the well-known approach of Refs.~\cite{Typel:1999yq,Typel:2005ba} where the coupling constants were assumed to be functions of the baryon density.
Therefore thermodynamical consistency of the former models is fulfilled automatically. The RMF models with the $\sigma$ scaling have been shown to satisfy many experimental constraints from heavy-ion collisions and from compact star observations. For example, the models fulfil constraints imposed by directed and elliptical particle flows measured in nucleus collisions~\cite{Danielewicz:2002pu}, describe very massive neutron stars  allowing admixtures of hyperons and $\Delta$s in neutron star matter, and shift the threshold of the most efficient direct Urca (DU) neutrino reaction to occur only in the sufficiently heavy neutron stars. The latter possibility is favored by the fit of the experimental data on cooling of pulsars, e.g.  see \cite{Blaschke:2013vma,Grigorian:2016leu,Grigorian:2017xqd,Grigorian:2018bvg}. Only for simplicity, the parameters of the models used in \cite{KV2005,KTV2007,KTV2008,MKVPhysRev2016,MKV2016} were chosen so that the new Lee-Wick minimum was not realized. By changing parameters of the models,  new solutions can be obtained, some of which would exhibit a sudden drop of the effective nucleon mass at a critical density, cf. discussion in Sect.~3 of Ref~\cite{KMV2017}. Such a possibility was realized in the model~\cite{KMV2017} at sufficiently strongly attractive $\Delta$ optical potential, see Fig.~\ref{fig:mkvstar-meff} below.

\section{Supercharged nuclei, nuclearites and nuclei--stars}\label{sec:Anomalous}

In Ref.~\cite{Migdal1974,Migdal:1976wm} three types of abnormal nuclei with a $\pi$ condensate were discussed.
These were named the ``supercharged'', ``neutron'' and  ``superheavy'' anomalous nuclei.
Later, it became clear that this gradation is very conditional. What is important is that these hypothetical objects may differ in their baryon number, interior density, isospin content, and  the type of the condensate ($\pi^{\pm}$ or $\pi^0$) that holds them together.

\subsection{Supercharged nuclei in a nutshell}\label{nutshell-sec}

We consider a nuclear system with a rather large baryon number $A$, approximately constant density $n\sim n_0$ at the proton number $Z\approx A/2$ (in this respect the nucleus with a large $Z\gg 1/e^2=137$ is supercharged although for $Z\gg 1/e^3\sim 10^3$ the charge is compensated by electrons produced from vacuum, see below). To demonstrate the key idea, we assume for a moment that there exists a negatively charged light scalar bosons (or pseudoscalar "light pions") with a mass $m_b$. Below we will show that in realistic situations, pions, which mass is changed in nuclear matter and in external fields, can play the role of such a light boson. One may ask whether these light bosons can bind supercharged nuclei, see~\cite{Voskresensky1977}, and~\cite{Migdal:1990vm}.

The energy of the nucleus with approximately constant density $n\simeq n_0$ at the proton number $Z\simeq A/2$ is given by
\begin{align}
\mathcal{E} &\approx -\mathcal{E}_0\, A +
\mathcal{E}_{\rm surf}
-\int n_p V \rmd^3 x
-\int \frac{(\nabla V)^2}{8\pi e^2}\rmd^3 x
\nonumber\\
&+ \int  \left(|\nabla \phi |^2 - V^2|\phi |^2 + m_b^2 + {\textstyle \frac12}\lambda |\phi|^2) |\phi|^2\right) \rmd^3 x\,, \quad
\nonumber\\
&\mathcal{E}_0=16\,{\rm MeV}.
\label{LagrSuperch}
\end{align}
Here the first term is the volume binding energy, the second term produces the surface energy, which is suppressed as $1/A^{1/3}$ compared to the first term and will be dropped when the nucleus atomic number $A$ is large as we for simplicity will assume. The proton density is $n_p\simeq n_0/2$. The third and fourth terms describe Coulomb interactions of protons with the electric field.
For ordinary atomic nuclei these terms produce the Coulomb energy $3Z^2 e^2/5R$, where the nucleus radius is $R\simeq r_0 A^{1/3}=(r_0/2^{1/3}) Z^{1/3}$ and $r_0\simeq 1.2$\,fm. The last term in Eq.~(\ref{LagrSuperch}) is the contribution of the classical charged boson field, the coupling of the pion-pion self-interaction $\lambda\sim 1$. With an increase in charge $Z\sim A/2$, the Coulomb energy repulsion always begins to exceed the volume energy attraction. So for the existence of extended systems with $ Z \sim A/2 >>1/e^3$ the charge must be essentially compensated by electrons and $\mu^-$. Indeed when $A$ and $Z$ grow, the electrical potential well, $V$, deepens. When it becomes $V \leq -m_b$, the charged light bosons may appear in the reactions $n\to p+b^-$ and compensate the charge of the protons. As the result, in a nucleus with $Z >>1/e^3$,  the charge-neutral interior is formed~\cite{Migdal:1977rn} surrounded by a thin (with thickness $\sim 10$ fm) surface layer, where the strength of the electrical field is  $E\sim 10^2  m_e^2 \sim 10^{18}$\,V/cm with $m_e=0.5$\,MeV being the electron mass.

The spatial structures of the electric potential $V(r)$ and the boson field $\phi(r)$ are governed by equations
\begin{align}
&\Delta \phi +(V^2 -m_b^2 +\lambda |\phi|^2)\phi =0\,,
\label{Vm}
\\
&\Delta V =4\pi e^2 (n_p + 2V|\phi|^2) \,.
\label{Vel}
\end{align}
Here we put $\lambda\to 0$ to simplify consideration. Hence From these equations in case of a spherical nucleus with large $A$ we get uniform solutions $V=-m_b$ and $|\phi|^2 =n_p/(2m_b)$ valid in the interior, except for a  boundary layer of the depth determined by the typical Debye screening length, being much shorter than the nucleus size. Thus, after the formation of the charged boson condensate the potential well stops growing keeping limiting depth of $V=-m_b$.

The volume part of the energy reads now
\begin{align}
\mathcal{E} = -\mathcal{E}_0 A -V Z=-\mathcal{E}_0  A + \frac12m_b A\,.
\label{lightboson}
\end{align}
We conclude that if negatively charged scalar bosons with a mass of $m_b <2\mathcal{E}_0\simeq(30\mbox{--}32)$\,MeV existed in nature, then stable supercharged nuclei and nuclearites of an arbitrary size (with $A$ up to $10^{56}$ and $n\sim n_0$) would also exist, being glued by strong and electromagnetic forces, see Refs.~\cite{Voskresensky1977,Migdal:1990vm}. For larger-$A$ systems, gravity comes into the game, and we deal with compact stars containing a nucleon--light-boson (e.g., light pion) component.

We discussed abnormal nucleonic systems glued by negatively charged bosons but certainly, on equal footing, we could consider antinucleonic systems glued by positively charged bosons. In early Universe, the appearance of the latter anti-systems could contribute to a solution of the problem of the observable charge asymmetry of the Universe.

In the literature the exotic objects have been given various names: anomalous nuclei, abnormal nuclei, nuclear drops, nuclearites, and nuclei-stars for heavy objects. Similar self-bound objects made of quarks  were  called --- quark nuggets  or strangelets, quark nuclearites, strange stars.  For objects build with the Witten's strange quark matter, we will use further on the names like strangelets, quark nuclearites and strange stars.

\subsection{Dibaryons and neutral light bosons}

The possibility of the existence of multiquark states was predicted by QCD-inspired models~\cite{Jaffe:1976yi,Matveev:1977xt}.
These works initiated a lot of experimental searches for six-quark states (dibaryons). Usually, one looked for dibaryons in the $NN$ channel.
Such dibaryons would have decay widths from a few up to a hundred MeV. The obtained results were not very conclusive and the background determination was questionable.   Enthusiasm has been revived in the field after experimental reports~\cite{Tatischeff:1997wj,Fil'kov:1999ig,Fil'kov:2000vz}.
Reaction $pp\rightarrow p\pi^+\, X$ was studied in~\cite{Tatischeff:1997wj}  and three narrow peaks (width $\sim 5$~MeV) in the
missing mass spectrum were seen at $M_X=1004$\,, $1044$ and $1094$ MeV with high statistical significance.
The further study in~\cite{Tatischeff:2002nq} indicated the narrow baryonic structures at 1004, 1044, 1094, 1136, 1173, 1249, 1277, and 1384 MeV (and possibly 1339 MeV) in the missing mass $M_X$ and in the invariant mass $M_{p\pi^+}$.
The reaction $pd \rightarrow pp X_1$ was studied in~\cite{Fil'kov:1999ig,Fil'kov:2000vz} and three peaks of the 5~MeV width were clearly observed in the  missing mass $M_{p X_1}$ spectrum at $M_{pX_1}=1904\pm 2$\,, $1926\pm 2$ and $1942\pm 2$~MeV.
In the missing mass $M_{X_1}$ spectrum the peaks are located at $M_{X_1}=966\pm 2$\,, $986\pm 2$ and $1002$~MeV.
The peaks in the $pd$ reaction can be interpreted as the signatures of either dibaryons or nucleon resonances. Theoretical and experimental bounds on the properties of nucleon resonances were discussed in Ref.~\cite{Azimov:2003bb}.
The resonance states below the pion-nucleon threshold and their consequences for nuclear systems were discussed in~\cite{KVres2003}.

Rather recently several very heavy compact stars were observed. The current highest precisely measured mass is $2.08(7)\,M_{\odot}$ for PSR J0740+6620~\cite{Cromartie2010,Fonseca2021}.
On the other hand appearance of hyperons and $\Delta$ resonances in neutron star interiors results in a decrease of the maximal neutron star mass. Nevertheless these so called the hyperon and $\Delta$ isobar puzzles can be resolved, see~\cite{KMV2017} and reference therein. Reference~\cite{KVres2003} demonstrated that in  presence of light dibaryons  the maximal possible mass of a neutron star would decrease below the observational limit.
Appearance of rather heavy dibaryons and baryon resonances although results in a decrease of the maximal neutron star mass does not contradict to observations.

An interesting alternative interpretation of the tower of resonance  states was given in~\cite{Walcher:2001nr}. It was noted that drawing  all new states (both dibaryonic and nucleonic ones) on one energy scale, one obtains a tower of almost equidistant states.
It was hypothesized the existence of a light pseudoscalar meson ($J^P = 0^-$) with a mass of $m_{{\rm light}\,\pi} \simeq 21\pm 2.6$\,MeV. The basic idea is that the tower of excited nucleonic states is formed by the nucleon in its ground state plus $1,2,3,\dots$ light pions as the quantum of excitation with the energy $m_{{\rm light}\pi}$. It was argued in \cite{KVres2003} that the presence of light charged pions would allow for  existence of abnormal nuclei ($A\gsim 10^3$) and ‘‘nuclei stars’’ of arbitrary size, bound by strong and electromagnetic interactions.

Recent possible discovery is also worth mentioning: a light neutral boson (X17) with a mass of 17\,MeV and a very narrow width (presumably less than $0.07$\,eV) was found at the accelerator of ATOMKI. Three significant anomalies were observed in electron–positron pairs emitted in the $^3$H $(p, e^-e^+)$ $^4$He, $^7$Li $(p, e^-e^+)$ $^8$Be, and $^{11}$B $(p, e^-e^ +)^{12}$C reactions \cite{Krasznahorkay1,Krasznahorkay2,Krasznahorkay3}. If confirmed, the existence of this new particle could be important for particle physics and  cosmology, see review \cite{Gustavino2024}.
Reference~\cite{Feng2016} assumed that X17 is the vector gauge boson coupled more strongly
to neutrons than to protons. Reference~\cite{Ellwanger2016} suggested another interpretation of the experimental results, assuming a light, pseudoscalar particle. In both cases the coupling of X17 with nucleons was assumed to be very weak.

On the other hand, independently of the value of the strength of the interaction it is worthwhile to ask, if existence of a light neutral boson does not contradict to  the neutron star properties. Reference~\cite{Veselsky2024} assumed that X17 can  mediate low-energy nucleon-nucleon interactions and attempted to find possible constraints on the hypothetical X17 boson from the analysis of the equation of state of neutron stars. It was shown that such a neutron star could have $M-R$ curve rather similar to that for the ordinary neutron star.

Note that since $m_{\rm X17}/m_\pi \sim 1/8$, one could expect presence of the nucleon--light boson polarization effects of the same order  as for pions, if the coupling of the light boson  with nucleons were $f_{XNN}\gsim f_{\pi NN}/8$. In the latter case a possibility would exist for  the  Bose condensation of such a particles in the compact star matter (similarly to $\pi^0$ condensation) and feasibly for existence of metastable or stable nuggets and light-boson enriched nuclei-stars. Nevertheless we should repeat that~\cite{Ellwanger2016} and~\cite{Feng2016} estimated X17-nucleon polarization effects as tiny.

\subsection{Rotating nuclear systems}\label{Sec-rot-nucl-syst}

The fastest known pulsar, PSR J1748-2446ad, makes one revolution in 1.4\,ms (angular frequency $\Omega\approx 4.5\times 10^{3}$\,Hz). Thus typical angular velocities of so far observed neutron stars are $\Omega R\lsim {0.15 c}$, where $c$ is the speed of light and $R\sim 10$\,km is the neutron star radius. The limiting rotation frequency of a compact star --- the Kepler frequency --- is determined by equation $m_N \Omega_{\rm K}^2 R^2/2 =  GMm_N/R$, where $G$ is the gravitational constant, $R$ is the radius of the object, and $M$ is its mass. For a compact star rotating with the frequency close to the Kepler frequency $\Omega_{\rm K}$,  the rotation affects  the equation of state resulting in an increase of the mass of the static star on the value  $\lsim (0.3-0.4) M_\odot$, cf. \cite{Farrell:2024bka}. When such a compact star collapses to form a black hole, its angular rotation frequency increases due to conservation of the angular momentum up to $\Omega< 1/r_G=1/(2MG)\sim 10^5$Hz for the compact star with the mass $M\sim M_{\odot}$, where $M_\odot\approx 2\times 10^{33}$\,g is the solar mass.
For primordial black holes with masses $M<10^{14}$\,g, this estimate would give $\Omega\lsim 10^{24}$\,Hz.

Rotational frequencies of excited  nuclei usually do not exceed $\Omega\sim 3\times 10^{21}$\,Hz, cf.~\cite{AfanasievNucl}. Estimates yield angular momenta $L\sim \sqrt{s_{NN}} A b/2\lsim 10^6\hbar$ in peripheral heavy-ion collisions of Au+Au  at the center of mass $NN$ collision energy $\sqrt{s_{NN}} = 200$ GeV, for the impact parameter $b = 10$ fm, where $A$ is the nucleon number of the ion~\cite{Chen2015}. The global polarization of $\Lambda (1116)$ hyperon observed by the STAR Collaboration in non-central Au-Au collisions~\cite{Adamczyk2017} indicated existence of a vorticity with  rotation frequency  $\Omega\simeq (9\pm 1) \times 10^{21}$ Hz $\simeq 7$\,MeV/$\hbar$.

Magnetic fields in ordinary pulsars, like  the Crab pulsar, reach values  $\lsim 10^{13}$\,G at their surfaces.
At the surface of magnetars magnetic fields are higher, $\gsim 10^{15}$\,G. In the interior the magnetic field might be even stronger (up to $\sim 10^{18}$\,G) depending on the assumed  mechanism of the formation of the magnetic field~\cite{Voskresensky:1980nk}.
Also, magnetic fields $\gsim (10^{17}\mbox{--}10^{18})$\,G, $h\sim h_{\rm VA}=Z e/R^2\sim H_\pi (Ze^6)^{1/3}$ (where typical time of the collision process was estimated as $\Delta t\sim R/c$) may exist in non-central heavy ion collisions  at collision energies $\sim $\,GeV/A, as it was evaluated in Ref.~\cite{Voskresensky:1980nk}, $Z$ is the charge of the fireball, $H_\pi =m_\pi^2/e\simeq 3.5\times 10^{18}$\,G.

As in Sect. \ref{nutshell-sec} we consider now negatively charged scalar bosons (or pseudoscalar  pions) with a mass $m_b$.
In a presence of a magnetic field and an electrical potential well, there exists a critical angular velocity, above which, from the vacuum in the rotation frame there arises a condensate of the charged bosons~\cite{Liu:2017spl,Guo:2021gbz,Voskresensky:2023znr,vosk25-111-036022,Bordag:2025bso}. In the frame rotating with the angular velocity $\vec{\Omega}\parallel z$, the solution for the charged condensate field has the form of a vortex, $e^{il\theta}\phi$, where $\theta$ is the angle in the cylindrical system  and $l$ is the quantum number of the angular momentum. To be specific let further $\Omega$ and $l$ be positive. In the presence of a uniform constant magnetic field $\vec{H}\parallel z$ and a rectangular potential well $V=-V_0=const$ for $\sqrt{x^2+y^2}<R$, the critical angular velocity is given by $l\Omega_{c}=-V_0+\sqrt{m^2_b +|eH|(1-2\alpha)}$  with
$-1\leq \alpha \leq 0$.

For the case of the rotating (for simplicity cylindric) nucleus equations for the boson field and the electric potential~(\ref{Vm}) and (\ref{Vel}) are to be replaced by the following ones~\cite{Voskresensky:2023znr,vosk25-111-036022},
\begin{align}
&\Delta_r \phi +[(V-\Omega l)^2 -m_b^2-(l/r -A_\theta(r))^2]\phi =0\,,
\label{VmR}
\\
&\Delta_r V =4\pi e^2 (n_p + 2(V-\Omega l)|\phi|^2) \,, \quad \Delta_r =\partial_r^2-\partial_r/r\,,
\label{VelR}
\end{align}
where  $A_\theta (r)$ is the angular component of the vector potential of the magnetic field, and a $\phi$-field self-interaction is for simplicity neglected. Equations (\ref{VmR}), (\ref{VelR}) should be supported by equation for the vector-potential $\vec{A}$. However with the ansatz that $A_\theta(r)=l/r$ in the vortex exterior we have $\mbox{curl}\vec{A}=0$ and equation for $\vec{A}$ is fulfilled identically.

Now in case of a  nuclear  system ($N\simeq Z$, $n\approx n_0$) of a large size solutions are $V=\Omega l-m_b\leq 0$, $|\phi|^2=n_p/(2m_b)$ and instead of (\ref{lightboson}) we find ~\cite{Voskresensky:2023znr,vosk25-111-036022}
\begin{align}
\mathcal{E}=-\mathcal{E}_0 A + \frac12(m_b -\Omega l)A + \mathcal{E}_{kin}\,,\, A=N+Z,
\label{lightbosonOmega}
\end{align}
where ${\cal{E}}_{kin}$ is the kinetic energy of the rotation of the nucleon subsystem. The term $l A/2$ is smaller or equal to  the initial angular momentum of the charged uniformly rotating nucleon sub-system $L\sim A m_N\Omega R^2$ (before appearance of the pion giant vortex, $l\gg 1$), $m_N$ is the  mass of the nucleon. At the rigid rotation $\Omega$ is determined by the initial value of the total angular momentum. The  critical  values of $l$ and $\Omega$ are determined by the condition $\Omega_c l_{c}=m_b-2\mathcal{E}_0$. The role of $m_b$ can be played by the effective mass of a charged pion, which decreases with an increase of the baryon density, see Sect.~\ref{sec:Superheavy} below. Here we considered the possibility of the second-order phase transition to a state with a giant vortex of the charged pion field. In reality, the phase transition can be of the first order with a jump of the condensate field at the critical point. Then the critical angular velocity is given by the condition $\Omega_c l_{c}=m_b-2\mathcal{E}_0+2\mathcal{E}_{\rm cond}(\Omega_c l_{c})$, where $\mathcal{E}_{\rm cond}\leq 0$ is a gain in the energy per baryon due to the formation of the vortex. In the rotating frame the nugget remains self-bound  until the system is not decelerated by the weak radiation force below $\Omega_c$. For a large-size system the deceleration time can be very long.

Now we mention another possibility. In the case of rectilinearly moving and rotating media, condensation of Bose excitations on a low-lying branch of the excitation spectrum $\omega(k)$, behaving similar to that in superfluid $^4$He and in Fermi liquids, was studied in Ref.~\cite{Voskresensky1993}, see also Ref.~\cite{Kolomeitsev:2016isb} and references therein.
As shown, when the speed of rectilinear motion exceeds a critical value (typically  Fermi velocity for a nucleon medium) a part of the  momentum can be transferred from the normal nucleon system to the superfluid condensate of Bose excitations with energy $\omega(k_c)$. Similarly when the rotation frequency exceeds some critical value a part of the angular  momentum  can be transferred from the normal nucleon system to the superfluid condensate of Bose excitations~\cite{Voskresensky1993,Kolomeitsev:2014gfa}. Consequently, the angular frequency of the normal component may decrease below $\Omega_{\rm K}$, whereas it would exceed $\Omega_{\rm K}$ otherwise.

When the baryon density in the interior of the star exceeds the value  $n_c^{\pi^+}\lsim n_0$, there appears a new branch of $\pi^+$ excitations with a negative frequency $\omega(k)$ and it may become energetically profitable to transport a part of the angular momentum  to the $\pi^+$ condensate droplets, for a more detail see Ref.~\cite{Voskresensky1993} and Sect.~\ref{cond-sec} below.
This allows to propose the following scenario. When during the initial compression stage of a supernova collapse, the density exceeds the critical density $n_c^{\pi^+}$ it becomes energetically favorable to transfer a part of the initial angular momentum from the nucleon subsystem to pion condensate drops. If the star is destroyed in the end, these drops will continue to rotate (with angular velocities $\Omega R_{\rm d} <1$, with $R_{\rm d}$ being a size of the drop) until the angular momentum has been completely radiated from the drop surface. These drops of $\pi$ condensate nuclear matter may be observed as abnormal nuclei and nuclearites propagating in cosmos.

Notice also that differentially rotating stars can support a larger star mass than their uniformly rotating or non-rotating counterparts~\cite{Morrison2004}. In the case of massive remnant stars of binary merger events, differential rotation is one of the mechanisms that may provide an  extra centrifugal support to stabilize the star well above the Tolman-Oppenheimer-Volkoff  (uniform rotation) mass limit.

\subsection{Anomalous supercharged  nuclei consisting dark matter}

In 1981 Cahn and Glashow~\cite{Cahn:1980ss} assumed existence of superheavy charged dark matter particles, cf. also~\cite{DeRujula:1989fe}.
In 2005 Glashow~\cite{Glashow:2005jy} conjectured that more specifically the dark matter may consist of the U tera-quarks and O-{UUU} (charge $-2$) baryons, and in 2006 Khlopov~\cite{Khlopov:2005ew} studied  $O$He-hidden dark matter with $m_{ O} \sim\!1 $\,TeV. Reference~\cite{Gani:2018mey} suggested existence of stable $O$-nuclearites of arbitrary size (up to $A\sim 10^{57}$), if there exist charged dark matter stable superheavy particles. In this case the proton charge of the nucleus is compensated by the heavy $O$-particles. Due to the large mass of the $O$ particles they yield a negligible contribution to the kinetic energy.  Therefore, the ordinary ISM at density $n\simeq n_0$ with embedded $O$-particles compensating the charge  would be absolutely stable. Similarly to estimates (\ref{LagrSuperch}), (\ref{lightboson}) performed above for supercharged nuclei, in case of  the nuclear--$O$ matter one may present the energy as
\begin{eqnarray}
\mathcal{E}\simeq \mathcal{E}_0  A -\int d^3 x (n_p -2n_O) V\,.
\end{eqnarray}
With $2n_O=n_p$ we get ${\cal{E}}\simeq \mathcal{E}_0 A=-16A\,{\rm MeV}$.

At present there exist various anomalous events, some of which may have a relation to the discussed above scalar fields, $\pi^-$ condensate, quark  or dark matter nuggets, see \cite{Zhitnitsky:2024ydy} and references therein and a discussion in Sect. \ref{sec:observation-sec} below.

\section{Pion mode and condensation in  dense nuclear systems}\label{sec:Superheavy}

The phenomenon of a pion condensation was proposed  in Refs.~\cite{Migdal1971,Migdal:1972,Scalapino:1972fu,Sawyer:1972cq} and investigated then in numerous works, see \cite{Migdal1978,Migdal:1990vm,V1993} and references therein.
The pion-nucleon interaction is strongly attractive in the p wave. Therefore, starting from the first works  one expected that a pion condensate is formed  in a state with a finite momentum $k\neq 0$ when the nucleon density exceeds some critical density $n_c^{\rm (p)}$~\cite{Migdal:1973PL,Migdal:1973zm,Baym:1973zk,MMM1974,Campbell:1974qt,Campbell:1974qu,Baym:1975tm,Dautry,Tamagaki77}.
In first works one anticipated that $n_c^{\rm (p)}$ might be smaller than $n_0$. Soon, after analyses of atomic nucleus data, it became clear that $n_c^{\rm (p)}>n_0$ because of a stronger repulsive effect of short-range nucleon-nucleon correlations, see ~\cite{Migdal1978,Migdal:1990vm} and references therein. In Ref.~\cite{Migdal1974} Migdal used the name ``superheavy anomalous nuclei'' and then ``superdense'' nuclei in Ref.~\cite{Migdal1978} and later in Ref.~\cite{Migdal:1990vm} simply ``abnormal'' pion condensate nuclei.
These objects are neutron-rich to reduce the repulsive Coulomb energy. The key idea is that for $n>n_c^{\rm (p)}\gsim n_0$ there may appear the $\pi^{+}$, $(\pi^{+}\pi^-)$, or $\pi^0$ condensates leading a decrease in the energy density by a value which can be roughly estimated as
$$E_\pi \simeq -\frac12\beta_0 (n) (n-n_c^{\rm (p)})^2\theta (n-n_c)$$
with  $\beta_0 \lsim 1/m_\pi^2$, compare with Eq.~(\ref{k-correl}) that was used above in the case of the Lee-Wick  $\sigma$ condensate.
The quantity $\beta_0$ depends on the value of the Landau-Migdal parameter $g'$, which is not well determined for $n\neq n_0$ and for isospin-asymmetric matter. For a smaller $\beta_0$, the pion condensate only slightly modifies the total energy density. For a larger $\beta_0$ the gain in the condensate energy may overwhelm the loss in the purely baryon repulsive contribution. In the latter case there could exist metastable or even absolutely stable abnormal $\pi$ condensate nuclei and  nuclei-stars.

\subsection{Pion-nucleon interaction amplitude and pion condensation}

Resummation of a nucleon-nucleon interaction in the particle--hole channel in the Fermi liquid approach yields the following representation of the $NN$ interaction amplitude in the spin-isospin channel \cite{Migdal1978,Migdal:1990vm}
%--------
\begin{eqnarray}
\parbox{7.5cm}{
\includegraphics[width=7.5cm]{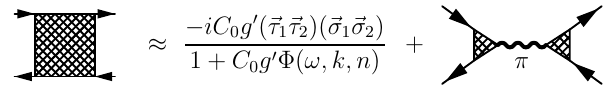}
}\,.
\label{gam-res}
\end{eqnarray}
Here $\tau$ and $\sigma$ are Pauli matrices acting in the isospin and spin spaces, respectively, $g'$ is the Landau-Migdal parameter of $NN$ correlations in the spin-isospin channel, $C_0^{-1}$ is the normalization factor -- the density of states at the Fermi surface at $n=n_0$. Similar equation holds for separated $nn(pp)$ and $np$ spin channels in isospin-asymmetric matter, where $g_{nn(pp)}$ and $g_{np}$ Landau-Migdal parameters enter instead of the $g'$ parameter. Note that the $\Delta$-nucleon Landau-Migdal parameter $g'_{\Delta N}\ll g'$ and thereby we for simplicity disregarded this contribution in (\ref{gam-res}), whereas the $\Delta$ contribution in the second term of Eq. (\ref{gam-res}) is taken into account, see \cite{Migdal:1990vm} for a more detail. $\Phi(\omega,k,n)$ stands for the famous Lindhard function corresponding to the particle-hole loop
\begin{align}\parbox{3cm}{
\includegraphics[width=3cm]{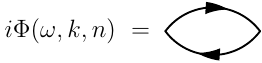}}.
\label{FigNN}
\end{align}
For nonzero temperature $T$ it also depends on $T$.
The first term in r.h.s. of Eq.~(\ref{gam-res}) can be cast as the $NN$ vertex correlation factor
\begin{equation} \label{gam-x}
\Gamma (x,\omega,k,n) =\frac{1}{ 1+2\,x\,\frac{p_\rmF(n)}{p_\rmF (n_0 )}\Phi (\omega,k,n)
}\,,
\end{equation}
which appears not only in the spin-isopsin channel (for $x=g'$) but in other channels with other Landau--Migdal parameters
$x=f$, $f^{\prime}$, $g$. The wavy bold line represents the full pion Green function. For $\omega\ll m_\pi$ and $k\lsim p_\rmF$ the main contribution to the $NN$ amplitude is determined by the second diagram in r.h.s. of (\ref{gam-res}).

Main contributions to the pion polarization operator come from the nucleon--nucleon-hole and $\Delta$--nucleon-hole terms corrected by the short-range correlations, for more details see \cite{Migdal:1990vm,V1993,Voskresensky2023}. The contribution of the $\Delta$--nucleon-hole is as important as that of the nucleon-nucleon hole. Indeed, the mass of the $\Delta$ isobar is only $2m_\pi$ larger than the nucleon mass but the degeneracy factor is four times larger and the attractive $\pi N \Delta$ coupling constant is twice as large as the $\pi N N$ one.

The retarded pion propagator is
\begin{align}
D_\pi^R(\om,\vec{k},n_n,n_p)=\frac{1}{\om^2-m_\pi^2-k^2-\Pi^R(\om,k,n_n,n_p)}\,,
\end{align}
where $\Pi$ is the pion polarization operator, $n_n$ and $n_p$ are the neutron and proton densities. For $\om =\mu_\pi$:
\begin{align}
-[D_\pi^R(\mu_\pi,\vec{k})]^{-1} &= \widetilde{\omega}^2(\mu_\pi ,k^2)
\nonumber\\
&=m^{2}_\pi +k^2+\Re\Pi^R (\mu_\pi, k^2)-\mu^2_\pi\,,
\end{align}
$\Re\Pi(\om,k)$ is  the real  part  of the pion polarization operator, the imaginary part $\Im\Pi(\mu_\pi,k)=0$, $f_{\pi NN}\approx m_\pi^{-1}$ is the $\pi NN$ coupling constant, and  $\Gamma(n)$ is the factor taking into account the $NN$ correlations, $\Gamma(n\to 0)\to 1$, $\mu_\pi$ is the pion chemical potential. In the case of the ISM, $\mu_{\pi^{\pm}}=0$ in accordance with reactions $p\leftrightarrow n+\pi^+$, $n\leftrightarrow p+\pi^-$; $\mu_{\pi^0} =0$  for the case of arbitrary isospin composition in accordance with reactions $N\leftrightarrow N+\pi^0$, $N=(n,p)$.
The quantity
\begin{align}
\widetilde{\omega}^2_0=\min_{k}\{\widetilde{\omega}^2(\mu_\pi ,k)\}=\widetilde{\omega}^2(k_0) \,,
\label{efpigap}
\end{align}
where the minimum is realized at a density dependent momentum  $k_0(n)$, has the meaning of the squared effective pion gap, see Ref.~\cite{Migdal:1990vm} and references therein. We notice that the in-medium $\pi^-$ can play a role of the light negatively charged boson considered in Sect.~\ref{sec:Anomalous} with the parameter $m_b^2$ used there replaced by the squared effective pion gap $\widetilde{\omega}^2_0$.

At densities $n\sim n_c^{\rm (p)}$ the pion gap behaves like  $\widetilde{\omega}^2_0\sim \beta(n_c^{\rm (p)}-n)$, with a constant $\beta(\mu_\pi,k_0)>0$ (at the neglect  of the quantum fluctuations) and the change in the sign of the gap indicates the onset of instability with respect to the pion condensation with the momentum $k_0$. The critical point $n_c^{\rm (p)}$ is the same for $\pi^\pm,\pi^0$ in ISM. In the neutron star matter, $n_c^{\rm (p)}=n_c^\pm$ for charged pions, and it is equal to  $n_c^0$  for $\pi^0$. Which value $n_c^\pm$ or $n_c^0$ is larger depends on the model under consideration, see \cite{Migdal1978,Migdal:1990vm}.

In the ISM  the momentum dependence of the pion gap,  $\widetilde{\omega}^2(0,k^2)$ develops  minimum at $k=k_0\neq 0$ at densities $n>n_c^{(1)}\simeq (0.5-0.7)\,n_0$, see \cite{V1993} and in its vicinity one may use parametriztion:
$$\widetilde{\omega}_0^2(k^2)\simeq \widetilde{\omega}_0^2+\gamma \frac{(k^2-k_0^2)^2}{4k_0^2} \,,$$
where $\gamma\sim 1$. The squared effective pion gap $\widetilde{\omega}^2_0$ changes sign at $n=n_c^{\rm (p)}$.
Considering the pion condensation in a finite system for $n>n_c^{\rm (p)}>n_0$, one may find the spatial structure of the classical pion field solving the equation
 $$\widetilde{\omega}^2(\mu_\pi ,\hat{k})\phi +\lambda |\phi|^2\phi =0\,,\quad  \hat{k}=-i\nabla \,,    $$
which replaces Eq.~(\ref{Vm}). Nuclear systems obeying the roton-like spectra have been extensively studied, see~\cite{Voskresensky:1980nk,Migdal:1990vm,Voskresensky2023} and references therein. From a recent time, some authors became to  name  such a spectrum the ``moat'' spectrum.

 %--------
\begin{figure}
\centerline{\includegraphics[height=8cm,clip=true]{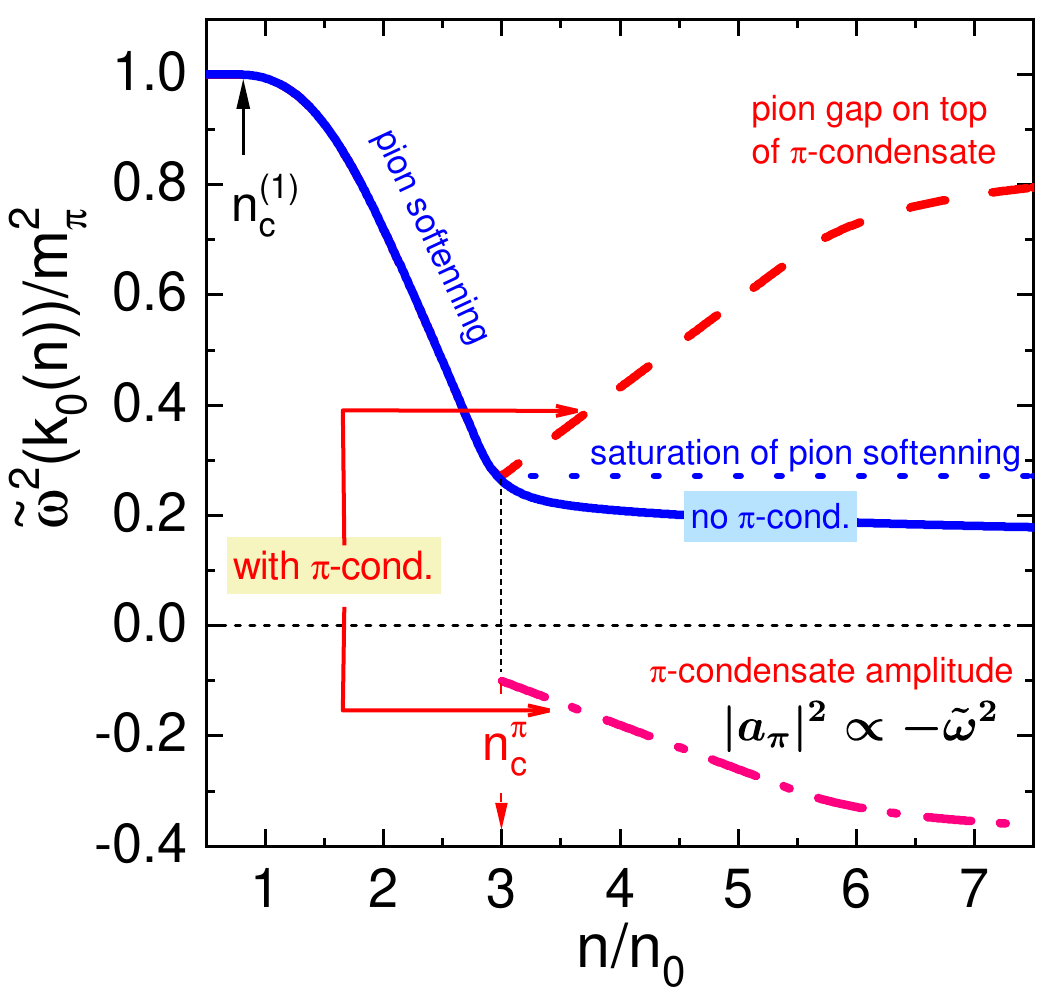}}
\caption{Typical density dependence of the squared effective pion gap for $\pi^{\pm,0}$ in ISM and for $\pi^0$ in neutron star matter.
Solid line shows $\widetilde{\omega}^2 (k_0(n))$ for $n_{c}^{(1)}<n<n_c^{\pi}$.
For $n>n_c^{\pi}$  it is continued for the case when the pion condensate is not formed and the system is in a metastable state. Dotted line corresponds to the case of a saturation of the pion softenning due to artificially strong increase of the Landau--Migdal parameter $g^{\prime}$ with a density growth. Dash-dotted line characterizes the amplitude of the pion condensate appearing in the first order phase transition at $n_c^{\pi}$. Dashed line shows the squared effective pion gap on top of the pion condensate.}
\label{piongap1}
\end{figure}

%%%%%%%%%%%%%%%%%%%%%%
A typical density dependence of the squared effective pion gap, $\widetilde{\omega}^2 (k_0(n))$, for $\pi^{\pm},\pi^{0}$ in ISM and for $\pi^0$ at any $N/Z$  is illustrated in Fig.~\ref{piongap1}, see also Ref.~\cite{Migdal:1990vm}. For $n>n_{c}^{(1)}$ the effective pion gap acquires a minimum at a finite momentum $k_0$. With a subsequent density increase, $\tilde{\omega}^2 (k_0(n))$ decreases (solid line). Various calculations done within a mean-field approximation (ignoring contributions of pion fluctuations to the pion self-energy) demonstrate that at $n>n_c^\pi$ the pion condensate with a liquid-crystal-like or a solid-like structure may appear via the second-order phase transition.  Actually at $n=n_c^\pi$  the classical pion field is developed, in the first-order phase transition due to the effect of fluctuations with $k\sim k_0\sim p_{{\rm F},N}$. Dashed-dotted line illustrates the sudden jump and the following growth of the pion condensate amplitude. The squared effective pion gap on top of the condensate is shown by the dashed line. It is growing with an increasing density. As we have noted, in the ISM the value $n_c^\pi$ is the same for $\pi^{\pm}$ and $\pi^0$. In the isospin-asymmetric matter the value $n_c^\pi$ depends on the sort of the pion. Actual values of the critical densities depend on poorly known density dependence of the Landau-Migdal parameters. Numerical analysis~\cite{Akmal:1998cf} performed within the ``variational theory of nuclear matter'', which employs  the realistic  two-nucleon potential in vacuum and then introduces three-nucleon interactions, gave the value $n_c^\pi \simeq 2n_0$ for the critical density of the $\pi^+ ,\pi^- ,\pi^0$ condensation in the ISM and the value $n_c^\pi\simeq 1.3n_0$ for the $\pi^0$ condensation in the purely neutron matter. Cooling of neutron stars is appropriately described within the ``nuclear medium cooling scenario" which takes into account of effects of the pion softening~\cite{Voskresensky:1984zzn,Voskresensky:1986af,Voskresensky:1987hm,Migdal:1990vm} provided $n_c^\pi \gsim (2-2.5)n_0$ for $\pi^{\pm}$ and $\pi^0$ condensates, cf.~\cite{Schaab:1996gd,Voskresensky:2001fd,Blaschke:2004vq,Grigorian:2005fn,Blaschke:2011gc,Blaschke:2013vma,
Grigorian:2016leu,Grigorian:2018bvg} and references therein. Continuation of the solid line for $n>n_c^\pi$ corresponds either to the possibility of a metastable phase in the system, when the ground state contains no liquid-crystal-like or solid-like pion condensates. Dotted line corresponds to the case of saturation of the pion softening because of an artificially strong increase of the Landau--Migdal parameter $g^{\prime}$ with the density.

Finally, we note that the resummation of the in-medium $NN$ interaction (\ref{gam-res}) imposes that at $n>n_0$ the $NN$ cross section has strong density dependence owing to the interplay of a moderately suppressing effect from the vertex correction (\ref{gam-x}) because of $NN$ correlations and a strong enhancing  effect of the softening of the pion mode with the increasing density and, therefore,
\begin{align}
\sigma_{NN}\propto \frac{\Gamma^4 (g',0,k_0,n)}{(\tilde{\omega}(k_0(n))/m_\pi)^4}.
\label{cross}
\end{align}

\subsection{$\pi^\pm$ condensate nuclear systems}\label{cond-sec}

Spectra of pions in the ISM and purely neutron matter were demonstrated in \cite{MMM1974}.
In the latter case for $\pi^\pm$  there are the $\pi^+$ and $\pi^-$ branches, $\om^{\pi^+}(k)$, $\om^{\pi^-}(k)$ (corresponding to $\om\to m_\pi$ for $n\to 0$) and the higher lying $\Delta$ branches. It proves to be that up to large densities $\om^{\pi^-}(k)>\mu_e$  \cite{Migdal:1973PL,MMM1974} and the reaction $N+e\to N+\pi^-+\nu$ is forbidden. If pion were free, $\om^2=m^2_\pi+k^2$, such a reaction were allowed for $\mu_e(n)>m_\pi$ resulting in efficient cooling of the neutron stars heaving such a densities in their interiors.
For  $n>n_c^+$ ($n_c^+<n_0$ according to evaluations of  \cite{MMM1974}) there appears extra branch with $\om^{\pi^+}_s(k)<0$ corresponding to appearance of the so called $\pi^+_s$ condensate. We have mentioned this possibility in Sect. \ref{Sec-rot-nucl-syst}. Accordingly, for $n>n_c^+$ the reaction $p\to n+\pi^+$ becomes possible resulting in the production of $\pi^+$'s occupying this branch. For $n=n_c^\pm >n_c^+$ the branches  $-\om^{\pi^+}_s(k)$ and $\om^{\pi^-}(k)$ coincide  at  $k=k_0\neq 0$, where  $\om^{\pi^+}_s (k_0)+\om^-_\pi (k_0)=0$. Thereby for $n>n_c^\pm$ the reactions $N\leftrightarrow N+\pi^+ +\pi^-$ become energetically favorable and there appears $\pi^\pm$ condensate. Appearance of $\pi^\pm$ condensation for $n>n_c^\pm$ results in the efficient neutrino cooling processes, e.g.  via $N+\pi_{cond}^-\to N+e+\bar\nu$ reaction.

At large densities the baryon subsystem is essentially rearranged. In the approach exploiting the chiral symmetry ~\cite{Campbell:1974qt,Campbell:1974qu,Baym:1975tm} it was shown that instead of two different Fermi seas for protons and neutrons, a common Fermi sea for baryon quasiparticles, being superpositions of proton neutron and  $\Delta$ isobar states mixed due to interactions with the developed pion condensate, is formed. At densities much larger than $n_c^{\rm (p)}$ the condensate becomes developed and the squared amplitude of the pion condensate reaches its maximum value $f_\pi^2/2$ for $\pi^{\pm}$, where $f_\pi\approx 92$ MeV is the pion weak decay constant and the electric charge density of the baryon subsystem becomes equal to $n/2$. Taking into account these effects, and  electron and muon populations, the energy density acquires the form~\cite{Voskresensky:1977mz}
\begin{align}
&{E}_{B+\pi+e+\mu}  = {E}_B + {E}_\pi + {E}_e + {E}_\mu\,,
\label{EB-picond}
\\
&{E}_{e,\mu} \simeq \intop_0^{\sqrt{V^2-m_{e,\mu}^2}}\frac{\rmd p p^2}{\pi^2}\sqrt{m_{e,\mu}^2+p^2}\,\Theta (|V|-m_{e,\mu})\,,
\nonumber\\
&{E}_\pi =\frac{V^2 f^2_\pi}{2} +\frac{1}{3}( m^*_\Delta-m^*_N)\,n - \frac{81}{50}f^2_{\pi NN}(1-g') n^2\,,\nonumber
\end{align}
where $E_B$ is the energy density of the baryonic subsystem, $g'$ is the isospin Landau-Migdal parameter, and $m_\Delta^*$ is the effective mass of the $\Delta$ isobar (one may put for simplicity $m^*_\Delta-m^*_N \simeq m_\Delta-m_N=293$\,MeV), $\Theta(x)$ is the step function. It is taken into account that in case of the developed condensate $\mu_n\simeq \mu_p$. For more details the reader may consult Refs.~\cite{Voskresensky:1977mz,Migdal:1990vm}.
The electric charge of protons is screened by the condensate of negative pions and by electrons and muons produced via the weak $\beta$ processes and from the vacuum  (at rather large  values of $Z$)~\cite{Voskresensky:1977mz}.
The electric potential is found from the Poisson equation for a spherical system of  radius $R$  and constant baryon density $n$,
\begin{align}
\Delta V &= 4\pi e^2 (n_e+n_\mu+n_h)\,,\label{PoisCond}
\\
n_h &= \big({\textstyle\frac12} n +f^2_\pi V\big)\Theta (R-r)\,,
\nonumber\\
n_{e,\mu} &= -\frac{1}{3\pi^2} \big(V^2-m_{e,\mu}^2\big)^{3/2}\Theta(|V|-m_{e,\mu})\,.\nonumber
\end{align}
Here $f^2_\pi V$ is the contribution of the developed $\pi^-$ condensate as it follows from the consideration within the $\sigma$ model and $n/2$ is the total baryon charge density in the presence of the developed condensate~\cite{Baym:1975tm}, $n_e$ and $n_\mu$ are contributions of electrons and muons. Nullifying  the r.h.s. of Eq.~(\ref{PoisCond}), similarly to that we did above for Eq.~(\ref{Vel}), demonstrates the charge neutrality condition in the interior of a large size system, for $R\ll l_V$ where $l_V$ is the Debye screening length.

\begin{figure*}
\includegraphics[width=12cm]{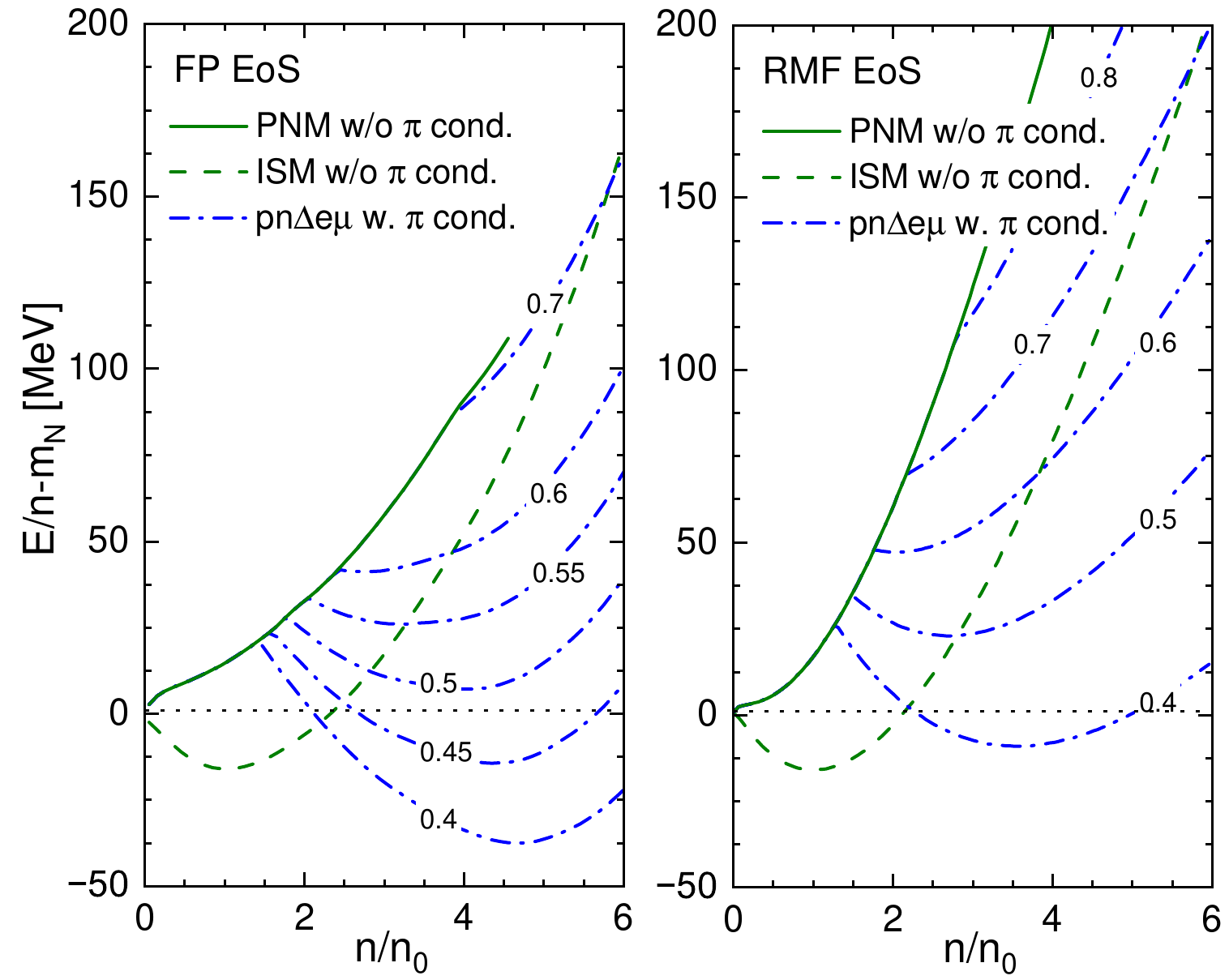}
\caption{Binding energies per nucleon for the PNM (solid lines), the ISM (dashed lines), and for the beta-equilibrium matter ($p n \Delta e\mu$) with the developed $\pi^-$ condensate (dash-dotted lines). The latter lines are calculated for several values of the Landau-Migdal parameter $g'$ characterising the strength of the short-range nucleon-nucleon correlations; the parameters are shown by labels on the lines.
On the left panel, the nuclear part of the energy is taken from the FP equation of state~\cite{Friedman:1981qw}. On the right panel we use the RMF model~\cite{Cubero1988}. Extra explanations are given in the paper body.   The choice of parameters is such that the RMF equation of state supports the maximum neutron star mass of $2.09\,M_\odot$~\cite{KV2005}.
\label{fig:dpicond}}
\end{figure*}

In Fig.~\ref{fig:dpicond} by solid curves  we show the binding energy per nucleon for the purely neutron matter (PNM). Other curves show calculations for the ISM with and without the developed condensate done with the help of Eq. (\ref{EB-picond}). Actually this expression does not hold for $n\lsim n_c^{(p)}$ since it was derived for $n\gg n_c^{(p)}$, see a discussion in  \cite{Migdal:1990vm}. However we conjecture that for  $n\gsim (2-3)n_0$ we already can deal with the developed pion condensate On left panel the nucleon  part (dashed line) is calculated with a soft Friedman-Pandharipande (FP) equation of state~\cite{Friedman:1981qw} and on right panel with a stiffer RMF equation of state, see Eq.~(\ref{Lee-W}). In the RMF case parameters are tuned to describe the saturation density $n_0=0.160$\,fm$^{-3}$, the binding energy $\mathcal{E}_0=-16.0$\,MeV, the compressibility modulus $K=273.6$\,MeV, the effective nucleon mass $m_N^*(n_0)=0.755$\,MeV, and the value of the coefficient of the nuclear symmetry energy at saturation, $J=32.0$\,MeV. With these  parameters the  maximum neutron star mass is $2.09\,M_\odot$, being larger $2\,M_\odot$ in agreement with the modern experimental data.
The attraction strength of the $\pi^{\pm}$ condensate contribution, $E_\pi$ in Eq.~(\ref{EB-picond}), is controlled by the value of the $g'$ Landau-Migdal parameter. The density dependence of this parameter is poorly known, therefore following \cite{Voskresensky:1977mz,Migdal:1990vm} we allow for its variation.
So, the energies  per nucleon of the baryon, $pn\Delta e\mu$ matter with the pion condensate (dash-dotted lines) are shown for several values of $g'$. Note that various fits of the atomic nuclei data produce   the value of $g'(n_0)\approx 0.7$. At this instance to avoid a possible misunderstanding we should mention that different authors use different normalizations of $g'$ and distinct expressions for the pion polarization operator. The smaller $g'(n)$ is the stronger is the pion  condensate attraction for $n>n_c^{\rm(p)}$. In \cite{Migdal:1990vm} there were given some arguments for a possible decrease of $g'(n)$ with increasing density. For the FP equation of state (left panel in Fig.~\ref{fig:dpicond}) the energy of the pion condensate matter becomes negative in some interval of density for $g'(n>n_c^{\rm (p)})<0.48$. In this case there may exist superdense abnormal pion condensate nuclei, nuclearites and nuclei-stars of arbitrary size bounded by nuclear forces rather than by gravity (for $10^3\lsim A\lsim 10^{56}$), as it was suggested in Ref.~\cite{Voskresensky:1977mz}. The maximal net charge of an abnormal nucleus with $A\gg 1/e^3$ observed at infinity behaves as $A^{1/3}$. In this case the extra charge is with necessity screened by electrons and muons occupying the vacuum shell. The pion condensate state becomes an absolute ground state of nuclear matter if $g'<0.45$, then $\min_n\{{E}_{B+\pi}\}< -16$\,MeV. For $0.48<g'< 0.62$ there is a minimum in energy ${E}_{B+\pi}$ that would correspond to a positive energy metastable state and for $0.45<g'< 0.48$ to a negative energy metastable state. For the RMF nucleon equation of state, which we used (see right panel in Fig.~\ref{fig:dpicond}) the abnormal negative energy states may exist for $g'<0.43$ and  the absolute ground state of matter appears for $g'<0.38$. The metastable states with positive energy are possible for $0.43<g'<0.63$.

\subsection{Dense $\pi^0$-condensate ferromagnetic nuclear systems}\label{Ferro}

Possibility of neutron-enriched abnormal  $\pi^0$ condensate nuclei was discussed  in~\cite{Migdal1974,Migdal:1976wm}.  Ferromagnetic behavior of $\pi^0$ condensate was  studied within the $\sigma$ model first in~\cite{Dautry}. The $\sigma\pi^0$ running wave condensate in the neutron matter corresponds to a state, in which all neutron spins are aligned, presumably with some macroscopic domain structure. Recently there were found that extra  contributions to the p-wave $\pi^0$ condensate term in the Lagrangian may arrive from the so-called  Wess-Zumino-Witten  axial anomaly term describing the anomalous interaction of the neutral pion field with the external electromagnetic field and the rotation, and a related pion contribution to the baryon current, cf.~\cite{SonStephanov,Hatsuda,Hashimoto2015,BraunerYamamoto,Yamamoto}. Comparison of the ferromagnetic phase of the $\pi^0$ condensate appearing due to the axial anomaly term and the alternating layer structure of the pion condensate~\cite{Tamagaki77} demonstrated that the latter state is probably energetically more favorable \cite{Hashimoto2015}. These effects were recently studied in Ref.~\cite{Voskresensky:2025gjy}.

\subsection{About the s-wave pion condensation}\label{swave}

Pion off-mass-shell effects are of primary importance for description of the pion spectra and a possibility of the pion condensation. The most important contribution to the pion polarization operator is given by the p-wave pion-nucleon and the pion-$\Delta$-isobar interactions. For an s-wave part of the pion polarization operator already first works  employed the Weinberg-Tomozawa expression, $\Pi^s_{\pi^-} =C (n_n-n_p)\,\om$, with $C\simeq 1/(2f_\pi^2)\simeq 1/m^2_\pi$, which does not contribute in case of  the ISM. Parametrization of the optical pion-nucleus potential
\cite{Migdal1978,Migdal:1990vm,V1993} used the fully off-mass-shell pion-nucleon amplitude, which fulfills the current algebra theorems and the canonical PCAC condition (see the MSTV model in \cite{Voskresensky:2025gjy}). Oppositely,
Refs. \cite{Delorme1992,Ericson1994,KKW2003,KKW2003b} (see the KKW model in \cite{Voskresensky:2025gjy}) used the on-mass-shell pion-nucleon amplitude, taking incoming and outgoing pion 4-momenta such that $q^2=q^{'2}=m^2_\pi$, that does not allow to fulfill the so
called Adler and Weinberg current algebra conditions. Subtleties associated with the current algebra theorems and field redefinitions were discussed in \cite{Voskresensky2022}. The s-wave pion-nucleon interactions in the linear sigma model, and in the Manohar-Georgi and Gasser-Sainio-Svarc models with finite number of terms in Lagrangians, as well as in a general phenomenological
approach were reviewed.

\begin{figure}
\centering
\includegraphics[width=8cm]{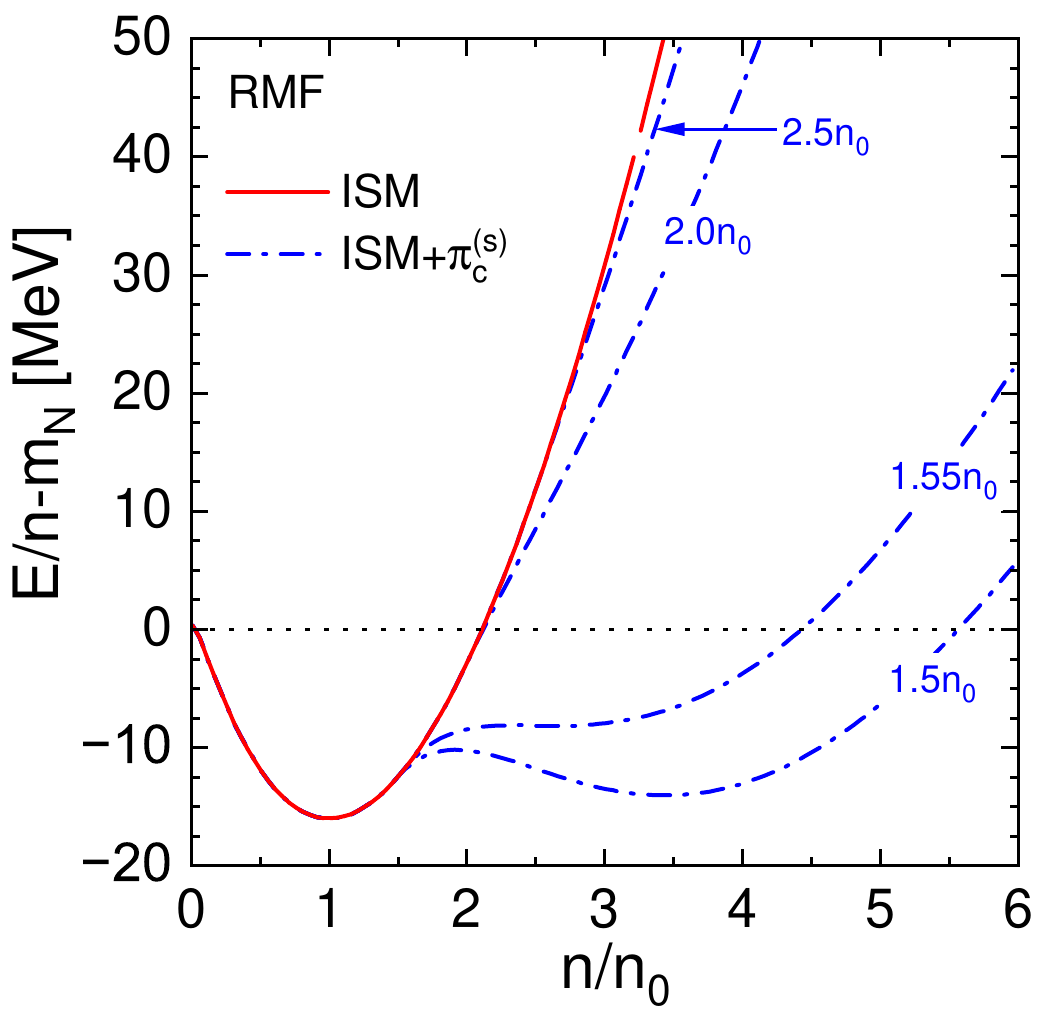}
\caption{Binding energy per nucleon for the ISM with and without the neutral $\pi^{\pm,0}$ condensate (solid and dash-dotted lines, respectively). The baryon part of the energy for ISM is calculated in the framework of the RMF model used on the right panel of Fig.~\ref{fig:dpicond}. The dashed dotted curves are calculated within the KKW model for different values of the critical density $n_c^{\rm (s)}$ indicated by the line labels. In calculations we used $\lambda=m_\pi^2/(2f_\pi^2)$.
\label{fig:spicond}}
\end{figure}

With the MSTV model for the s-wave pion-nucleon interaction the s-wave pion condensation in the ISM does not occur at least up to very high densities. Oppositely, if the KKW model \cite{Delorme1992,Ericson1994,KKW2003,KKW2003b}
were valid, the s-wave pion condensation in the ISM would be expected to occur already for $n>n_c^{\rm (s)} \sim (1.4-2.5)n_0$, if the nucleon-nucleon correlations are not taken into account, resulting in the energy density gain
\begin{align}
E_\pi^{\rm KKW}(n)\simeq -\frac{m^4_\pi}{2\lambda} \left(\frac{n}{n_c^{\rm (s)}} - 1 \right)^2\Theta (n-n_c^{\rm (s)})\,.
\label{E-PiC-s}
\end{align}
In Fig.~\ref{fig:spicond} we plot the energy per particle of the ISM with and without  contribution of the s-wave pion condensation (\ref{E-PiC-s}). We use the same RMF model as on the right panel of Fig.~\ref{fig:dpicond}. The pion-pion self interaction parameter is chosen to be  $\lambda=m_\pi^2/(2f_\pi^2)$. The value of the softening of the equation of state essentially depends on the value of the critical density. As we see, for $n_c^{\rm (s)}\gsim 2.5\,n_0$ the softening is weak, whereas for $n_c^{\rm (s)}\leq 1.55\,n_0$ the effect is strong and there appears  possibility of a metastable state with the s-wave pion condensate. Thereby one could expect to observe  experimental manifestations of the s-wave pion condensate in heavy-ion collisions. We must, however, emphasize that the estimate of the value of the critical density $n_c^{\rm (s)}$ remains strongly model-dependent even within the on-mass shell based model \cite{Delorme1992,Ericson1994,KKW2003,KKW2003b}. For example, the $NN$ correlation effects in the $s$-wave pion polarization operator, which cannot be constrained from the experiments at present, could shift the condensation critical point to higher values~\cite{Voskresensky:2025gjy}.

\section{$\Delta$ resonance  matter}\label{sec:Delta-sec}
Above we  indicated that  the pion-nucleon-$\Delta$ attraction promotes a decrease of the effective pion gap and occurrence of the  pion condensation at smaller baryon density. Now we will
review the idea of $\Delta$ resonance matter.

The idea of a stable or metastable baryon-resonance matter was introduced in Ref.~\cite{Troitsky:1979ch}.
The Fermi momentum of the nucleon is $p_\rmF\propto n^{1/3}/\nu^{1/3}$, the total Fermi energy of baryons grows as $E_\rmF\propto n^{2/3}\nu^{2/3}$ where $\nu$ is the degeneracy factor. When the nucleon Fermi energy, $E_\rmF$, reaches the difference between effective masses $m^*_\Delta -m^*_N$, it becomes energetically favorable to create $\Delta(1232)$ isobars. The appearance of any new degree of freedom results in a softening of the equation of state. The nucleon--nucleon-hole contribution to the pion polarization operator is $\Pi_{\pi NN}\propto -\nu p_{\rm F}k^2\propto \nu^{2/3} n^{1/3}$,  Therefore, the latter quantity effectively corresponds to the $\nu^2$ times higher density compared to the case of $\nu =1$. The $\Delta$--nucleon-hole contribution  is $\Pi_{\pi \Delta N}\propto -nk^2$, see~\cite{Voskresensky:1982vd}. The extra $\Delta$--$\Delta$-hole attractive term to the pion polarization operator leads to a further decrease of the critical density of the pion condensation. Moreover, let us recall that the spin-isospin degeneracy factor for $\Delta$ is $\nu =16$ compared to 4 for nucleons and that $f_{\pi N\Delta}\simeq 2 f_{\pi NN}$.

\begin{figure*}
\centering
\includegraphics[width=8cm]{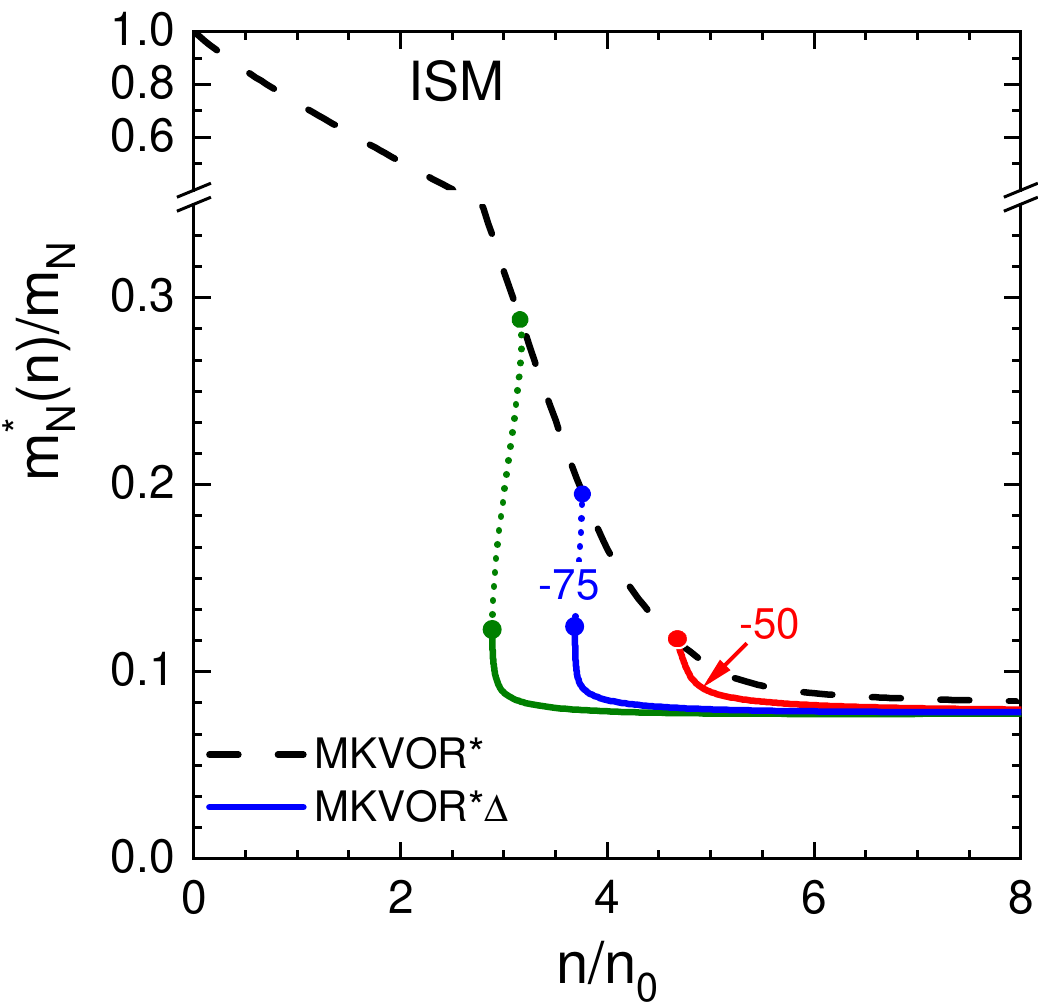}
\includegraphics[width=8cm]{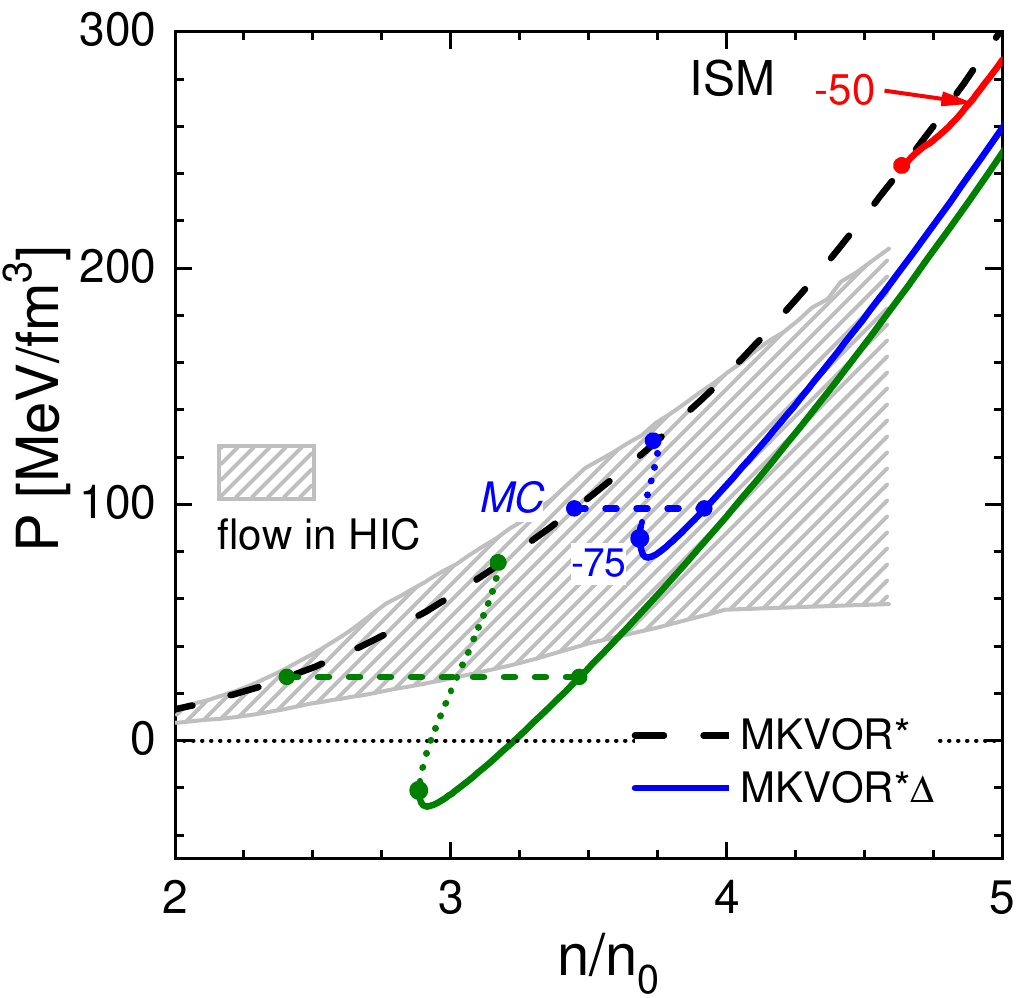}
\caption{Left panel:
Effective nucleon mass as a function of density in the ISM  without $\Delta$ isobars (dashed line) and with the allowance for $\Delta$s  (solid continuations). Right panel: Pressure as a function of density without and with $\Delta$s shown by dashed line and dashed line with dotted--solid continuations, respectively. The hatched region indicates the constraint obtained from the analysis of the momentum  distributions (directed flow) of particles produced in heavy-ion collisions~\cite{Danielewicz:2002pu}. The calculations are done within the RMF models MKVOR* and MKVOR*$\Delta$ for several values of the $\Delta$ potential at saturation density $U_\Delta =-50,-75$, and $-100$\,MeV, see Ref.~\cite{KMV2017} for details
}
\label{fig:mkvstar-meff}
\end{figure*}
$\Delta$ isobars are produced copiously in relativistic heavy-ion collisions forming a resonance matter~\cite{Metag}. A possibility of density isomer states was discussed in~\cite{Boguta1982,Waldhauser1987,Cubero1988,Diakonov:1987zp} within RMF models and the chiral model of the nucleon.

Reference~\cite{KMV2017} employed the RMF models of the baryonic matter, in which all involved hadron masses are assumed to be functions of the scalar $\sigma$ mean field. Below we consider two specific examples of such a model: the MKVOR* model not including $\Delta$ isobars and the MKVOR*$\Delta$ model including $\Delta$ isobars. The key ingredient of the RMF model with $\Delta$ isobars is the magnitude of the $\Delta$ optical potential at the saturation density, $U_\Delta$. It controls the balance between attraction ($\sigma\Delta$ coupling constant) and repulsion ($\om\Delta$ coupling constant). Although, one typically expects $U_\Delta$ to be of the order of the nucleon optical potential $\sim -50$\,MeV, uncertainties in the extraction of the $\Delta$ properties in nuclear matter allow for variations of $U_\Delta$ in a broad interval.
In the left panel of Fig.~\ref{fig:mkvstar-meff} we show the results from~\cite{KMV2017} for the effective nucleon mass $m_N^*(n)$ in ISM as a function of the density at zero temperature for MKVOR*  model and for MKVOR*$\Delta$ model for different values of the $\Delta$ optical potential. In Ref.~\cite{KMV2017} it was shown that for $U_\Delta>-67$\,MeV,  $m_N^*$  decreases monotonously in the MKVOR*$\Delta$ model with an increase of the baryon density $n$ and reaches a small value in the high-density limit. For deeper  potentials the $\Delta$ appears in matter abruptly at some critical density depending on the potential $U_\Delta$ that leads to a characteristic jump in the $m_N*$ density dependence. This behaviour resembles the sharp change of the nucleon mass in the Lee-Wick model, see Fig.~\ref{fig:Lee-mass}.
On the right panel of Fig.~\ref{fig:mkvstar-meff} we show the pressure $P(n)$ at zero temperature as a function of the density with and without $\Delta$ isobars (models MKVOR* and MKVOR*$\Delta$, respectively). The hatched region indicates the constraint of the pressure extracted from analysis of the momentum distributions (directed flow) of particles produced in  the heavy-ion collisions~\cite{Danielewicz:2002pu}.  For  $U_\Delta =-50$\,MeV, the derivative of the pressure gets a discontinuity  that corresponds to a second-order phase transition owing to the appearance of $\Delta$ isobars.  For $U_\Delta <-67$\,MeV, $P(n)$ demonstrates the behaviour typical for the first-order phase transition, except that in our case there is a specific back banding in the $P(n)$ dependence. The horizontal lines connecting points ($P(n_1^{\rm MC})=P(n^{\rm MC}_2)$, $\mu (n_1^{\rm MC})=\mu(n_2^{\rm MC})$) show the Maxwell construction (MC). For $U_\Delta <-91.4$\,MeV, $P(n)$ crosses zero at two values of the density. One of these zeros corresponds to unstable state, other (right one), to metastable or even stable state corresponding to the $\Delta$ resonance matter. The presence of a pion condensate would result in a further enhancement of the $\Delta$ abundance.

Note that the occupation of the new $\Delta$ Fermi sea occurs at nucleon densities exceeding
a critical value $n>n_c\sim (2-4)n_0$, which depends on the $N/Z$ ratio and parameters of the model under consideration. Appearance of $\Delta$ isobars makes the nuclear matter more stable. Besides this, the occupation of the $\Delta$ Fermi sea additionally promotes a decrease of the effective pion gap and a pion condensation at smaller baryon density.

In addition to the occupation of the $\Delta$ Fermi sea, in the neutron star matter (at $N\gg Z$), $\Lambda$, $\Sigma^-$ and even $\Xi^-$ hyperon Fermi seas can be occupied at approximately the same densities, $n>(2-3)n_0$. The appearance of any new fermion degree of freedom decreases  the energy and, consequently, the maximum compact star mass. Such a decrease of the maximum neutron star mass has been observed in many models~\cite{Schaffner-Bielich:2008zws,Djapo:2008au,Schulze-Rijken,Drago:2014oja,Drago:2015cea}.
These problems are known as the ``hyperon puzzle'' and the ``$\Delta$ puzzle''. They have been resolved in the RMF models, in which the hadron masses are assumed to be functions of the scalar $\sigma$ mean field~\cite{MKVPhysRev2016,MKV2016,KMV2017}.

\section{Strangelets,  strange and hybrid stars}\label{sec:Strangelets-sec}

The idea of stable strange quark matter proposed by Witten in 1984~\cite{Witten1984} is in some sense similar to the mentioned above idea of stable $\Delta$ resonance matter by Troitsky and Chekunaev from year 1979~\cite{Troitsky:1979ch}. A quark nugget (strangelet or nuclearite) is a baryonic state composed of quarks in a deconfined phase. There is a correspondence between this state and nontopological solitons with baryon constituents studied by Lee and collaborators and then by Sidney Coleman~\cite{Sirlin,Coleman1986,Lee1987,Lee1989,Lee1992}. In Ref.~\cite{Witten1984} Witten estimated the energy per quark for a three-flavor system (strange quark matter) to be $\sim 90\%$ of the energy of a two-flavour system. This additional energy gain could be enough to compensate the energy loss due to the heavier strange quark mass leading, thereby, to a stable quark system. The strange quark matter can be stable or metastable for a wide range of strong interaction parameters. If the strange quark matter is more stable than the nucleus of Fe and the baryon number is $A\ll 10^{56}-10^{57}$ one speaks about strangelets and quark nuclearites, and if the baryon number is large, $A\sim 10^{56}-10^{57}$, about strange stars. In the latter case, the gravity essentially contributes to binding. The surface layer of these objects has typically a microscopic length $\sim 10$\,fm.
Provided the gravity is switched off some models of quark matter do not allow for the absolutely stable strange quark matter but allow for a metastable state. Other models use such parameters that the strange and non-strange deconfined matter is unstable. In these both cases, the quark matter, nevertheless, may exist in interiors of compact stars, named in this case hybrid stars. All such stars have macroscopic hadron shells. Sometimes some authors conflate mentioned terms calling hybrid stars as strange stars or quark stars. From our point of view, the most important is that strangelets, quark nuclearites and strange stars could be stable already without assistance of gravity. The difference is only that in the case of strange stars the baryon number typically is $A\sim 10^{56}-10^{57}$, and then the gravity contributes essentially to binding. For object with a smaller $A$ the gravity is less important. However, the important difference exists between strange stars and hybrid stars, since the latter contain hadronic shells of macroscopic size. Also we should note that in all  quark nuggets with $A\gg 1/e^3$  up to a strange star size, the net electric charge is screened mainly by electrons and muons similarly to the case of the pion condensate nuggets and nuclei-stars, cf. results of Ref.~\cite{Alcock1986} and \cite{Migdal:1976wx,Migdal:1977rn}.

For strange matter to form, there must initially be a sufficient number of strange quarks. Since the probability of multiple weak processes occurring simultaneously is enormously suppressed, the large amount of strangeness required to satisfy conditions of Witten's hypothesis can only be obtained under very specific conditions. As far as we know, these conditions can only be met in the Early Universe and/or in the cores of massive stellar objects.
Discussions of hypothetical strangelets and hybrid and strange stars began with the works~\cite{Ivanenko:1965dg,Itoh:1970uw,Bodmer:1971we,Witten1984,Farhi:1984qu,Haensel:1986qb} and remain a hot topic till these days, see, e.g., Refs.~\cite{Madsen1998,Weber2005,Farrell:2024bka,Lugones2025,Clemente2025} and references therein.

\subsection{Quark models}

The most commonly employed model of quark matter is the MIT bag model \cite{Chodos1974}. By balancing the vacuum pressure outside the bag with the pressure of quarks inside it, one mimics phenomena of quark confinement and asymptotic freedom~\cite{Farhi:1984qu}. The quark matter is stable, metastable or unstable in these models in dependence mainly on the assumed value of the bag constant $B$ and the strange quark mass $m_s$.  A large value of $B$ excludes the existence of hybrid stars. With a decrease of $B$, first the most massive neutron stars acquire a quark core, becoming hybrid stars. For a small $B$ and a smaller value of $m_s$, strange quark matter becomes absolutely stable, i.e., more stable than $^{56}$Fe.

In addition to the MIT-bag based approaches, the Nambu--Jona-Lasinio (NJL) model is frequently employed, see Ref.~\cite{Klevansky1992}. However the latter does not describe quark confinement, and its preferred sets of parameters generally do not allow for stable strange quark matter~\cite{Buballa1999}. Competition of inhomogeneous chiral phases and superconducting quark phases in the NJL model was studied in \cite{Lakaschus:2020caq}.

The quark objects are anticipated to be color superconductors. The attraction comes either from the one-gluon exchange, or from a nonperturbative four-point interaction motivated by instantons, or from nonperturbative gluon propagators. Various quark-paring patterns were investigated: the color-flavor-locked (CFL)  phase, the two-flavor color superconductivity (2SC) phase, the color spin-locking (CSL) phase, etc., see Refs.~\cite{Alford2008,Farrell:2024bka}. The 2SC phase allows for unpaired quarks of one color. The CFL  phase deals with very large values of the baryon chemical potential. In this case the color superconductivity is complete in the sense that the diquark condensation produces a gap for quarks of all three colors and flavors. The value of the gap for CFL and 2SC phases can be of the order of 100\,MeV.
Since the CFL phase is expected to occur at ultra-high-density, its presence in compact stars is more questionable than presence of the 2SC phase. Owing to strong interactions, a fluctuation region near the critical point of the color superconductivity is expected to be broad and it is not excluded that some effects of the color superconducting fluctuations can be manifested in heavy-ion collisions~\cite{V2004}.

Depending on the value of the diquark coupling  a region of the phase diagram was found where the
inhomogeneous chiral and the 2SC condensates may coexist, see Ref.~\cite{Buballa:2014tba} and references therein. Let us add here that at large densities QCD may exhibit a moat regime in the scalar-pseudoscalar sector~\cite{Pisarski2021b} similar to the occurrence of the minimum in the effective pion gap, see Sect.~\ref{sec:Superheavy}. Also, one uses the quark models with quark masses varying with the baryon number density, see Ref.~\cite{Xia2014}.
The RMF approach proposed in Ref.~\cite{Blaschke2017} gives a simple way to model confinement of quarks via introducing the particular density dependence of quark masses divergent for low baryon densities and a density-dependent screening effect. The screening is described analogously to an excluded volume effect for a hadronic matter model with density-dependent coupling constants. Effects of various pasta phases occurring in the quark-hadron phase transition~\cite{Voskresensky:2002hu} were then investigated in Ref.~\cite{Maslov2019} for a set of RMF equations of state for both the hadron matter and the quark one. Parameters used in these works  allow for hybrid stars but exclude stable quark star configurations.

Some constraints can be gained from the lattice QCD calculations. The standard bag and NJL models cannot  reproduce the lattice data without
significant modifications. However, the lattice results at zero and small finite chemical potential are appropriately described within phenomenological quasiparticle models employing massive quarks and gluons with masses dependent on temperature~\cite{Rischke:1992rk,Goloviznin:1992ws,Peshier:1994zf,Gorenstein:1995vm}. These models were also used
to extrapolate the lattice equation of state to the case of cold baryon matter. This has been done in Ref.~\cite{Peshier:1999ww} and then in Refs.~\cite{Peshier:2002ww,Ivanov:2004gq,Schulze2009} and in other works.

\begin{figure}
\centering
\includegraphics[width=8cm]{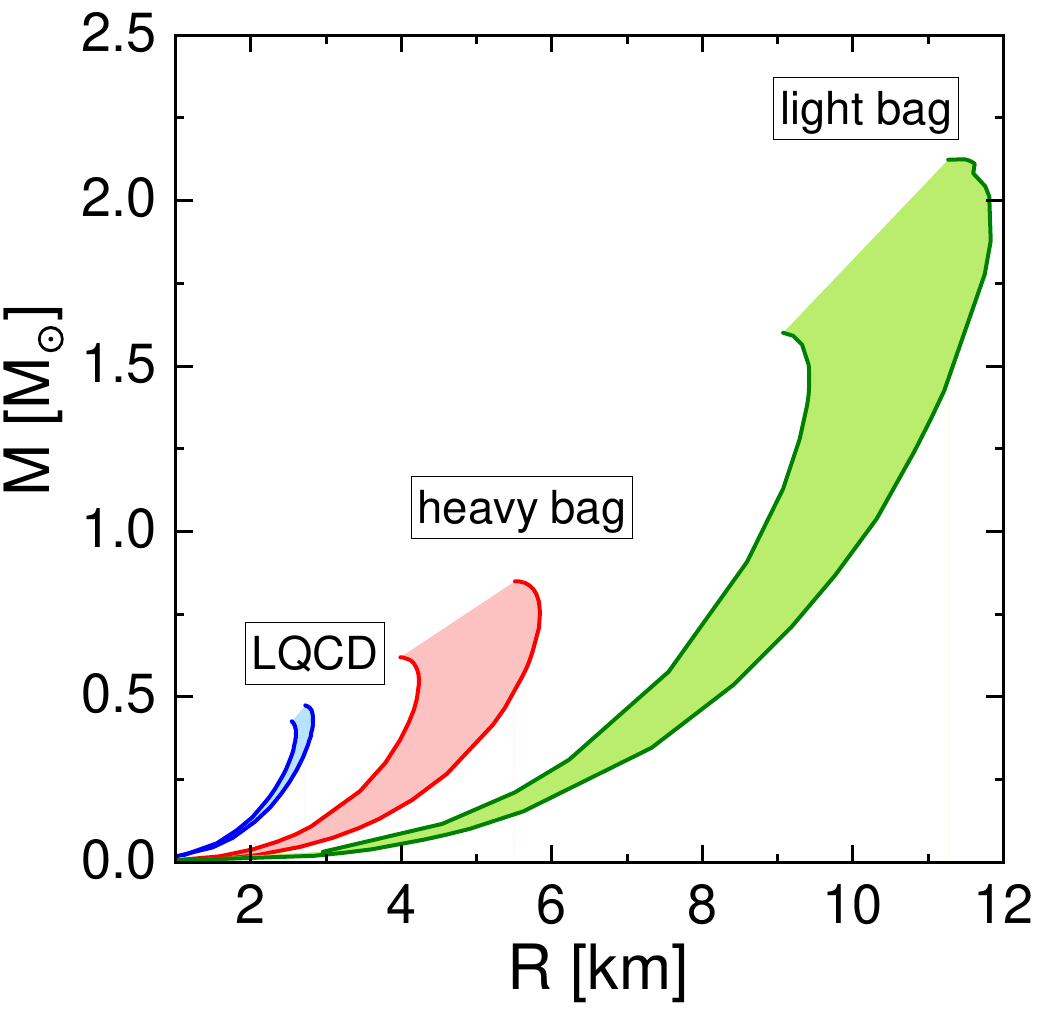}
\caption{Strange star mass-radius relation for the light-bag and heavy-bag models and the quasiparticle model~\cite{Ivanov:2004gq} tuned to the lattice QCD data.
For details see text and Ref.~ \cite{Ivanov:2005be}. The bands correspond to variations of pairing gap $\Delta$  between 0 ad 100\,MeV for the CFL type of quark superconductivity.
\label{fig:MRstrange}}
\end{figure}

The quasiparticle model~\cite{Ivanov:2004gq} for the equation of state of the quark-gluon matter at finite temperatures and baryon chemical potential, which is tuned to the lattice QCD data~\cite{Fodor:2002sd,Csikor:2004ik}, was used in Ref.~\cite{Ivanov:2005be} to study the possibility of a hadron-to-quark transition in the neutron star matter. For the hadronic equation of state the RMF model~\cite{KV2005} was used. For comparison two ''light-bag'' and ''heavy-bag'' models were also considered. The ``light-bag'' models with conventional current quark masses ($m_{u} = 5, m_{d} = 7, m_{s} = 150$\,MeV) and vanishing gluon mass and values of the bag constant $B^{1/4} = 155$\,MeV
and $B^{1/4} = 200$\,MeV were used. These values  of $B^{1/4}$\,MeV are typical for models applied to the treatment of hybrid and strange stars~\cite{Alford:2004pf,Heiselberg:1999mq}. It proved to be that the  ``light-bag'' models  cannot describe the equation of state obtained in the lattice QCD calculations. Also it was considered a ``heavy-bag'' model with
parameters $m_u = m_d = 330$ MeV, $m_s = 450$ MeV, $m_g = 600$ MeV, and $B^{1/4} = 183$\,MeV. This model can reasonably reproduce the lattice results at temperatures not too close to the critical temperature. Also a more specific quark quasiparticle model \cite{Ivanov:2004gq} was  introduced, which fits well the lattice QCD calculations above the critical temperature.
The general result of Ref.~\cite{Ivanov:2005be} is that the families of ordinary neutron stars and pure quark stars are well separated. The hadron-quark interface can be constructed only between the light-bag model and unrealistically stiff Walecka RMF model~\cite{Walecka1974}. For the quark matter model fitted to the lattice data the hybrid star cannot be constructed.
The application of these models to the quark stars is illustrated in Fig.~\ref{fig:MRstrange}.
The light-bag model produces neutron stars with masses up-to 1.5--2.1\,$M_\odot$ and radii $9<R<12$\,km. For the heavy-bag model, neutron star becomes lighter and more compact, and the model tuned to the lattice QCD supports only stars with very low masses ($M\lsim 1M_{\odot}$), high densities ($n\gsim 10 n_0$), and small radii ($R \lsim 6$ km).
However the situation remains rather uncertain. Some studies provide arguments for stability or metastability of two-flavor quark matter, while others argue against the stability of quark matter, e.g. see \cite{Holdom2018,Bai2025} and references therein.

We also should mention the possibility of a quarkyonic phase~\cite{McLerran:2018hbz}, which may occur between hadron and quark phases. Thus the temperature--baryon density, $T-n$, phase diagram of the nuclear matter proves to be very rich. With increasing density at a moderate temperature there may appear region, in which we may deal with the stable or metastable $\pi$ or $\sigma$ condensate matter or the strange quark matter.

Besides finite size strangelets and strange stars, also a branch of strange white dwarfs was also discussed~\cite{Glendenning1995}. Such dwarfs would have conventional nuclear crusts and strange quark matter cores. Possible candidates for strange dwarfs were suggested in Ref.~\cite{Kurban2022}. Potential implications for gravitational wave astronomy are highlighted in Ref.~\cite{Perot2023}.

Recently, other state of deconfined matter was proposed in Ref.~\cite{Zhitnitsky:2024ydy} based on the model described in Ref.~\cite{Zhitnitsky:2002qa}. The compact soliton-like object  consists of a core of quarks (antiquarks) in a color-superconducting state surrounded by an electrosphere of positrons (electrons), and enclosed by an axion domain wall. Large axion domain walls can form large bubbles filled with $u$, $d$ and $s$ quarks. Earlier, the idea that soliton-like configurations could serve as dark matter arose in Refs.~\cite{Sirlin,Coleman1986,Lee1987,Lee1989,Lee1992}. Among other conditions, the possibility of existence of these objects is based on the feasible existence of a conserved charge associated with an ungauged unbroken continuous internal symmetry. In Ref.~\cite{Coleman1986} such objects were called $Q$ balls. Large $Q$ balls may form so-called boson stars.
%%%%
There is a scenario that the dark matter is associated with slow light bosons with a mass $m\sim 10-22$\,eV and the huge de Broglie wavelength of $\sim 1$\,kpc. This is called the fuzzy dark matter. A soliton is surrounded by an envelope resembling a cold dark matter hallo~\cite{Hui}. Recent work~\cite{McDonald:2025rfm} proposes a model for the cosmological formation of superheavy $Q$ balls as MACHOs (massive compact halo objects), the objects in the mass range $(10^{-7}-10^{-6})\,M_{\odot}$. Also it shows that it is possible to produce asteroid mass $Q$ balls. A review of these topics can be found in~\cite{Shnir,Nugaev,McDonald:2025rfm}.

Finally, if the mass distribution of strangelets is obtained, it is
possible to formulate well-focused campaigns to search for these
objects, cf. \cite{Clemente2025}.

The most promising  signals of the scalar and pion condensate nuclei-stars and strange stars seem to be an
abnormal mass-radius relation \cite{Migdal:1990vm,Bombaci:2001ry}, an $r$ mode spin-down \cite{Madsen:1999ci,Drago:2004vu,KV2016},
and specific stellar cooling properties \cite{Voskresensky:2001fd,Iwamoto1982,Pizzochero1991,Blaschke:1999qx,Blaschke:2000dy,
Grigorian:2004jq}.

\subsection{Abnormal $\pi$ condensate and strange compact stars and the star mass-radius relation}

For quite some time there has been a consensus that most neutron stars being produced in supernova explosions have masses near $1.4M_{\odot}$  close to the neutron star maximum mass $M_{\rm max}\simeq 1.5 M_{\odot}$~\cite{Thorsett:1993sz,Brown-Bethe-ApJ423}. For $M>M_{\rm max}$ one deals with the black hole. Most of the existing that time hadronic equations of state could describe neutron stars with such masses. However then compact stars of different masses were observed.   By now it is well established that the compact star masses vary in a broad interval~\cite{Lattimer-ARNPS62}. At present the minimal value of the accurately measured mass is $1.174(4)\,M_{\odot}$ for the pulsar PSR~J0453+1559~\cite{Martinez2015}. There was the claim that the object HESS CCO XMMU~J173203.3-344518 has a very small mass of $0.77^{+0.20}_{-0.17} M_{\odot}$~\cite{Doroshenko-light-HESS}.
The analysis of Ref.~\cite{Doroshenko-light-HESS} was
criticized in Ref.~\cite{Alford-Halpern-2023} with the conclusion that a small mass source is not needed for the description of observed spectra. Thereby one should consider the result ~\cite{Doroshenko-light-HESS}  with a caution.

Today, the highest precisely-measured mass of a neutron star is $2.08(7)\,M_{\odot}$ for PSR J0740+6620~\cite{Cromartie2020,Fonseca2021}.
Recent analyses of the NICER data indicate that there is no significant variation of star radii for stars with masses varying between
1.4 and 2.0\,$M_\odot$,  $R\simeq (12-13)$ km. A more detailed discussion can be found in Ref.~\cite{Kolomeitsev:2024gek}.
Although any strong softening of the equation of state leads to a decrease of the radius for the objects with  fixed mass, e.g. see Ref.~\cite{Ivanov:2005be}, the maximum mass of a pion-condensate nucleus-star, a hybrid star, or a strange star can be smaller or larger than the purely hadronic star, depending on the utilized model.
In this respect we should mention that recently, the mass and radius of the Type-I X-ray burster and accretion-powered millisecond pulsar XTE J1814-138
were inferred in Ref.~\cite{Kini2024} to be $M = (1.21 \pm 0.05)\,M_{\odot}$ and $R = 7.0 \pm 0.4$\,km ( with 68.3\% credibility).
If confirmed, the compact object of such a small radius could not be described by a conventional purely hadronic equation of state. Then, this object would be a candidate for a strange star, a pion condensate nucleus-star, a compact object with the Lee-Wick $\sigma$ condensate, or something else.

Reference~\cite{Migdal:1979je} suggested that a massive cold neutron star can undergo the first-order phase transition to the pion condensate state with a blowing off a part of matter or at least a giant glitch. Then a similar idea was applied to the first-order phase transition to a hybrid (quark-hadron) star configuration \cite{Kampfer:1981yr}. In majority of papers the later term ``hybrid star'' is associated with the star heaving a quark interior and an extended  hadron exterior although consequences of any strong first-order phase transition are approximately the same,  see further discussion in \cite{Migdal:1990vm}. Then, an interesting idea appeared that a neutron star and a star, where  the first-order phase transition to the state with a superdense interior (pion condensate, $\sigma$ condensate, quark matter or something else) occurred, can have the same mass at very different radii. If mass-twins were observed, it would be an evidence for the existence of compact stars different from ordinary neutron stars  (e.g. hybrid stars, strange stars, pion-condensate nuclei-stars, or something else),
see Refs.~\cite{Gerlach:1968zz,Glendenning:1998ag,Schertler:2000xq,Benic:2014jia,Maslov2019} and references therein.

\subsection{Cooling of $\pi$ condensate abnormal nuclei, nuclei-stars, strangelets and  strange stars}

Below we show that nuclei-stars held together by a pion condensate and neutron stars containing a broad pion condensation region in their interiors  as well as hybrid stars with an extended region of a quark 2SC phase are probably rather massive rapidly cooling stars, with $M\gsim 1.5\,M_\odot$.

At present time  there exist data on the surface temperatures of some  pulsars as function of their age, see \cite{Ioffe-stars,Potekhin2020}. Therefore, it is worthwhile to check whether existing models can describe them. The data can be separated on three groups: slow, intermediate, and  rapid coolers. The data impose that the averaged neutrino emissivity for the slow and rapid coolers should differ by factor $\gsim 10^3$. The minimal cooling scenario, in which important in-medium effects are disregarded, cannot appropriately describe the data.  Shortcomings of this scenario are discussed in Refs.~\cite{Voskresensky:2001fd,Blaschke:2013vma,Grigorian:2016leu}.

The most efficient processes of the neutrino production in neutron stars are one-nucleon processes because of their large phase-space volume. These are the DU processes, e.g. $n\to p + e + \bar{\nu}_e$ and $ p + e \to n+ \nu_e$, and the processes going on the mesonic condensates, e.g. on the pion condensate, such as $n+\pi^-_{\rm cond}\to n+e+\bar{\nu}_e$. The emissivity of the DU process can be roughly estimated as
\begin{align}
\epsilon_\nu^{\rm{DU}} \sim 10^{27} (n_e/n_0)^{1/3}e^{-\Delta/T} T_9^6\Theta(n-n_c^{\rm{DU}})\frac{\mbox{erg}}{\mbox{cm}^3\cdot \mbox{s}},\label{DU-neut}
\end{align}
where $T_9$ is the temperature measured in $10^9$ K, $n_e$ is the density of electrons. The suppression factor $e^{-\Delta/T}$ appears for temperatures below the critical temperature of a nucleon pairing $T_\Delta$, and a pairing gap $\Delta\propto (T_\Delta-T)\Theta (T_\Delta-T)$, where $\Theta (x)$ is the step-function. The DU process occurs for $n>n_c^{\rm DU}$, when the fraction of protons exceeds $\sim 10\%$. The DU process is so efficient that, if it occurred, the stars with $M>M_{\rm DU}+0.1 M_{\odot}$ would cool so rapidly that they would not be manifested in the cooling data~\cite{KV2005,Klahn:2006ir}.  Therefore, the deficiency of the ``minimal cooling + DU'' scenario is that it would require all stars in the cooling diagram (temperature vs. time) to have
a mass close to the threshold for the DU processes to occur in the star center, with an uncertainty of only $0.1 M_{\odot}$.
However now it is known that masses of neutron stars proved to be essentially different.
Moreover, there are equations of state in the literature, in which the DU process either does not appear in the whole available density interval or it occurs only for $M>1.5 M_{\odot}$. For example, in case of the famous variational equation of state~\cite{Akmal:1998cf} the threshold mass for the DU process is $M_{\rm DU}\simeq 2 M_\odot$.

In works~\cite{Schaab:1996gd,Voskresensky:2001fd,Blaschke:2004vq,Grigorian:2005fn,Blaschke:2011gc,Blaschke:2013vma,
Grigorian:2016leu,Grigorian:2018bvg} the ``nuclear medium cooling scenario'' was developed. It relied on the results of Refs.~\cite{Voskresensky:1984zzn,Voskresensky:1986af,Voskresensky:1987hm,
Voskresensky:2001fd} where it was assumed that neutron stars with measured surface temperatures (first data fixed only upper limits on surface temperatures) may have very different masses, and that neutrino emissivity of the star strongly depends on the nucleon
density and, consequently, on the neutron star mass. This dependence appeared since the in-medium modification of the pion exchange was taken into account that significantly affected two-nucleon reaction rates. Within the ``nuclear medium cooling scenario'' the DU reactions were not allowed, at least for stars with masses $M<1.5 M_{\odot}$.

Another process of the one-nucleon origin is a reaction going on the pion condensate, e.g. $n+\pi_{\rm cond}^-\to n+e+\bar{\nu}$. Its emissivity is roughly estimated as
\begin{align}
\epsilon_\nu^{\pi} \sim 10^{27}\frac{\phi^2}{m_\pi^2}\Big(\frac{n_e}{n_0}\Big)^{1/3} e^{-\Delta/T} T_9^6\Theta(n-n_c^{\pi})\frac{\mbox{erg}}{\mbox{cm}^3\cdot \mbox{s}},\label{pi-neut}
\end{align}
where $n_c^{\pi}\sim (1.5-4)n_0$ depending on the model, and $\phi$ is the amplitude of the pion condensate field, $\phi\sim (0.1-1)m_\pi$.
This process would be too efficient as the DU process, if $n_c^{\pi}$ were smaller $2n_0$, $M\lsim (1.3-1.35) M_{\odot}$. It would affect most of the neutron star population and lead to a contradiction with the measured Log(N) - Log(S) distribution.
We should also mention the DU-like processes occurring on hyperons and $\Delta$ isobars at densities exceeding the corresponding critical densities. Values of these emissivities are similar to those for the $\pi$ condensate processes, cf. \cite{Grigorian:2017xqd,Grigorian:2018bvg}

In the presence of nucleon pairing (for $T<T_\Delta$) a new important process of the quasi-one-nucleon origin occurs. It is associated with breaking and formation of nucleon pairs, so the name --- the pair-breaking-formation (PBF) process~\cite{Flowers:1976ux,Voskresensky:1987hm,Leinson:2006gf,Kolomeitsev:2008mc}, e.g. $n_{\rm pair}\to n+\nu+\bar{\nu}$. Its emissivity is estimated as
\begin{align}
\epsilon_\nu^{\rm{PBF}} \sim 10^{28}\big(\Delta/{\mbox{MeV}}\big)^7 (T/\Delta)^{1/2}e^{-2\Delta/T}\frac{\mbox{erg}}{\mbox{cm}^3\cdot \mbox{s}}\,.\label{PBF-neut}
\end{align}
The in-medium renormalization of the vertices is crucially important for the PBF processes~\cite{Kolomeitsev:2008mc}.

Note that processes going on $\pi$ condensate and the PBF processes accounting the renormalization of the vertices are purely in-medium processes, which would not be consistent to include in the minimal cooling scheme disregarding effects of the polarization of the medium. However the PBF process was included, whereas the $\pi^-$ condensate processes were disregarded, even though the latter would occur already in the stars with $M\gsim 1 M_{\odot}$, since the electron chemical potential would then reaches the bare pion mass if in-medium effects were switched off. The \emph{pionization} of the neutron star matter can only be precluded if the effects of the medium on pions are taken into account~\cite{Migdal1978}.

In the absence of the DU process and PBF process in the minimal cooling scheme the main role in the neutron star cooling is taken by the modified Urca processes, e.g. $n+n\to n+p+e+\bar{\nu}_e$. Its emissivity is roughly estimated as
\begin{align}
\epsilon_\nu^{\rm{MU}} \sim 10^{21} \Big(\frac{n_e}{n_0}\Big)^{1/3}e^{-2\Delta/T} T_9^8\frac{\mbox{erg}}{\mbox{cm}^3\cdot \mbox{s}}.
\end{align}

Within the ``nuclear medium cooling scenario'' it was shown that the emissivity of two-nucleon reactions, e.g. $n+n\to n+p+e+\bar{\nu}_e$, --- the medium-modified Urca processes (MMU) --- significantly increases with an increasing of the central density (and the mass) of the object due to strong modification of the $NN$ interaction as demonstrated in Fig.~\ref{FigNN}. Already in Refs.~\cite{Voskresensky:1984zzn,Voskresensky:1986af,Voskresensky:1987hm} it was suggested that (i) the central densities (masses) of cooling neutron stars are different, and (ii) the nucleon-nucleon scattering is significantly enhanced in denser matter, cf. Eq.~(\ref{cross}). The appropriate formalism for calculation of the reaction rates in medium is the non-equilibrium closed diagram formalism employing the exact Green's functions. It was developed in the application to the neutrino emmisivity problem in Ref.~\cite{Voskresensky:1987hm} using the quasiparticle approximation and then in Ref.~\cite{Knoll:1995nz} was extended by inclusion of finite nucleon widths, see also \cite{Voskresensky:2001fd}.
In the absence of the DU processes, the MMU processes are dominant  already for  $n>n_c^{(1)}$, where $n_c^{(1)}<n_0$, see Fig.~\ref{piongap1}. Its emissivity is roughly estimated as
\begin{align}
\epsilon_\nu^{\rm{MMU}} \sim 10^{21} \frac{F(n)}{F(n_0)} \Big(\frac{n_e}{n_0}\Big)^{1/3}e^{-2\Delta/T} T_9^8\frac{\mbox{erg}}{\mbox{cm}^3\cdot \mbox{s}},\label{MU-neut}
\end{align}
where the factor $F(n)$ appears due to a strong density dependence of the $NN$ interaction amplitude for $n>n_c^{(1)}$,
\begin{align}
F(n) \simeq 3\Big(\frac{n}{n_0}\Big)^{10/3}\frac{[\Gamma (n)/\Gamma (n_0)]^6}{[\tilde{\omega}(k_0(n))/m_\pi]^8}.
\label{MMU-neut}
\end{align}
Here $\Gamma^{-1}\simeq 1+C(n/n_0)^{1/3}\Phi (0,k_0)$
is the $NN$ correlation factor, see (\ref{gam-x}),  $\Phi (0,k_0)\simeq 1$, and $C\simeq 1.4-1.6$ is expressed via the Landau-Migdal parameter in the spin channel, $\tilde{\omega}(k_0(n))$ is the effective pion gap introduced in (\ref{efpigap}). Since $\tilde{\omega}(k_0(n))$ decreases with increasing density one easily recovers the required factor from one to  $\gsim 10^{3}$ in emissivity for stars with masses from 1 $M_\odot$ to
2 $M_\odot$ required to describe the cooling data, appropriately. The pion condensation is not needed for that but, if occurred at $n\gsim (2.5-3)n_0$, it would not contradict the cooling data. The neutron stars with the pion condensate in their interiors and the pion condensate nuclei-stars are then rather massive rapidly cooling stars with $M\gsim 1.5 M_\odot$.

A modification of the reaction rates due to multiple nucleon scattering, --- the Landau-Pomeranchuk-Migdal effect,--- was evaluated in Ref.~\cite{Knoll:1995nz}. Accounting for it, the neutrino emissivities acquire extra pre-factors of the type
\begin{align}
\frac{\omega_\nu^2}{\omega_\nu^2+\Gamma_N^2(n,T)}\,,
\end{align}
where $\om_\nu$ is the neutrino energy and $\Gamma_N$ is the nucleon width, $\Gamma_N\sim \epsilon_{\rm F}(1-\pi^2T^2/\epsilon_{\rm F}^2)$ and $\epsilon_{\rm F}$ is the nucleon Fermi energy. The pre-factor essentially deviates from unity only for the neutrino energy $\omega_\nu \lsim \Gamma_N$, with
$\omega_\nu\sim (3-5)T$ depending on the reaction type. Thus, the multiple scattering effects are important for the description of mergers and the initial stage of the cooling of the compact stars \cite{Knoll:1995nz,Voskresensky:2001fd,Kolomeitsev:2010pm,Alford:2024xfb}.

There are two stages in the cooling of compact objects: the initial stage, when neutrinos are trapped, and the stage of a long-time cooling when neutrinos are radiated from the compact object by the direct reactions, cf.~\cite{Migdal:1990vm}.
At the initial stage of the compact star evolution (lasts from minutes to  several hours depending on the mass of the object and the model)  the neutrinos and antineutrinos are captured inside the star until their mean-free path will  not reach the radius of the object.
%For pion or sigma condensate nuclei-stars  as well as the strange stars $R\sim 10$ km is their radius.
In case of hybrid stars there are two quantities playing the role of $R$. One is the radius of the quark or pion or sigma condensate interior core and other is the radius of the object  (the core plus  the hadronic exterior).
The mean free paths of the neutrino and antineutrino are determined by  the most efficient processes. Roughly in the process under consideration the ratios of the inverse  mean free pathes are proportional to the ratio of the emissivities~\cite{Sawyer:1977ja}. As function of the temperature the  emissivity of the DU process is $\propto T^6$, see Eq.~(\ref{DU-neut}) and for the MU and MMU processes it is $\propto T^8$, see Eq.~(\ref{MU-neut}) and (\ref{MMU-neut}). The ratio of the emissivities of the DU and MU processes is $\epsilon_{\rm DU}/\epsilon_{\rm MU}\sim 10^5/T_9^2$. The emissivity of the MU process exceeds that of the DU one for $T\gsim 30$\,MeV.
For the in-medium modified Urca processes, MMU, the emissivity increases sharply with grows of the density.  Therefore the MMU processes become more efficient than the DU processes at a smaller temperature.  For the compact stars with
$T>T_{\rm opac}(R)$ neutrinos and antineutrinos are trapped  inside the compact star. The heat flows to the surface due to the heat conductivity. The path length of neutrinos is much larger than for quarks and hadrons and the heat conductivity is therefore governed by neutrinos. In case of hybrid stars the neutrino transport changes on the boundary between the quark and hadron phases. For the hadron phases this stage was studied in Refs.~\cite{Sawyer:1977ja,Sawyer:1978qe,Friman:1979ecl,Sawyer1980,Voskresensky:1986af,Migdal:1990vm}. In-medium effects proved to be  very important, see Refs.~\cite{Voskresensky:1984zzn,Voskresensky:1986af,Voskresensky:1987hm,Migdal:1990vm,Knoll:1995nz,Voskresensky:2001fd,Kolomeitsev:2010pm}. As it was mentioned, appropriate optical theorem technique for the calculation the reaction rates within the nonequilibrium diagram formalism was developed  within the quasiparticle approximation in  \cite{Voskresensky:1987hm} and with inclusion of the baryon width effects in Ref.~\cite{Knoll:1995nz}. Similar effects should be included in the description of a quark phase. With inclusion of in-medium effects associated with a decrease of the effective pion gap at an increasing baryon density, the value $T_{\rm opac}$ diminishes. For more massive neutron and hybrid stars one may expect that it is below 1\,MeV.

The neutrino emissivity of the quark matter was first calculated by Naoki Iwamoto in Ref.~\cite{Iwamoto1982}.  In Ref.~\cite{Pizzochero1991} the surface temperature of the strange star was found to be lower than that of a neutron star of the same age (ordinary or with mesonic condensates).  The effect of the color superconductivity on the cooling of strange and hybrid stars for $T<T_{opac}$ has been first studied in Refs.~\cite{Blaschke:1999qx,Blaschke:2000dy,Grigorian:2004jq} and in some subsequent works. Some models show that cooling can be so fast that strange stars may presently have temperatures smaller than the cosmic microwave background radiation~\cite{Shaulov2022,Shaulov2023}.

In the case of the 2SC superconducting phase, the most efficient process is the DU process going on unpaired quarks. The cooling evolution is then very rapid that does not agree with the cooling data. Thereby, it is important to have a model, which does not allow for unpaired quarks.  Since there exist other attractive quark pairing channels with a rather weak attraction ($X$ channels), Ref.~\cite{Grigorian:2004jq} introduced a hypothetical 2SC+$X$ phase of two-flavor quark matter, where all the quarks are paired, some strongly in the 2SC channel and some weakly in the $X$ channel. Typical values of pairing gaps $\Delta_X$ in the $X$ channel should be in the range 10 keV–-1 MeV, as for example in the case of
the CSL pairing~\cite{Schafer:2000tw,Schmitt:2002sc}.
Nevertheless, Ref.~\cite{Grigorian:2004jq} did not identify the $X$ channel with a spin one pairing channel like the CSL one simply because this phase does not coexist with the 2SC phase, which was assumed as the dominant pairing structure. Furthermore, Ref.~\cite{Grigorian:2004jq} supposed that there are no Goldstone bosons in the 2SC+$X$ phase. As the result it was demonstrated that the cooling data can be appropriately described at such assumptions.

Since we have mentioned the axion-based model of nuggets, we should also mention a possible axion contribution to cooling of neutron stars. The emission of axions by neutron stars was, first, discussed in Ref.~\cite{Iwamoto1984} without the inclusion of $NN$ interactions in the hadron phase. Then in Ref.~\cite{Voskresensky:1986af} it was shown that axion emission should be strongly affected by in-medium baryon-baryon interactions in the hadron phase similarly to the corresponding reaction of the neutrino production. Recent discussion of the axion  contribution to the cooling of the quark matter can be found in \cite{Sedrakian:2018kdm}.

\section{Dilute stable nuclear systems}\label{sec:dilute-sec}

\subsection{Again, a scalar condensate}\label{dilute-secScalar}

The analysis of two independent nuclear emulsion experiments exposed to beams of $^{16}$O and $^{56}$Fe ions with 2A\,GeV kinetic energy allowed for the interpretation of some events as the relatively rare occurrence of anomalous nuclear fragments interacting with an unexpectedly large cross section as described in Ref.~\cite{Friedlander:1982qn}. These objects were called anomalons~\cite{Bayman1987}.
Although these results were not reproduced by subsequent measurements,  uncertainty remains.

Below we discuss a possibility of existence of very dilute stable or metastable nuclear systems.
We will consider the nuclear matter at zero temperature as a normal Fermi liquid. The quasiparticle part of the nucleon Green's function close to the Fermi surface can be presented as
\begin{align}
G_{N}(\epsilon,\vec{p},n) &= \frac{a}{\epsilon-\xi_{\vec{p}}+i\, 0\, {\rm
sign} \epsilon}\,,\quad
\nonumber\\
\xi_{\vec{p}} = \frac{p^2-p_{\rm F}^2}{2\,m_{N}^*} &\approx v_\rmF (p-p_\rmF)\,,\quad v_\rmF=p_\rmF/m_N^*\,.
\label{Gn-QP}
\end{align}
The non-pole part of the Green's function is hidden in the quasiparticle renormalization factor $a$, effective nucleon mass $m_N^*$ and the quaisparticle interaction amplitude in the particle-hole channel, $\Gamma_0^\omega =f_0/(a^2 N)$, which we are interested here. The interaction is expressed here through the scalar Landau-Migdal parameter $f_0$ and the density of states at the Fermi surface
$N=\nu m_N^* p_{\rm F}/\pi^2$. For ISM  the degeneracy factor  $\nu=2$. The parameter $f_0$ is related to the compressibility of the matter as follows
\begin{align}
K_N=n\frac{\rmd^2 E_N}{\rmd n^2}= \frac{p_\rmF^2}{3m_N^*} (1+f_0) \,,
\label{K-f0-relat}
\end{align}
where $E_N$ is the energy density of the nucleon system. Parameter $K_N$ is related to the standard compressibility modulus of the nuclear matter at saturation that we mentioned in Section~\ref{sec:Lee-Wick} as $K=9 K_N$.

The particle-hole scattering amplitude in the spin zero channel on the Fermi
surface can be written throught a propagator of the scalar bosonic mode \cite{KV2016}:
\begin{align}
T_{\rm ph,0}= -{\rm sgn}(f_0) D_{\phi,0}(\om,k) \,,
\label{Tph}\\
\parbox{5cm}{\includegraphics[width=5cm]{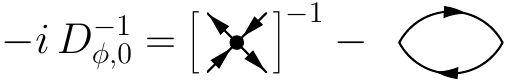}} \,.
\nonumber
\end{align}
The particle-hole loop calculated with the Green's functions (\ref{Gn-QP}) can be expressed through the Lindhard's function $\Phi$,
\begin{align}
&\intop_{-\infty}^{+\infty}\frac{\rmd \epsilon}{2\pi i}\intop\frac{2\rmd^3p}{(2\pi)^3} G_+G_-=-a^2N \Phi(s,x)\,,
\nonumber\\
&\qquad s=\frac{\om}{kv_\rmF}\,,\,\,x=\frac{k}{p_\rmF}.
\end{align}
In the limit $x\ll 1$ the Lindhard's function is
\begin{align}
\label{PhiLow-x}
\Phi (s, x)\approx
1 + \frac{s}{2} \log\frac{s-1}{s+1} -\frac{x^2}{12 \left(s^2-1\right)^2}\,.
\end{align}
It attains an imaginary part for $s<1$, and in the limit $s\ll 1$ we have
\begin{align}\label{PhiLow}
\Phi (s, x)\approx 1  +i \frac{\pi}{2}s -s^2 -\frac{x^2}{12}\,.\end{align}

The spectrum of excitations in  the scalar channel, $\omega(k)$, is determined from the pole of $T_{\rm ph,0}$,
\begin{eqnarray}
f_0^{-1}=-\Phi(s,x)\,.
\label{phiLind}
\end{eqnarray}
For $-1<f_{0} < 0$, this equation has only damped solutions with $\Re s<1 $ and $\Im s <0$.  In the
limit of $s,x\ll 1$, using the expansion of  (\ref{PhiLow})
we find for the low-lying mode
\begin{align}
\label{omnegf}
\om_{\rm d}(k) \approx -i\frac{2}{\pi}kv_\rmF \frac{1-|f_{0}|}{|f_{0}|}\,,
\end{align}
which is valid for $1-|f_{0}|\ll 1$ and $k\ll p_\rmF$.
If $f_0<-1$, i.e. the compressibility of the system becomes negative according to Eq.~(\ref{K-f0-relat}), one has $\Im\om_{\rm d}(k)>0$, and the scalar excitation modes grow exponentially with time as $\propto\exp(+\Im\om_{\rm d} t)$. This instability is called the Pomeranchuk instability of a Fermi liquid. It may result in the spinodal instability which appears in systems with a van der Waals-like equations of state. Another possibility is the formation of the Bose condensate of a scalar field, in analogy with  the Lee-Wick state, but in dilute matter~\cite{KV2016}.

We assume that growing unstable bosonic modes may form the Bose condensate  described by a scalar boson complex field $\phi$. Such a scalar field can be introduced by means of the Hubbard--Stratonovich transformation or by a formal replacement of the contact interaction with an exchange of an artificial heavy scalar boson~\cite{KV2016}. The effective Lagrangian density of the condensate field, taken in the simplest form $\phi =\phi_0 e^{-i\omega_c t+i\vec{k}_0\vec{r}}$, can be presented as
\begin{align}
&\mathcal{L}_\phi = -{\rm sgn}(f_0)D_{\phi,0}^{-1}(\om_c,k_0)|\phi_0|^2 -{\textstyle\frac12}\Lambda(\omega_c, k_0) |\phi_0|^4\,,\quad
\label{Lphidil}\\
&D_{\phi,0}^{-1}(\om,k) = (\Gamma_0^\omega)^{-1} + a^2 N\Phi\Big(\frac{\om}{kv_\rmF},\frac{k}{p_\rmF}\Big)\,.
\nonumber
\end{align}
The self-interaction term $\Lambda$ is determined in the leading order by the loop integral of four fermion Green's functions
\begin{align}
i\Lambda(\om,k)=\parbox{4cm}{\includegraphics[width=4cm]{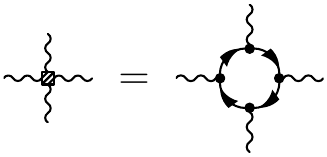}}+\dots\,
\label{Lamda-diag}
\end{align}
It was  evaluated in~\cite{Brovman,Brovman1} and \cite{Dyugaev:1982ZHETF} in case of one-component fermion system. Energetically preferable is the state with $\omega_c =0$ as argued in Ref.~\cite{KV2016}. Then we can estimate for $k_0\ll p_{\rm F}$,
\begin{align}
\Lambda(0,k_0)\approx a^4 \lambda\, \big( 1 + k_0^2/2p_{\rm F}^2 \big)\,,\quad \lambda = \xi\frac{\nu m_N^{*\,3}}{\pi^2p_{\rm F}^3}\,.
\label{Brovman1}
\end{align}
The factor $\xi$ is introduced to demonstrate  uncertainty in the value of the self-interaction parameter $\lambda$, which may arise from other diagrams beyond the leading one in (\ref{Lamda-diag}). Below we  vary it to illustrate  sensitivity of our numerical results.

The equation for the condensate amplitude following from the Lagrangian (\ref{Lphidil}) reads
\begin{align}
& a^2 N\widetilde{\omega}^2_s(k_0)\phi_0 +\Lambda(0,k_0)\,|\phi_0|^2\phi_0 = 0\,,\quad
\nonumber\\
& \widetilde{\omega}^2_s(k_0)=|f_0(k_0)|^{-1}-\Re\Phi(0,k_0) \,.
\label{Eq-phi}
\end{align}
The quantity $|\widetilde{\omega}_s(k_0)|$ plays the role of an effective gap in the excitation spectrum of the scalar quanta similarly to that we considered above in the case of the pion-nucleon interaction.
The  amplitude of the condensate field, $|\phi_0|$, and the Bose condensate energy-density term, $E_{\rm B}$, become for $\widetilde{\omega}^2_s(k_0)<0$,
\begin{align}
|\phi_0|^2 &= - \frac{N\widetilde{\omega}^2_s(k_0)}{ a^2\lambda}\,
 \Big(1+\frac{k_0^2}{2p_{\rm F}^2}\Big)^{-1}
\,,
\nonumber\\
E_{\rm B} &= - \frac{N^2\widetilde{\omega}^4_s(k_0)}{ 2 \lambda}
\Big(1+\frac{k_0^2}{2p_{\rm F}^2}\Big)^{-1}.
\label{EB-MF}
\end{align}
Simplifying consideration we further put $k_0=0$. The case $k_0\neq 0$ should be considered separately.

The energy of the system per nucleon including the nucleon part and the contribution from the Bose condensate is as follows
\begin{align}
\mathcal{E}_N+\mathcal{E}_B = (E_N + E_{\rm B}[f_0])/n \,.
\label{Enb-f0}
\end{align}

Reference \cite{KV2016} employed a simple phenomenological parametrization for the volume part of the energy per nucleon,  $\mathcal{E}_{N\rm }$ as an expansion in powers of $p_{\rm F}/m_N$ proposed in Ref.~\cite{KFW02},
\begin{eqnarray}
\mathcal{E}_{N\rm }(n)=\frac{3p_\rmF^2}{10\,m_N} - c_1\frac{p_\rmF^3}{m_N^2}
+ c_2 \frac{p_\rmF^4}{m_N^3} + c_3 \frac{p_\rmF^5}{m_N^4}
\,,
\label{weise-exp-2}
\end{eqnarray}
where  parameters $c_1$, $c_2$ and $c_3$ are expressed through values of the binding energy $\mathcal{E}_0=-\mathcal{E}_{N\rm }(n_0)$, compressibility modulus $K$ of the nuclear matter and the nuclear saturation density $n_0$.
Taking $m_N=939$\,MeV for the free nucleon mass and appropriate values $n_0=0.17\,{\rm fm}^{-3}$, $\mathcal{E}_0=-16$\,MeV and $K=285$\,MeV we find~\cite{KV2016}, $c_1 = 3.946$\,,  $c_2 = 3.837$\.,  $c_3 = 13.10$\,. From Eq.~(\ref{weise-exp-2}) the volume part of the nucleon contribution to the scalar Landau-Migdal parameter $f_0$ follows as
\begin{align}
f_{0}(n) &= \frac{3m_N^*}{p_\rmF^2}n\frac{\rmd^2(n\mathcal{E}_{N,\rm v})}{\rmd n^2} -1
= \frac{m_N^*}{m_N}-1
\nonumber\\
 & + \frac{m_N^*}{m_N}\frac{p_\rmF}{m_N}
\Big( -6 c_1 +c_2 \frac{28 p_\rmF}{3 m_N} + c_3 \frac{40 p_\rmF^2}{3 m_N^2}
\Big)\,.
\label{f0_N}
\end{align}

\begin{figure*}
\centering
\includegraphics[width=6.05cm]{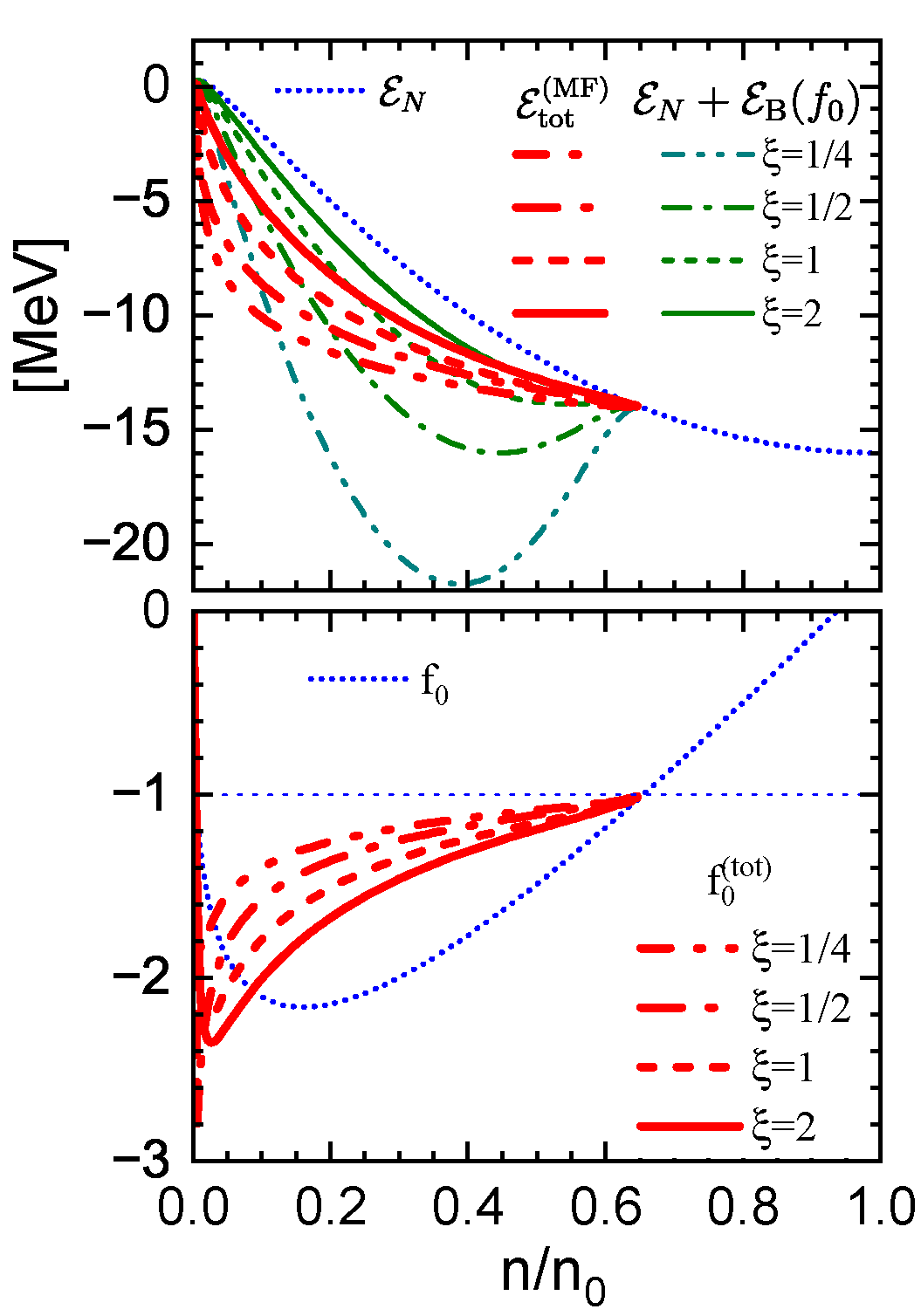}
\includegraphics[width=6cm]{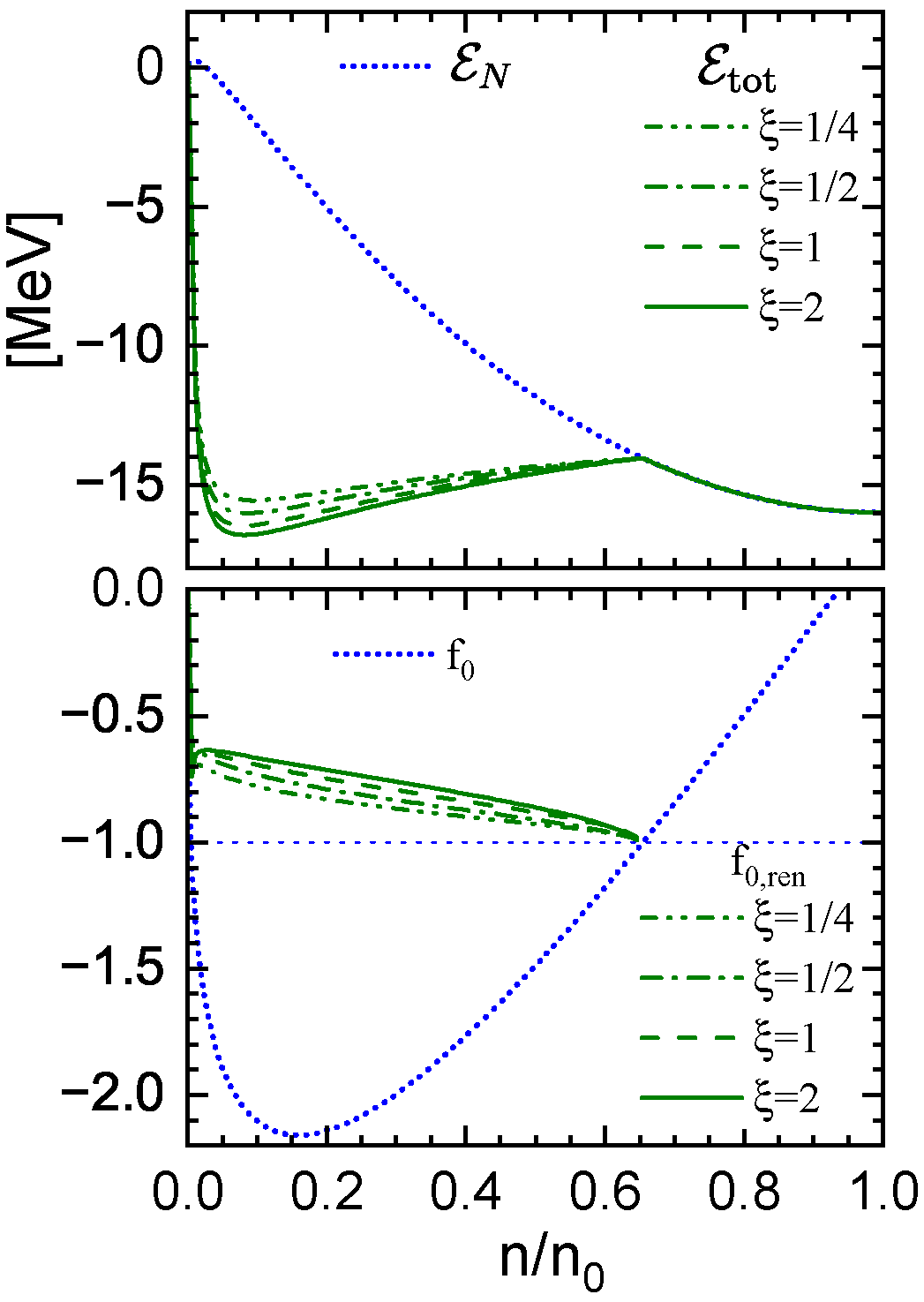}
\caption{Energy per particle (upper panel) and the scalar Landau-Migdal parameter (lower panel) as functions of the nucleon density.
\label{fig:SC}
}
\end{figure*}

The energy per particle  and the scalar Landau-Migdal parameter are shown in Fig.~\ref{fig:SC} as functions of the nucleon density. The nucleon volume part of the energy per particle and the parameter $f_{0}$ given by Eq.~(\ref{f0_N}) are presented on the left panel in Fig.~\ref{fig:SC} by the dotted lines. There is a broad  interval of densities, $n_{c}^{\rm (l)} < n < n_{c}^{\rm (u)}$ where $f_0<-1$ with $n_{c}^{\rm (l)} =0.422\times 10^{-2}\,n_0$ and $n_{c}^{\rm (u)} = 0.655\, n_0$. Similar dependence $f_0(n)$ is obtained for the DD2 equation of state, see Fig. 3 in~\cite{Ropke:2017pvg}. Notice that taking into account of the clustering effects~\cite{Ropke:2008qk,Ropke:2014fia} may affect this dependence, see  Fig.~9 in~\cite{Ropke:2017pvg}. We will focus on the alternative taking into account effect of possible Bose condensation.
The condensate contribution proves to be strongly attractive creating a minimum  at $n\sim 0.4\,n_0$ for a reduced value of $\lambda$ for $\xi<1/2$. The energies per nucleon $\mathcal{E}_N+\mathcal{E}_B$ are shown on the left upper panel by thin lines labeled by values of the parameter $\xi$. They are calculated according to Eq.~(\ref{Enb-f0}) with $f_{0}(n)$ given by Eq.~(\ref{f0_N}).

In the presence of the condensate the incompressibility of the system changes as $K = K_{N} +K_{\rm B}$, $K_{\rm B}=n\frac{d^2 E_{\rm B}}{d n^2}$, and, therefore, the scalar Landau parameter changes too
\begin{eqnarray}
f_0\to f_{0}^{\rm tot} = f_{0}+\frac{3}{2\epsilon_{\rm F}}K_{\rm B}[f_{0}^{\rm tot}]\,.
\label{diff-ftot}
\end{eqnarray}
%In  case of a weak condensate (for $|f_{0} +1|\ll 1$) one can use $E_{\rm b}[f_{0}^{\rm tot}]\approx E_{\rm b}[f_{0}]$. For a developed condensate the perturbative analysis does not work and one should solve Eq.~(\ref{diff-ftot}) self-consistently.
%The self-consistent solutions, ${\cal{E}}_{\rm tot}^{(\rm MF)}$ and  $f_{0}^{\rm tot}$ is shown on the left lower panel in Fig.~\ref{fig:SC} by thick lines.
We used the variational approach to find the solution of Eq.~(\ref{diff-ftot}), see Ref.~\cite{KV2016}, and we fixed the values of $f_{0}^{\rm tot}$ at the ends of the instability interval $f_{0}^{\rm tot}(n_c^{(l)})=f_{0}^{\rm tot}(n_c^{(u)})=-1$. Thereby we postulated that the condensate arises by the second-order phase transition. Taking into account the possibility of a first-order phase transition, the energy could be further decreased. The energy per particle in our case is constructed using Eq.~(\ref{K-f0-relat}):
\begin{align}
\mathcal{E}_{\rm tot}^{\rm (MF)} (n) = \frac{1}{n}\intop_{n_c^{\rm (l)}}^n\rmd n'\!\! \intop_{n_c^{\rm (l)}}^{n'}\rmd n''
\frac{2\epsilon_{\rm F}(n'')}{3 n''} (f_0^{\rm tot}(n'')+1)
+C\,.
\label{E-C-MF}
\end{align}
The constant $C$ is fixed by requiring $\mathcal{E}_{\rm tot}^{\rm (MF)} (n_c^{\rm(u)})= \mathcal{E}_N(n_c^{\rm(u)})$.
The  energy (\ref{E-C-MF}) corresponds to a mean-field approximation when the reconstruction of the excitation spectrum due to an interaction with the condensate field is not taken into account.

The result of calculations using (\ref{E-C-MF}), (\ref{diff-ftot})  is shown on the left panel in Fig.~\ref{fig:SC} by tick lines for different value of the $\xi$ parameters. The self-consitency requirement changes the condensate contribution essentially. The mean-field energy $\mathcal{E}_{\rm tot}^{\rm (MF)}$ becomes now only a smoothly decreasing function of the density and no minimum appears. Nevertheless, as it was expected, the decrease of the $\lambda$ parameter  (smaller $\xi$) leads to the decrease of the energy. The density dependence of $f_{0}^{\rm tot}$ is flatter than that of $f_{0}$ for densities $0.2<n<n_c^{\rm(u)}$ and $f_{0}^{\rm tot}> f_{0}$. At smaller densities $n\lsim 0.1\,n_0$ we have oppositely
$f_{0}^{\rm tot}< f_{0}$. We have to notice that our solutions are accurate at the level of $0.1\%$ for $0.01\,n_0\lsim n\lsim 0.55\,n_0$ but the accuracy is getting worse when density approaches $n_c^{\rm (u)}$. Then our solution can be viewed as a smooth extrapolation to $n_c^{\rm (u)}$.

%The typical EoS of ISM for the realistic values of the saturation parameters is shown in Fig.~\ref{fig:SC} calculated in the model described in Appendix~\ref{app:EoS}. The corresponding ,

The interaction of  the condensate with over-condensate excitations  leads to appearance of a new term in the excitation propagator,
$$D^{-1}_{\phi,0}\longrightarrow D^{-1}_{\phi}(\om,k) =  D^{-1}_{\phi,0}(\om,k) -\delta\Sigma_\phi(\om,k) \,,$$
where the additional term is
\begin{align}
-i\delta\Sigma_\phi(\om,k) =
\parbox{1.5cm}{\includegraphics[width=1.5cm]{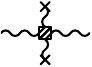}}
 = -i 2\Lambda(\omega,k)\,|\phi_0|^2 \,.
\end{align}
In such a way we include both mean-field and quadratic fluctuation contributions.

The new spectrum of excitations on top of the condensate is determined now by the equation  $D^{-1}_{\phi}(\om,k)=0$.
For  $k_0=0$ and $-1\ll\tilde{\omega}^2_s(0)<0$ we find
$$
\omega_{\rm d} \approx i\frac{2}{\pi} k v_{\rm F} \tilde{\omega}^2_s(0)  = i\frac{2}{\pi} k v_{\rm F} \frac{1-|f^{\rm tot}_{0}|}{|f_{0}^{\rm tot}|} \,,$$
where we used Eqs.~(\ref{EB-MF}) and (\ref{Brovman1}). We pay attention to opposite sign compared to Eq.~(\ref{omnegf}) for  $|f_0^{\rm tot}|>1$. Thus, we have now $\Im \omega_{\rm d}<0$. Hence, in the presence of the condensate, excitations are damped and the Fermi liquid becomes free from the Pomeranchuk instability.

The particle-hole interaction is also changed in the presence of the condensate. There appears a new term in equation for the particle-hole amplitude (\ref{Tph}) after the replacement $D^{-1}_{\phi,0}\longrightarrow D^{-1}_{\phi}$.   The new term
can be included in the renormalized local interaction
\begin{align}
\frac{1}{f_{\rm 0,ren}}= \frac{1}{f_{0}^{\rm tot}} + 2 \, \tilde{\omega}^2_s(0) =-\frac{f_{0}^{\rm tot}}{1+2f_{0}^{\rm tot}} \,,
\label{fren}
\end{align}
where we used $\tilde{\omega}^2_s(0)= -1-1/f_{0}^{\rm tot}$.  Thus, if originally $f_{0}<-1$ and therefore $f_{0}^{\rm tot}<-1$, the renormalized interaction yields $-1<f_{\rm 0,ren} < -1/2$. Hence, the first-sound modes become stable in the Fermi liquid with the condensate. The quantity $f_{\rm 0,ren}$ is plotted on the right lower panel in Fig.~\ref{fig:SC}.
Knowing the value $f_{\rm 0,ren}$ one can reconstruct the energy per particle $\mathcal{E}_{\rm tot}(n)$ of the system using Eq.~(\ref{E-C-MF}) after the replacement $f_0^{\rm tot}\to  f_{\rm 0,ren}$ in the integrand.  The results are shown on the right upper panel in Fig.~\ref{fig:SC}.
We observe the appearance of a minimum of the function $\mathcal{E}_{\rm tot}(n)$. For $\xi=1$ the minimum is at $n_{\rm min}=0.085\,n_0$ with $\mathcal{E}_{\rm min}=-16.01$\,MeV. Variation of the parameter  $\xi$ in interval $\xi=1/4\mbox{--}2.0$ leads to variations of the values $n_{\rm min}=(0.82\mbox{--}0.93)\times 10^{-1}\,n_0$ with $\mathcal{E}_{\rm min}=-(16.8\mbox{--}15.6)$\,MeV. So the minimum in energy can correspond to a new stable or metastable state of the dilute nuclear matter compared with the usual saturation energy of $-16$\,MeV used to specify the equation of state of the ISM.

Finally, we should mention that the obtained results are valid, if on the one hand  $\phi_0$  is rather small and on the other hand  fluctuations on the top of the condensate yield a yet smaller contribution, since the self-interaction of excitations on top of the condensate and feedback of fluctuations on the mean field were disregarded.

Considering finite system one must take into account the surface and Coulomb energies. We parameterize the surface energy as in  \cite{Migdal:1976wm,Migdal1978},
\begin{align}
\mathcal{E}_{\rm S} A &= a_{\rm S} A^{2/3}\,,\quad a_{\rm S} \approx C_{\rm S} \Big(\frac{n}{n_0}\Big)^{1/3}
\frac{\mathcal{E}(n,z)}{\mathcal{E}(n_0,z=1/2)}\,,\quad
\nonumber\\
 C_{\rm S} &= 18\,{\rm MeV}\,,
\label{Esurf}
\end{align}
where $z=Z/A$.

With the account for the surface energy (\ref{Esurf}) the Landau-Migdal parameter, $f_0^{\rm S}$, takes the following form
\begin{align}
f_{0}^{\rm S}(n) &=f_{0}(n)+ \frac{C_{\rm S}}{A^{1/3}}
\frac{1}{\mathcal{E}_{N,\rm v}(n_0)} \Big(\frac{n}{n_0}\Big)^{1/3} \frac{m_N^*}{m_N}
\nonumber\\
&\times \Big( \frac{3}{5} - c_1 \frac{28 p_\rmF}{9 m_N} + c_2 \frac{40 p_\rmF^2}{9 m_N^2} + c_3 \frac{6 p_\rmF^3}{m_N^3}
\Big)\,.
\label{f0_Ns}
\end{align}

Following ~\cite{KV2016}, as an example, we consider a system of 125 nucleons at $N\approx Z$. Then the energy per particle $\mathcal{E}_N+\mathcal{E}_{\rm S}$ saturates at density $0.95\,n_0$ with the minimum $\mathcal{E}_{0}^{\rm S}=-12.4$\,MeV.
With taking into account the surface energy and solving Eq.~(\ref{diff-ftot}) with $f_{0}^{\rm S}$ instead of $f_0$ we find that
the new minimum appears at $n_{\rm min}= 0.74\times 10^{-1}\,n_0$ with the depth $\mathcal{E}_{\rm min}=-13.04$\,MeV.
So, $\mathcal{E}_{\rm min}<\mathcal{E}_{0}^{\rm S}$ and, hence, a new stable state of matter is possible in the finite size object with account for the surface energy. To analyze its stability against fusion we must include the Coulomb energy
\begin{align}
\mathcal{E}_{\rm C} A\simeq a_{\rm C} z^2 A^{5/3}\,,\,\, a_{\rm C} = C_{\rm C} \Big(\frac{n}{n_0}\Big)^{1/3}
\,,\,\, C_{\rm C}=0.7\,{\rm MeV}\,.
\end{align}
A spherical nucleus is stable against an ellipsoidal deformation (stretching) if
\begin{align}
\frac{Z^2}{A}< 2\frac{a_{\rm S}(n_{\rm min})}{a_{\rm C}(n_{\rm min})}=2\frac{C_{\rm S}}{C_{\rm C}}\frac{\mathcal{E}_{\rm min}}{\mathcal{E}_{0}}\simeq 42\,.
\end{align}
We see that, because of the condensate formation, a new metastable dilute nuclear state is stable against spontaneous  fission for $A\lsim 170$.

Reference \cite{MV2019} studied the equation of state of isospin-asymmetric matter and demonstrated that even for the asymmetry parameter $n_p/n_n\lsim 0.2$ there remains a broad region of proton and neutron densities where $f_0<-1$ and, therefore, the spinodal instability may develop. For a nugget of a size $R$ much larger than the Debye screening length $l_{D}\simeq \sqrt{\pi}/(2|e|V_0)$, the Coulomb contribution from protons must be screened by electrons, $n_p=n_e$, cf.~\cite{Migdal:1977rn}. Here we for simplicity assume that the scalar field is electrically neutral. The electron Fermi momentum is then $p_{\rmF,e}=V_0=(3\pi^2 n_p)^{1/3}$ and the electron kinetic energy density is
$E_e=p_{\rmF,e}^4/(4\pi^2)$. In the isospin-asymmetric matter, the nuclear symmetry energy must be included. A simple parametrization was proposed in \cite{Heiselberg:1999fe}, which is a fit to microscopic calculations
\begin{align}
E_{N,{\rm sym}}(n_p,n_n) &= J\, \frac{(n_p+n_n)^\gamma}{n_0^\gamma} \frac{(n_p -n_n)^2}{n_p+n_n}\,,\quad
\nonumber\\
J &= 32\,{\rm MeV}\,,\quad \gamma=0.6\,.
\label{Esym}
\end{align}
Let $n_p=\chi n_n$, then the total energy per particle of the system is
\begin{align}
&\mathcal{E}(\chi) = \mathcal{E}_{\rm min} + J\, \Big(\frac{n_{\rm min}}{n_0}\Big)^\gamma\frac{(1-\chi)^2}{(1+\chi)^2}
\nonumber\\
& + \frac34 (3\pi^2n_{\rm min})^{1/3}\frac{\chi^{4/3}}{(1+\chi)^{4/3}}\approx\mathcal{E}_{\rm min}
\nonumber\\
& + 7.3\,{\rm MeV}\frac{(1-\chi)^2}{(1+\chi)^2} +149\,{\rm MeV} \frac{\chi^{4/3}}{(1+\chi)^{4/3}} \,,
\label{E-chi}
\end{align}
where $n_{\rm min}=0.85\times 10^{-1}\,n_0$ is used. If we take $\chi\sim 0.2$ we obtain
$\mathcal{E}(\chi\sim 0.2)\approx \mathcal{E}_{\rm min} + 16.9\,{\rm MeV}$. Minimizing $\mathcal{E}(\chi)$ with respect to $\chi$ we find that the minimum is realized for $\chi_{\rm min}\approx 3.1\times 10^{-3}$ and in the minimum $\mathcal{E}(\chi_{\rm min})\approx \mathcal{E}_{\rm min} +7.3\,{\rm MeV}$. Thus if $\mathcal{E}_{\rm min}$ is of the order $-(10-17)$\,MeV in the case of the isospin asymmetric nuclear matter, one may rise the question about a possibility of the existence of neutron-rich dilute compact stars glued by an collective scalar field. However, the above estimates crucially depend on the value of symmetry energy of nuclear matter at small densities. The smaller $E_{N,{\rm sym}}$ is, the smaller will be repulsive contributions in Eq.~(\ref{E-chi}).

Above we only demonstrated a possibility of the scalar field condensation in the dilute nuclear systems, which may result in occurrence of new metastable or stable states. A more detailed analysis is still required to justify this hypothesis.

\subsection{Bose condensates of nuclear clusters}

At densities when $f_0(n)<-1$ the incompressibility of nuclear matter is negative and, as we have mentioned, not only  the scalar particle-hole mode condensati on  may occur but also the spinodal instability of matter. In the low energy heavy-ion collisions the  latter instability manifests itself in an abrupt increase of the production of $\alpha$ particles and other nuclear clusters, see \cite{Schulz1982,Curtin1983,BS1983,SVB1983,MV2019,Borderie}.
With a binding energy of 7.07\,MeV per nucleon $\alpha$ particles are bound as strongly as the most tightly bound nucleus of $^{52}$Fe. Under certain conditions $\alpha$ particles can form a stable Bose-Einstein condensate state (see Ref.~\cite{Typel2010,Ropke:1998qs,Schuck2016} and references therein).
Reference~\cite{Clark2023} argued that the spectrum of excitations in the system of interacting $\alpha$ particles is similar to the spectrum of He$^4$ and has the roton minimum at $k\neq 0$. If so, the inhomogeneous $\alpha$ condensate occupying the roton minimum could occur in the moving (e.g. rotating) pieces of the $\alpha$ particle system. Similar effects were studied in Refs.~\cite{Voskresensky1993,Voskresensky2023,KV2017}, see discussion in Sect.~\ref{Sec-rot-nucl-syst}. Also, one considers the possibility of deuteron condensation in dilute nuclear systems as discussed in Ref.~\cite{Lombardo2001}, despite the fact that deuterons are seven times weakly  bound than $\alpha$ particles.

Consider an extended nucleon system of large size and study whether it is possible to form a Bose-Einstein condensate from bosonic clusters consisting of an even number of nucleons. We must take into account an electric field and $\beta$ equilibrium conditions, as it was done in Ref.~\cite{Voskresensky1977,Migdal:1990vm} and discussed above in Section~\ref{sec:Anomalous}.

Let $A$ is the total number of nucleons in a piece of dilute nuclear matter containing the Bose-Einstein condensate of clusters consisting of
$A_{\rm cl}$ nucleons each (e.g. $A_{\rm cl}=4$ for the $\alpha$ particle). The cluster condensate is described by a complex field $\phi$. The Lagrangian density of a stationary, homogeneous condensate field $\phi$ can be written as
\begin{align}
\mathcal{L}_{\rm cl} = (\omega +V)^2|\phi|^2 -m^2_{\rm cl}|\phi|^2+n_e V-\frac{V^4}{4\pi^2}\,,
\label{Lag-cluster}
\end{align}
where $\omega +V$ is the energy of a positively charged cluster quasiparticle shifted by the electric potential $V$.
Here, considering a large system, we dropped gradient terms with $\nabla \phi$ and $\nabla V$ and neglected $\phi$ self-interaction terms for simplicity, similarly to that we have done finding solutions in Sects.~\ref{nutshell-sec} and \ref{Sec-rot-nucl-syst}.
The equation of motion for the scalar field following from the Lagrangian (\ref{Lag-cluster}) is $(\omega +V)^2\phi=m^2_{cl}\phi$.
The condition of the charge neutrality of the system requires that the positive charge of clusters is compensated by the charge of electrons, therefore $-n_{\rm cl}=n_e=V^3/(3\pi^2)$, where the number density of cluster quasiparticles is
$$
n_{\rm cl}= \frac{\partial \mathcal{L}_{\rm cl}}{\partial \omega} =-2(\omega +V)|\phi|^2\,.
$$

The energy density of a cluster is $A_{\rm cl}\big(m_N - \mathcal{E}_{\rm cl}^{\rm (bind)}\big) n_{\rm cl}$,
where $\mathcal{E}^{\rm (bind)}_{\rm cl}>0$ is the cluster binding energy per nucleon, e.g., $\mathcal{E}_{\rm cl}^{\rm (bind)}\simeq 7.07$\,MeV for the $\alpha$ particle. Then the energy density of the condensate of the bosonic clusters can be presented as
\begin{align}
\Delta E &= E_{\rm tot}- A_{\rm cl} m_N n_{\rm cl}\simeq \frac{V^4}{12\pi^2} - A_{\rm cl}\mathcal{E}_{\rm cl}^{\rm (bind)} n_{\rm cl}
\nonumber\\
&=
\Big(\frac14 (3\pi^2n_{\rm cl})^{1/3}-A_{\rm cl} \mathcal{E}_{\rm cl}^{\rm (bind)}\Big)n_{\rm cl} \,.
\end{align}
Thus, in this simplified model, the system is bound ($\Delta E<0$) at densities
$$n<n_{\rm Mott}= (4A_{\rm cl}\mathcal{E}_{\rm cl}^{\rm (bind)})^3/(3\pi^2)$$
 in the large-system limit, and it is unbound for $n>n_{\rm Mott}$.
For $\alpha$ particles we find $n_{\rm Mott} \simeq  0.04\,n_0$.

\subsection{Dilute nonequilibrium ``anomalous'' states}

The production of pion-enriched matter occurs in ultrarelativistic heavy-ion collisions at highest RHIC and LHC energies. At  expansion of the system, the temperature of the pion gas  decreases. At a certain temperature, $T_{chem}$,  the particle populations stop changing, reaching a state called chemical freeze-out, though the particles still remain in the thermal equilibrium. Then, at a later stage of the expansion up to a thermal freeze-out,  till $T(t)>T_{th}$, we are dealing with a system with a dynamically fixed number of pions. At $T<T_{BEC}$,
provided $T_{th}<T_{BEC}$, there may appear the Bose-Einstein condensate of pions characterized by the dynamically fixed particle number
~\cite{Voskresensky:1994uz,Kolomeitsev:2019bju}.

Another interesting situation is possible: in peripheral heavy-ion collisions, the colliding nuclei can be considered as almost freely interpenetrating beams of nucleons. At the collision energy $\gsim 300$\,MeV per nucleon, the nucleon Fermi spheres from each beam do not overlap in momentum space. Since at low densities, the main contribution to the pion polarization operator is $\Pi_{\pi NN}\propto -\nu p_{\rm F}(n)k^2$, $\nu p_\rmF\propto (n\nu^2)^{1/3}$, see Sect.~\ref{sec:Delta-sec}. This corresponds to a density, which is effectively four times higher than in the case of overlapping Fermi spheres. Thus, there is a possibility of observing the effects of the p-wave pion condensation in peripheral heavy-ion collisions~\cite{Pirner:1994tt,Voskresensky:2025gjy}.

\subsection{Dilute quantum droplets}

The phenomenon of Efimov states in Fermi systems opens another interesting possibility. For particular values of the diluteness parameter $n|a|^3$, where $n$ is the fermion density, $a$ is the scattering length, a three-body interaction can provide a dominant contribution in the energy-density functional~\cite{Bulgac2002}. if this happens, both bosonic and fermionic systems of various sizes and densities could become self-bound. In this relation, we note that the nucleon-nucleon scattering length has an anomalously large magnitude, $a_{NN}\simeq -20$ fm. A discussion of the role of this circumstance for description of a dilute neutron matter can be found in \cite{Kolomeitsev:2022rjk}. Finally, we notice that some experiments claimed an observation of a tetraneutron ($n^4$) signal, which may be associated with a bound state or an unbound resonance state, see discussion in~\cite{Faestermann:2025our}.

As we have mentioned, in the spectrum of $\pi^+$ in neutron matter  in the interval of densities  $n_c^{\pi^\pm}>n>n_c^{\pi^+}$, where  $n_c^{\pi^+}<n_0$ according estimate of Ref.~\cite{MMM1974}, and $n_c^{\pi^\pm}\gsim n_0$, there appears the branch with $\omega^{\pi^+}_s (k)<0$ at $k\neq 0$, as we have discussed in Sect.~\ref{cond-sec}. Appearance  of such a branch may result in some important consequences, see Ref.~\cite{V1993}. For instance, at the formation of the neutron star in a supernova explosion a part of the initial   angular momentum can be absorbed by the $\pi^+$ condensate subsystem. Metastable $\pi^+$-multi-neutron droplets   with a large angular momentum and a density $n>n_c^{\pi^+}$ can be formed, as it was discussed in Sect.~\ref{Sec-rot-nucl-syst}. They may take away part of the total angular momentum of the star. Metastable $\pi^+$ condensate neutron-rich droplets also  can be formed with some probability  during heavy-ion collisions. Since density in such droplets might be smaller than $n_0$ they may behave as anomalons  mentioned above, see Ref.~\cite{V1993}. Also there a possibility of appearance of condensate fragments with a large angular momentum in peripheral heavy-ion collisions was discussed.

\section{Observations}\label{sec:observation-sec}

Search of abnormal nuclei, nuclearites and nuclei-stars composed of absolutely stable matter is an intriguing prospect in astrophysics. Detection of
such objects is not a trivial problem due to numerous observational signatures, which are superimposed on those of ordinary neutron stars and white
dwarfs. Distinguishing features may be found in ultra-fast rotation (well below 1\,ms), star radii less than 9-10\,km, specific photon emission from a star surface, etc. Possible candidates among compact objects were identified: RX~J1856.5-3754, PSR~J1614-2230, PSR~J0348+0432, 4U~1820-30,  HESS~J1731-347, XTE J1814-138. However, definitive conclusions whether these objects are anomalous or not, could not be drawn. Additionally, particular measurements of the low-mass and small-radius compact star HESS J1731-347 depend heavily on the assumed atmosphere model and are sensitive to errors in distance and inferred age. Let us add here the latest results of NICER \cite{Mauviard:2025dmd} for PSR J0614-3329 giving  rather small equatorial radius  $10.29^{+1.01}_{-0.86}$\,km for the mass $M = 1.44^{+0.06}_{-0.07}\,M_{\odot}$ (median values with equal-tailed $68\%$ credible interval).

Experimental search for abnormal nuclei was carried out in natural samples among the products of interactions between high-energy particles and matter, in fission products, in heavy-ion collisions, etc., see Refs.~\cite{Frankel:1975fa,Holt:1976af,Aleshin:1976ww,Kulikov:1976bw,Price1975,Frankel1976,Karnaukhov1977,Avdeev:1982xd,Anikina:1983ypp,Anikina:1983ypp,Avdeyev1988,Buchwald}.
Reference \cite{Jacobs:2014yca} accounted for the absence of traces found in the sample previously considered in Ref.~\cite{Price:1986ky} and excluded strangelets with a mass $M<55$\,g. More details can be found in \cite{Clemente2025}.

Although no abnormal states were found, presently a number of observations demonstrate
anomalies that defy conventional physics interpretations. May be at least some of them are related to the anomalous nuclear states. We mention several such anomalies. A very heavy particle was detected passing through a balloon-borne stack of Cherenkov film, emulsion, and Lexan sheets~\cite{PriceShirk1975}. It could be a neutron anomalous nucleus. The spectra of nuclei obtained in the aerostatic JACEE experiment with an emulsion chamber show a peculiar bend at the cosmic ray energy of 3\,PeV. The change in the slope of the spectrum was suggested in Ref.~\cite{Shaulov2022} to be explained by the appearance of an anomalous component of cosmic rays, which is much heavier than the ordinary nuclear component and could be associated, e.g., with quark nuggets or abnormal $\pi$ or $\sigma$ condensate nuclei. In the transition region, both nuclei and strangelets can exist, that leads to large fluctuations in the events observed in the experiments with X-ray emulsion chambers.

Recent works~\cite{Bertolucci2017,Zioutas2020,Zioutas2023} argued that there are other observations, which can be hardly explained in the framework of conventional physics. These anomalies include: unexpected correlations with temperature variations in stratosphere, the total electron content of the Earth atmosphere, correlations between earthquake activity and relative positions of planets, unexpected seasonal stratospheric temperature variations,
ionospheric anomalies, and their correlations with seismic activity, as well as solar phenomena like the coronal heating problem, the origin of sunspots, and the trigger mechanism of the solar flares, see also \cite{Maroudas2022,Zioutas2022}.
The statistical significance of the observed correlations is high. Reference \cite{Zhitnitsky:2024ydy} suggested that in mentioned events one may deal with  hypothetical very dense and microscopically large composite objects with the mass of the order of grams and sub-micrometer size consisting of nuggets of strongly interacting matter. For instance, it was suggested to look for so-called solar nano-flares with the axion quark nugget annihilation events in the solar corona. {{Note that, if instead of quark nuggets one considered the Lee-Wick abnormal states or the Migdal pion condensate  abnormal nuclei, at least a part of physical consequences would be similar to those for strangelets.}}

%%%%%%%%%
We now mention another possibilities.
In a uniform electric field with the strength of $\gsim (10^4-10^5)$\,V/cm
the depth of the potential $V_0$ exceeds $2m_e\approx 10^6$\,eV when the distance between plates is  $\gsim 10$\,cm. For  a 300\, times larger distance between plates $V_0$ exceeds two pion masses, $2m_\pi\approx 280$\,MeV. For comparison, the breakdown voltage for the dry air at atmospheric pressure is $(2\mbox{--}3)\times 10^4$\,V/cm and for some types of glasses it reaches $3\times 10^6$\,V/cm. Similar estimates can be found for the cylindrical and spherical capacitors.
However we must stress that despite the potential inside the capacitor may readily exceed $2m_e$  and $2m_\pi$ respectively, the probability of the spontaneous  pair production from the vacuum is exponentially small in both cases, $\propto e^{-m^2/E}$. Nevertheless, electrons and charged pions and, maybe, kaons can be produced with the larger probability near the walls of the empty capacitor or in reactions, if capacitor is filled by a medium, provided the electric potential well is sufficiently deep. In the broad potential well for $\pi^-$ in the region, where $-V>m_\pi$, the $\pi^-$ ground state energy level crosses $\om =0$ and  formation of the static  $\pi^-$ condensate field  becomes energetically favorable. In the region with $V>m_\pi$ the static $\pi^+$ condensate field can be formed, if the time needed for the formation of the condensate is sufficient. Note that in the presence of a charged condensate, photons become effectively massive particles via the Anderson-Higgs mechanism. Therefore, in the case of empty capacitor the condensate region should look dark. This circumstance can help in identifying the effect.

It is worth noticing that deep potential wells can be formed in thunderstorms on Earth. During tens of minutes before a discharge, the electric voltage may reach values of $\gsim 10^8$\,V at distance $\gsim (1-10)$\,km  between clouds, cf.~\cite{Marshall2001}. Hence the typical potential well reaches values  $V_0\gg m_e$ at such distances. An excess of neutrons during thunderstorms has been indeed detected, however it was associated with photonuclear reactions~\cite{Babich}.
Note that pions, which could form a condensate,  can also be produced in photonuclear reactions. Strong thunderstorms may also occur in atmospheres of other planets such as Jupiter and Venus. If a  nugget of a new phase of the matter passes through the thunderstorm clouds at the ultra-high voltage ($\sim 10^9$\,V) between them, the nugget will be accelerated and its electron shell will be destroyed with a release of the sufficient energy for the production of baryons, pions, quarks, etc. In these processes the nugget can be destroyed that could be observed via the significant energy release.

We should notice that 10 anomalous bursts correlated with lightnings have been observed in thunderstorms occurred during  5 years~\cite{Telescope,Okuda}. The estimated energy from individual events within the bursts is five to six orders of magnitude higher than the energy estimated by an event rate. The multiple air showers occurred within 1\,ms, defying explanation by conventional high-energy cosmic rays. Similar ``exotic events'' were recorded by the AUGER collaboration. The ANITA experiment detected two anomalous upward-propagating events with non-inverted polarity. The list of anomalous atmospheric events also includes the Multi-Modal Clustering Events observed by HORIZON 10T.
Among various possible explanations we indicate the possibility of the energy release due to the destruction of the previously bounded nuggets arrived from cosmos.

\section{Conclusion}\label{conclusion-sec}

In 1971  Migdal suggested a possibility of existence of metastable or stable abnormal pion condensate nuclei and Bodmer assumed possibility of collapsed quark nuclei. In 1974 Lee and Wick proposed scalar (sigma) condensate abnormal nuclei. In 1984 Witten suggested quark nuggets. These pioneering suggestions paved the way for numerous subsequent theoretical and experimental investigations of various anomalous nuclear objects. New more detailed models were constructed. The review is dedicated 115-year  anniversary of A. B. Migdal  and 100-year anniversary of T. D. Lee.  Our aim was to present the development of their ideas and discuss the modern status of the subject, demonstrating the interrelation between the different phenomena under discussion.

In Sect.~\ref{sec:Lee-Wick} we reviewed the original Lee-Wick model of scalar condensate abnormal nuclei. The model describes the coupling of nucleons with a scalar mean field. Typical behaviors of the effective potential and the energy of the model as function of a scalar field $\phi$ are shown in Fig.~\ref{fig:Lee-potential} and Fig.~\ref{fig:DeltaE-n}. The role of the scalar field can be played by the $\sigma$ meson and the model then can be treated as a $\sigma$ model permitting chiral symmetry breaking in a dense nuclear matter.
The same year when Lee and Wick proposed their model, Walecka formulated a relativistic mean-field  $\sigma-\omega$ model~\cite{Walecka1974,Walecka1975} with two coupling constants fitted to describe the nuclear binding energy and the value of the nuclear saturation density. Various modifications of the Walecka model were introduced for the better description of nuclear matter properties. The density dependencies of the effective nucleon mass within the Lee-Wick model and in the  RMF model are shown  in Fig.~\ref{fig:Lee-mass}. In the Lee-Wick model the effective nucleon mass undergoes a jump to a tiny value at the critical density, whereas it decreases smoothly with the density increase in the standard RMF models. With parameters, which were fitted to describe the nuclear system at $n=n_0$, at larger densities  in these RMF models   the Lee-Wick abnormal state does not appear. However  information about chiral symmetry is lost in the standard RMF based models, whereas it is present in the Lee and Wick model. Although  not permitting to describe properties of the ordinary atomic nuclei and nuclear matter at $n\lsim n_0$, the Lee-Wick model demonstrates  a principle possibility  of existence of the new stable (or metastable) abnormal nuclear state at $n$ significantly above $n_0$. As a possibility to reconcile these approaches, we indicated the RMF models employing the $\sigma$ field scaled hadron masses and coupling constants~\cite{KV2005,KTV2007,KTV2008,MKVPhysRev2016,MKV2016}, which can be treated as models using the idea of the partial chiral symmetry conservation.  For simplicity the parameters of the models used in \cite{KV2005,KTV2007,KTV2008,MKVPhysRev2016,MKV2016} were chosen such that the new Lee-Wick minimum is not realized.  Changing parameters of the models we could get new solutions, some of them with a jump to a low value of the effective nucleon mass at a critical density. Such a possibility is realized in the model \cite{KMV2017} at some choice of the attractive $\Delta$ optical potential, see Fig.~\ref{fig:mkvstar-meff} in Sect.~\ref{sec:Delta-sec}. This question should be more carefully studied in the future.

Section~\ref{sec:Anomalous} considered various aspects of the model of supercharged nuclei, nuclearites and nuclei-stars glued together by light charged bosons, if the latter existed. In reality the role of the light charged boson can be played by the ordinary charged pion since in the nuclear medium and in external fields the pion effective mass can be essentially decreased. For example, it was demonstrated that a rapid rotation of a nugget acts in favor of the decrease of the effective pion mass and appearance of the stable state.

Section \ref{sec:Superheavy} analysed the behavior of the pion degree of freedom in dense nuclear matter and the possibility of the Migdal's p-wave pion condensation, cf. Ref.~\cite{Migdal:1990vm}. Equation (\ref{gam-res}) is the final result of the resummation of the nucleon-nucleon interaction amplitude in the spin-isospin channel in the dense nuclear matter done within the Fermi liquid approach of Migdal allowing for the explicit treatment of soft (quasi-Goldstone) pion degree of freedom. For $n>n_0$ the in-medium pion exchange term in (\ref{gam-res}) becomes  dominant and the nucleon-nucleon scattering cross section appreciably increases with increasing density towards the pion condensation point, see Eq.~(\ref{cross}). The $\Delta$ isobar plays very  important role because the mass of the $\Delta$ isobar is only $2m_\pi$ larger than the nucleon mass and degeneracy factor is 4 times larger. Moreover the pion-Delta-nucleon  attraction coupling constant is twice larger  than that of the pion-nucleon-nucleon one.

Various types of the p-wave pion condensation ($\pi^+$, $\pi^{\pm}$, $\pi^0$) as well as a possibility of the s wave pion condensation were mentioned.  The typical effective pion mass (better saying the effective pion gap) in the dense medium depends on the pion momentum and decreases with increasing baryon density. It is shown in Fig.~\ref{piongap1}. The description of the pion degree of freedom can be formulated within the linear $\sigma$-model \cite{Campbell:1974qt,Campbell:1974qu,Baym:1975tm,Voskresensky:1977mz,Migdal:1990vm,Voskresensky:1982vd}, in which both the pion and sigma field condensations can be treated simultaneously. Also the $\sigma (n)$ dependence is related to the change of the quark condensate in baryon matter. The possibility of self-bound pion condensate abnormal nuclei, nuclearites and nuclei-stars found within the chiral $\sigma$-model of pion condensation used in Ref.~\cite{Voskresensky:1977mz} is demonstrated in Fig.~\ref{fig:spicond}.

Section~\ref{sec:Delta-sec} reviewed the idea of $\Delta$ resonance matter. Occupation of the new $\Delta$ Fermi sea with increasing baryon density occurs within many models for the equation of state of neutron star matter already at $n\gsim (2-3)n_0$. Also, $\Delta$ isobars
may appear in symmetric nuclear matter. Appearance of $\Delta$ isobars results in an energy decrease making the nuclear matter more stable. It  may result in a metastable or even stable ground state of the $\Delta$ resonance matter if the attractive $\Delta$ optical potential is sufficiently strong. Fig.~\ref{fig:mkvstar-meff} (left) demonstrates an abrupt decrease of the nucleon effective mass to a small value occurring  in the isospin-symmetric matter at a critical density $n\gsim (3-4)n_0$ in a RMF model~\cite{MKV2016} employing the $\sigma$ scaled hadron masses. We note that such a behavior is also typical for the Lee-Wick model.

Section~\ref{sec:Strangelets-sec} reviewed ideas of Bodmer, Witten and others on quark strangelets or nuclearites and strange stars as stable baryonic states composed of deconfined quarks. These states are interconnected with nontopological solitons containing baryon constituents studied by Lee and collaborators as well as by Coleman~\cite{Sirlin,Coleman1986,Lee1987,Lee1989,Lee1992}. The idea of stable strange quark matter by Witten 1984~\cite{Witten1984} is somewhat similar to the idea of stable $\Delta$ resonance matter by Troitsky and Chekunaev 1979~\cite{Troitsky:1979ch}. If a model predicts the strange quark matter to be more stable than an iron nucleus, one speaks about strangelets and quark nuclearites, provided the baryon number $A\ll 10^{56}-10^{57}$, and about strange stars for $A\sim 10^{56}-10^{57}$. In the latter case the gravity essentially contributes to the binding. The surface layer of these objects has a typical length of $\sim 10$\,fm. Some other models do not allow for the absolutely stable strange quark matter but allow for metastable matter at switching off the gravity. Other models use  parameters that render strange and not strange ($u,d$ quark) deconfined matter unstable. Nevertheless, we may deal with the quark matter in interiors of compact stars named hybrid stars. All hybrid stars have macroscopic hadron shells.

Various models describing quark deconfined state were reviewed. Among different models we indicated the heavy quark quasiparticle model  \cite{Ivanov:2005be} capable to describe the lattice data~\cite{Fodor:2002sd,Csikor:2004ik}. At some choice of the parameters the model \cite{Ivanov:2005be}  allows for the stable quark matter but does not permit the hybrid stars.

At $T<T_\Delta$ for the quark-quark pairing the quark matter should form a color superconductor. We briefly discussed this phenomenon. We focused on a possible manifestation of the pion and $\sigma$ condensate nuclei-stars and strange stars via the measurement of the $M-R$ relation and via the cooling of these stars.

Section~\ref{sec:dilute-sec} discussed the possibility of the existence of very dilute stable or metastable nuclear systems. For this the Fermi liquid Green's function approach was employed as  for the description of dense nuclear systems. However, in the case of pion-condensate nuclear objects we used nucleon-nucleon interactions in the spin channel (or in spin-isospin one), whereas in Sect.~\ref{sec:dilute-sec} we focus on the scalar channel of an $NN$ interaction. In this channel, the interaction is expressed via the scalar Landau-Migdal parameter $f_0(n)$. First we showed that in case of approximately isospin-symmetric matter at low densities we have $f_0(n)<-1$, and there appears the Pomeranchuk instability. Then we argued in~\cite{KV2016} that this instability may result in the formation of a static scalar-field condensate (compare with the Lee-Wick scalar field condensation). We demonstrated that a decrease in the energy due to the condensation can be sufficient to allow for  the existence of very dilute metastable or even stable abnormal dilute nuclei held together by the condensate of a scalar field, see Fig.~\ref{fig:SC}.

Then we studied if it is possible to form a Bose-Einstein condensate from bosonic clusters consisting of an even number of nucleons, e.g. of the $\alpha$ clusters or deutrons. We showed that the charged Bose cluster condensates can be described by a complex field $\phi$ similarly to the description of supercharged nuclei performed in Sect.~\ref{sec:Anomalous}. We estimated the energy for the dilute $\alpha$ condensate extended systems, which charge is compensated by the  electrons. For $n>n_{\rm Mott}\simeq 0.04n_0$ according to our estimate the system is destroyed.

Finally we made some speculations about formation of metastable pion Bose-Einstein condensate during the course of the heavy-ion collisions.
Section~\ref{sec:observation-sec} discussed the situation with observations of exotic phenomena considered in the review.

Concluding, in spite of more than fifty years were past, as in early days of this field, there remain essential uncertainties in theoretical predictions due to a lack of our knowledge of the behavior of the strongly interacting dense and dilute nuclear matter. In spite of abnormal nuclear states were not   observed, the number of observations grows, which are hardly  explained by conventional physics. Some of such events could be associated with manifestation of anomalous nuclear systems. Let us be patient. Time flows not in vain.

\section*{Acknowledgment}

EEK acknowledges the support by the Slovak grant VEGA~1/0521/22
and the support within the framework of the scientific program of the National Center for Physics and Mathematics in Sarov, topic No. 6 "Nuclear and Radiation Physics" (stage 2023-2025).

\bibliography{refs-symmetry.bib}
\end{document}